\documentclass[aps,pre,twocolumn,showpacs,amsmath,amssymb,amsfonts]{revtex4}
\usepackage{epsfig}
\usepackage{graphicx}
\usepackage{dcolumn}\usepackage{braket}
\usepackage{bm}
\usepackage{soul, xcolor}
\usepackage{xcolor,cancel}
\usepackage[titletoc,title]{appendix}

\usepackage{color}
\usepackage[normalem]{ulem} 

\begin{document}
\setstcolor{red}
\title{Density and tracer statistics in compressible turbulence: phase transition to multifractality}

\author{Itzhak Fouxon}\email{itzhak8@gmail.com}
\author{Michael Mond}\email{mondmichael@gmail.com}
\affiliation{Department of Mechanical Engineering, Ben-Gurion University of the Negev,
P.O. Box 653, Beer-Sheva 84105, Israel}

\begin{abstract}

We study the statistics of fluid (gas) density and concentration of passive tracer particles (dust) in
compressible turbulence. We raise the question of whether the fluid density which is an active field that reacts back on the
transporting flow and the passive concentration of tracers must coincide in the steady state, which we demonstrate to be crucial
both theoretically and experimentally. The fields' coincidence is provable at small Mach numbers, however at finite Mach numbers
the assumption of mixing is needed which we demonstrate to be not self-evident due to the possibility of self-organization. Irrespective of whether the fields coincide
we obtain a number of rigorous conclusions on both fields. As ${\rm Ma}$ increases from small or moderate values the density and the concentration in the inertial range go through a phase
transition from a finite continuous smooth to a singular multifractal spatial distribution. The concept of sum of Lyapunov exponents in the multifractal phase is generalizable to the inertial
range implying a singular tracers' distribution. We discuss various concepts of multifractality and
propose a way to calculate fractal dimensions from numerical or experimental data. We derive a
simple expression for the spectrum of fractal dimensions of isothermal turbulence and describe limitations of lognormality. The expression
depends on a single parameter: the scaling exponent of the density spectrum. We propose a mechanism for the phase transition of concentration to multifractality. We demonstrate that the pair-correlation function is invariant under the action of the probability density function of the inter-pair distance that has the Markov property. This implies applicability of the compressible version of the Kraichnan
turbulence model. We use the model to derive an explicit expression for the tracers pair correlation that demonstrates their smooth transition to multifractality and confirms the transition's mechanism. The obtained fractal dimension explains previous numerical observations.
Our results are of potentially important implications on astrophysical problems
such as star formation as well as on technological applications such as supersonic combustion. As an example we demonstrate strong increase of
planetesimals formation rate at the transition.

\end{abstract}
 \maketitle
\section{Introduction}

Supersonic turbulence plays a crucial role in such processes of fundamental nature as star formation in dense molecular clouds, density fluctuations in the solar
wind as well as in advanced technological applications such as combustion processes in scramjets and drag and stability of supersonically moving bodies in the terrestrial
atmosphere. This type of turbulence differs qualitatively from its incompressible counterpart due to the presence of a unique supersonic inertial range. The velocities that
characterize the turbulent eddies in this range are larger than the speed of sound so that their compressible and incompressible components are strongly coupled. This causes a transition to a new regime: above a critical
Mach number (defined as the ratio of the rms velocity of turbulence and the speed of sound), which is not much larger than one, the fluid density in the supersonic inertial range becomes a singular field supported on a multifractal set. Thus the five fluid mechanical fields of mass, momentum and energy densities are
all supported on the multifractal producing a singular fluid mechanical problem.

The picture becomes even more convoluted and perplexing when the compressible turbulence is seeded with passive tracer particles. These are particles that are carried by the flow but have no back reaction on it.
Since the tracers and the gas satisfy the same continuity equation it is tempting to assume that both settle on the same structure in space and actually information concerning the latter's spatial distribution may be inferred
from following the former. If so then we get insight in the dust-to-gas ratio of molecular clouds: approximating dust particles by tracers, which neglects their inertia and other effects \cite{dust}, the ratio is constant. This ratio plays key role in many astrophysical applications including absorption of light in the interstellar medium, evolution of galactic composition and ISM tracking \cite{lhs}. Its non-constancy can have far-reaching consequences \cite{dust}.

The equality of density and concentration however deserves a critical examination (herafter concentration $n$ refers to tracers and density $\rho$ to the fluid). The fluid density is an active scalar field that reacts back on the transporting flow. For instance this implies that the density and its gradients must be finite. The finiteness entails degeneracy of the flow where the sum of Lyapunov exponents, defined as the average logarithmic growth rate of infinitesimal volumes of the fluid, is zero, see \cite{review} and below. In contrast, the concentration is a passive field the values of whose gradients are unrestricted. Many examples of striking difference between active and passive scalars are known, see e. g. \cite{review,actpass} and below. That there can be a difference in our case is probably seen most easily by representing the fluid as a collection of a large number of discrete fluid particles, see \cite{monag,difference} for smoothed particle hydrodynamics (SPH). The particles move with the local flow velocity however their interact via their back reaction on that velocity. Particles' configurations are then possible that would become stable due to the interaction and would not be stable for non-interacting tracers (self-organization), cf. \cite{legras}.
It must be stressed that the fields' difference is a real, so far disregarded, possibility: in all chaotic systems besides the natural measure (steady state concentration) there exist other steady state solutions of the continuity equation that are reached from initial conditions of measure zero \cite{grass} and could represent the density. Thus a careful consideration of the fields' equality seems necessary.
Here we undertake this consideration, deriving a number of rigorous results on both the density and the concentration. These results hold irrespective of the coincidence of density and concentration in the steady state and are of their own interest. We demonstrate that despite that one can get close to demonstrating $n=\rho$, the complete proof is elusive. There is no uniqueness of steady state solutions of the continuity equation, see above, and the mixing cannot be taken for granted (the question of mixing of density that reacts back on the flow seemingly was not considered in mathematical literature so far). We propose to compare numerically the statistics of $n$ and $\rho$, provide a number of testable predictions and describe the subtle points that could bring erroneous conclusions in future studies.

The currently existing numerical data, inconclusive though, indicates a finite difference between the fields. Spatial distribution of tracers for three different values of the Mach number ${\rm Ma}$ was studied in \cite{grauer}. Multifractal distribution in the inertial range of turbulence was conclusively found at ${\rm Ma}=4.6$. At the same time the observations of \cite{krit2007} revealed density spectrum decay that is faster than $k^{-1}$ at ${\rm Ma}=6$. This decay corresponds to non-fractal distribution in space as we prove in detail below. Thus seemingly multifractality can hold earlier in ${\rm Ma}$ for the concentration than for the density. This implies a finite difference of $n$ and $\rho$. The difference was observed also in the simulations of \cite{difference} where it was attributed to numerical artefacts.

In this paper we present theoretical results that indicate that concentration must be multifractal under the conditions of the simulations of \cite{krit2007}. We demonstrate in
three different ways that if the scalings of the solenoidal and the potential components of the velocity are approximately equal then the concentration is multifractal. In fact, this seems necessary: multifractality implies scaling which demands scaling of the transporting flow. This provides a simple view of the concentration's transition to multifractality. The scalings' difference is of order one at ${\rm Ma}\lesssim 1$ where the concentration is a large-scale nonfractal field which is approximately constant in the inertial range. The Mach number's increase closes the scalings' gaps, see e. g. \cite{krit2007}, resulting in a multifractal distribution (the Mach number at which the scalings become identical with considered accuracy will be designated below by ${\rm Ma}_s$). This seems to be the situation of \cite{krit2007} where the components' scalings are very similar. Unfortunately the presently existing data does not seem conclusive. Our considerations here provide useful guidance for the future studies by providing measurable predictions that allow to resolve the question of equality of density and concentration experimentally. It seems that if there is a difference then it would not be large.
For instance we manage to pinpoint which properties of the flow determine the fractal correlation dimension of the concentration. We propose that similar properties determine the dimension of the density and demonstrate that this explains the observed dependence of the fractal dimension of the density on the compressibility of the forcing \cite{comprfr}.

We remark that the question of the difference between the fluid density and the concentration of passively transported particles was previously raised in \cite{dust}. In that case however the passively transported dust particles detach from the carrying flow due to inertia and their concentration does not obey the same continuity equation as the fluid density. It was observed that dust particles form a multifractal that is more rugged than that formed by the gas particles, see however \cite{crit}.

Density multifractality is a qualitative property of supersonic compressible turbulence that holds above a critical Mach number ${\rm Ma}_{cr}$ (the question of whether ${\rm Ma}_{cr}={\rm Ma}_s$ is one of the main questions raised here). The phenomenological theory of compressible turbulence existing today has been constructed in \cite{fleck96} on the basis of \cite{von}. It assumes that density fluctuations grow downscales as a negative power law. This growth is assumed to be self-similar and thus governed by a single scaling exponent $\alpha$ which is the fitting parameter of the theory. This assumption is true for fractal but not multifractal support of the density. A kinetic energy cascade has been further assumed \cite{lighthill} which implies that velocity weighted by the density in power one third behaves as the velocity of incompressible turbulence. Indeed, it was observed \cite{krit2007,kri2009} that for the weighted velocity the third order structure function scales linearly with the distance and the spectrum is close to the spectrum of incompressible turbulence \cite{kaneda} if the forcing is solenoidal \cite{fedor2013}. However, using these assumptions, $\alpha$ derived from the scaling of the third order structure function of the velocity would be different by about factor of two from that derived from the scaling of the first order structure function \cite{krit2007}. Introducing the notion of multifractality of the density, due to which $\alpha$ is non-constant and fluctuates in space, is therefore a necessary step towards unravelling the complex nature of supersonic turbulence.

Despite the crucial role it plays in compressible turbulence there is currently no consistent description of the multifractal properties of the gas density and tracers concentration. It is unknown what determines
the critical Mach number for the transition to multifractality and whether it is independent of the degree of compressibility of the stirring force (universality). Moreover, the physical mechanisms that lead to the transition to multifractality are currently not well understood. The present work aims at filling this gap. To this end we present a thorough discussion of the spectrum of the multifractal dimensions of density and concentration distributions and propose a viable method to calculate it from numerical and experimental results. In particular, for density of isothermal turbulence, which is characterised by a log-normal probability distribution function, we derive an explicit expression for the multifractal dimension spectrum that depends on a single parameter, the exponent of the pair correlation function. In addition, various physical processes and mechanisms that lead to the transition to multifractality are discussed. In particular, we show by employing a heuristic cascade model that as the scalings of the compressive and solenoidal components of the velocity grow closer as the Mach number is raised, the concentration transforms from a large-scale smooth field to a small-scale multifractal.

Carrying out the calculations that are proposed in the current work such as multifractal dimensions and following
Lagrangian trajectories as well as examining the
various mechanisms that are introduced and discussed
in the next sections pose a great challenge due to the
essential inability to obtain analytical solutions of the
Navier-Stokes equations (NSE) in a turbulent flow regime. We demonstrate that in fact this is unnecessary for the study of pair correlations of tracers. 
We demonstrate in Section \ref{mod} that the pair-correlation function of concentration obeys a closed stationarity condition.
This is derived by considering the conservation of the pair correlation function in the steady state, that holds at any times, at large times. This 
brings simplifications. We find that the correlation function must be invariant under the action of integral operator whose kernel is the propagator of the inter-pair distance, that is the transition 
probability for the distance between a pair of particles separating in the supersonic inertial range. Thanks to considering the invariance at large times the distance has the Markov property. Therefore instead of the full solution of the NSE we can use an effective velocity
which dependence on time is described by white noise. This passage is rigorous and parallels the description of complex molecular forces by the Langevin white noise terms in the theory of Brownian motion. If we further make an assumption of, at least qualitative, validity
of the eddy diffusivity approximation, then we find that the propagator can be described using the famous Kraichnan's model \cite{review}. We stress that this is the model 
of turbulent transport valid at large times and not the model of turbulence itself (for models of turbulence see e. g. \cite{gs,cl}). The modelled quantity of the NS flow is the large-time propagator of two tracers and not the NS flow itself which, of course, is not a white noise in time. These are different conceptually. We also observe the transport's robustness: many properties of the transported
fields hold irrespectively of the details of the transporting flow \cite{review}. Here we employ the compressible version of Kraichnan's model in order to shed light on the physical processes that participate in the transition to multifractality. In particular, we show by detailed analytical calculations that the simultaneous convergence of the scalings of the compressible and solenoidal components of the velocity to a single limit does indeed underlie the transition from a large-scale smooth distributions to multifractal measure. Also, we unravel the important role played by the effective attractive force that is generated by negative velocity divergence. Thus, the analytical calculations based on Kraichnan's model reveal that as the Mach number increases the number of times that two tracers collide with each other grows until a critical Mach number is reached for which and beyond the number of collisions is infinite (return with probability one). The relation of this phase transition to the phase transition of the density is yet unclear.

The transition of the tracers distribution within a gas to multifractality has significance to the chemical interaction between a fuel and an oxidant. The process rate depends naturally on the second power of the various relevant tracers' concentrations. Fluctuations of the latter are significantly enhanced due to the transition to multifractality where clusters and filaments of practically infinite concentration are formed. Therefore a dramatic increase in the process rate is expected at the transition. As an example we demonstrate that the transition to multifractality indeed results in a significant decrease of the formation time of planetesimals.

Recent years saw much research of the long-standing problem of supersonic turbulence \cite{fractal,dust,krit2007,kowal,fedor1,kri2009,fedor,ffo,aluie2011,bs1,aluie2012,kritfalk,bs2,aluie2013,bs3,fedor2013,krit2015}.
The progress was previously hampered by the lack of an exact scaling law that would be the counterpart of the four fifths law of incompressible turbulence \cite{frisch}. It was unclear whether the concepts of energy
cascade and approximate self-similarity, that proved indispensable in the theory of incompressible turbulence, can be extended to the compressible case \cite{aluie2011,aluie2012,aluie2013}. Despite the made progress,
see the references above, the problem is still unsolved.

In the next Section we introduce the fluid mechanical equations and their fundamental properties. This includes the study of the consequences of the finiteness of density and its gradients. Section \ref{rel} is devoted to
the study of different arguments that could be used for proving that in the steady state $n=\rho$. In Section \ref{forma} we describe multifractal growth of volumes of tracers in the supersonic inertial range. The growth obeys a power law whose scaling exponent is realization-independent in the limit of large times. This provides a non-trivial extension of the notion of sum of Lyapunov exponents to the rough flow in the inertial range. We introduce the cascade picture of the growth. In Sec. \ref{phen} we describe the evolution of smooth initial conditions to the multifractal steady state distribution. Section \ref{s1} is devoted to the description of the multifractal formalism and R\'{e}nyi dimensions whose use might simplify the measurement of fractal dimensions.  Section \ref{f} derives the spectrum of fractal dimensions of isothermal turbulence. The result is of high interest due to many simulations of this type of turbulence that have been performed. We correct some previous misconceptions. We introduce the cascade model of formation of fluctuation of concentration in Section \ref{cas}. In contrast with similar models for turbulent flow \cite{frisch} the model is close to being rigorous. Fully rigorous theory is constructed in Section \ref{mod}. We derive the stationarity condition on the pair correlation function of concentration, study when it has power-law solutions describing multifractality and relate the scaling of the higher-order correlation functions to the famous zero modes \cite{review,krzysztof}. This study is done in a model-independent way. We use the applicability of the Chapman-Kolmogorov equation for introducing the Kraichnan model in Section \ref{Kraic}. We derive the pair-correlation function and the smooth transition to multifractality. We confirm by concrete calculation the general connection of Section \ref{mod} between the properties of pair dispersion in the supersonic inertial range and the correlation dimension of the multifractal. Section \ref{acceleration} describes the acceleration
of formation of planetesimals due to the transition to multifractality. This parallels similar acceleration phenomenon in the rain formation \cite{nature}. The final Section resumes main results and describes open questions.
Some more detailed calculations are put in the Appendices.


\section{Fundamentals}\label{setting}

The continuity equations for the fluid density $\rho$ and the tracers' concentration $n$ read,
\begin{eqnarray}&&\!\!\!\!\!\!\!\!\!\!\!
\partial_t \rho+\nabla\cdot(\rho \bm v)=0,\ \ \partial_t n+\nabla\cdot(n \bm v)=0, \label{con}
\end{eqnarray}
where $\bm v$ is the same transporting flow that obeys,
\begin{eqnarray}&&\!\!\!\!\!\!\!\!\!\!\!
\rho\left(\partial_t \bm v+(\bm v\cdot\nabla)\bm v\right)=-\nabla p+\rho \bm a+\nabla\sigma'.\label{basic}
\end{eqnarray}
Here $\bm a$ is the (random) driving acceleration and $\sigma'_{ik}$ is the viscous stress tensor. The coupling of the velocity to the density  can be the fifth, energy (or entropy)
equation of fluid mechanics or the barotropic equation of state $p=p(\rho)$ that describes the dependence of the pressure $p$ on $\rho$. We consider stationary homogeneous turbulence.
The equations can be characterized by the dimensionless Reynolds number $Re=UL/\nu$ and the Mach number ${\rm Ma}=U/c_s$. Here $U$ is the rms of the flow where the mean flow is assumed to be zero, $L$
is the scale of stirring force, $\nu$ is the kinematic viscosity and $c_s$ is the speed of sound. The Reynolds number is assumed to be large.
We use units where the volume of the flow is one and also $\langle \rho\rangle=\langle n\rangle=1$. Here and below angular brackets stand for spatial averaging. Thus $\rho$ and $n$ are normalized dimensionless
non-negative functions. This allows to consider them as probability density functions as is useful sometimes.
The case of isothermal fluid,
\begin{eqnarray}&&\!\!\!\!\!\!\!\!\!\!\!
\rho\left(\partial_t \bm v+(\bm v\cdot\nabla)\bm v\right)=-c_s^2 \nabla \rho+\rho \bm a+\nabla\sigma', \label{con1}
\end{eqnarray}
where $c_s$ is constant, presents high interest because of its astrophysical applications, see e. g. \cite{krit2007} and cf. \cite{eyink}. In this case the equations are invariant with respect to rescaling of the density by a constant, cf. \cite{bsk}. The internal energy $E$ is proportional to the Gibbs entropy $S$,
\begin{eqnarray}&&\!\!\!\!\!\!\!\!\!\!\!
E=-c_s^2S,\ \  S=-\int \rho \ln\rho d\bm x=-\langle\rho \ln\rho  \rangle.\label{gibbs}
\end{eqnarray}
The entropy's properties imply that $E$ is minimal at the state of constant density $\rho=1$ where it is zero \cite{ff}.

\subsection{Finiteness of the density and its gradients}

It seems necessary that no infinite accelerations arise in the fluid. Thus the solutions must have finite pressure gradients (here we exclude the unlikely possibility of compensation by infinite viscosity terms). This implies that for barotropic flows the density gradients must be finite, cf. Eq.~(\ref{con1}). The density itself must also be finite for having finite internal energy and entropy, cf. Eq.~(\ref{gibbs}). 
The single-point PDF of density (obtained necessarily with finite resolution) is a regularly measured object which describes a finite quantity, see Sec. \ref{f}.  
This finiteness implies anticorrelations of the velocity divergence along the Lagrangian trajectory that will be described later.

The density evolves due to the combined action of convection and volume compression/rarefaction. This is described by the
solution of the continuity equation,
\begin{eqnarray}&&\!\!\!\!\!\!\!\!\!\!\!
\rho(t, \bm q(t, \bm x))=\rho(\bm x)\exp\left(-\int_{0}^tw(t', \bm q(t', \bm x))dt'\right),\label{solv}
\end{eqnarray}
where $t>0$ or $t<0$; $w(t, \bm x)=\nabla\cdot\bm v(t, \bm x)$; $\bm q(t, \bm x)$ are Lagrangian trajectories labeled by their $t=0$ positions,
\begin{eqnarray}&&\!\!\!\!\!\!\!\!\!\!\!
\partial_t \bm q(t, \bm x)=\bm v(t, \bm q(t, \bm x)),\ \ \bm q(0, \bm x)=\bm x. \label{Lagrangian}
\end{eqnarray}
Here and below we use the notation $\rho(\bm x)\equiv \rho(0, \bm x)$. The mixing in Eq.~(\ref{solv}) is described by the spatial proximity of $\bm q(t, \bm x)$ for $\bm x$ from space regions with very different $\rho(\bm x)$.
The compression is described by the exponential factor.

Finiteness of the density gradients implies fine tuning between the convection and the compression. Indeed, convection creates sharp contrasts of the transported field thus developing infinite gradients. In turn, compression tends to develop infinite densities as is well-known from the theory of dissipative dynamical systems, see e. g. \cite{review,ruelle,dor,gp}. How this fine tuning is realized can be described in detail at small
${\rm Ma}$, see below. The smallest spatial scale of variations of density is the viscous scale $\eta$ determined by the viscosity. This scale can be determined by shocks or structures similar to incompressible turbulence \cite{frisch}.

It must not be thought however that since $n$ obeys the same continuity equation as the density, the concentration transport also does not create indefinitely large gradients. In fact, at small Mach numbers the approximately incompressible
transport by turbulence creates infinitely fast variations of $n$ that are associated with infinite $|\nabla n|$, despite that the density gradients remain finite. Yet the fields coincide after coarse-graining, see below.

For non-barotropic fluids there are more possibilities. For ideal gas the density can become infinite without violating the finiteness of the pressure if the temperature drops to zero. This would not happen in ordinary fluid mechanics however can happen in the
presence of cooling terms. These terms describe local dissipative processes such as radiation and do appear in applications. Finite-time blowup of density is possible in this case keeping the pressure and its gradients finite,
see e. g.
\cite{dis,dis1} for ideal hydrodynamics in one and three dimensions, respectively, and \cite{dis2} for non-ideal case. In this work we will not consider these cases confining ourselves to the ordinary fluid mechanics. Finally the
possibility of  infinite temperature gradients that combine with infinite gradients of density for producing finite $\nabla p$ is also excluded by the thermal conductivity. Thus we assume that back reaction of the density on the
transporting flow causes the density and its gradients to be finite.

\subsection{Vanishing sum of Lyapunov exponents}

If $w(t, \bm q(t, \bm x))$ has a finite correlation time then the sum of Lyapunov exponents is generically negative that is an unrestricted smooth compressible vector field would have the sum which is negative. The steady state solutions of the continuity equation with this flow are then singular \cite{ff}. Correspondingly finiteness of the density implies that the sum is non-generic. It must equal zero as remarked in \cite{review} and is considered in detail here.

Since the density is finite taking the logarithm of Eq.~(\ref{solv}) yields:
\begin{eqnarray}&&\!\!\!\!\!\!\!\!\!\!\!\!\!\!\!
\lim_{t\to\pm \infty}\!\! \frac{1}{t}\!\ln\left(\!\frac{\rho(0, \bm x)}{\rho(t, \bm q(t, \bm x))}\!\right)\!\propto \!\!\lim_{t\to\pm\infty}\!\! \int_0^t\!\! w [t', \bm q(t', \bm x)]\frac{dt'}{t}\!=\!0,\label{identity1}
\end{eqnarray}
that holds for all $\bm x$. The last limit for $\infty$ and $-\infty$ represents sums of Lyapunov exponents of the flow and its time-reversal, respectively.
Indeed, the sum of Lyapunov exponents $\sum_{i=1}^3\lambda_i$ is defined as the limiting value of the growth exponent of infinitesimal volume of the fluid \cite{review}.
The growth exponent is the logarithm of the jacobian divided by time which gives,
\begin{eqnarray}&&\!\!\!\!\!\!\!\!\!\!\!\!\!\!\!
\sum_{i=1}^3\lambda_i\!=\!\lim_{t\to\infty}\!\! \frac{1}{t}\!\ln\det\frac{\partial \bm q(t, \bm x)}{\partial \bm x}\!=\!\lim_{t\to\infty}\!\! \int_0^t\!\! w [t', \bm q(t', \bm x)]\frac{dt'}{t}.\label{sums}
\end{eqnarray}
Similar relation holds for the sum of Lyapunov exponents of time-reversed flow $\sum_{i=1}^3\lambda_i^-$.
Both sums must vanish due to the density finiteness.
It was demonstrated in \cite{ff,fouxon1} that the sums can be written as time integrals of the different time correlation function of the flow divergence,
\begin{eqnarray}&&\!\!\!\!\!\!\!\!\!\!\!\!\!\!\!\!
\sum_{i=1}^3\!\lambda_i\!=\!-\!\!\int_0^{\infty}\!\!\!\!\!\!\langle w(0)w(t)\rangle dt,\ \ \sum_{i=1}^3\!\lambda_i^-\!=\!-\!\!\int_{-\infty}^0\!\!\!\!\!\!\langle w(0)w(t)\rangle dt,
\nonumber\\&&\!\!\!\!\!\!\!\!\!\!\!\!\!\!\!\! \langle w(0)w(t)\rangle=\int w(0, \bm x)w [t, \bm q(t, \bm x)]d\bm x.\label{sums01}
\end{eqnarray}
Here $\langle w(0)w(t)\rangle$ is generally not even function of $t$ since spatial averaging does not correspond to the average over the steady state density.
Thus $w(t)$ is anticorrelated with its own initial condition so that the integral of $\langle w(0)w(t)\rangle$ over positive or negative times is zero. The physical interpretation of this result is as follows  \cite{arxiv}: the divergence in the fluid particle's frame, $w[t, \bm q(t, \bm x)]$, is not stationary since $\bm q(t, \bm x)$, that is distributed uniformly at $t=0$, accumulate with time in compression regions that are characterised by negative $w$. Thus at $t=0$ we have $\langle w(0)\rangle=\langle \nabla\cdot\bm v(\bm x)\rangle=0$ while at small times Eq.~(\ref{sd}) implies that $\langle w [t, \bm q(t, \bm x)]\rangle=-t\langle w^2\rangle<0$. This accumulation however is transient and occurs at times of the order of the correlation time of $w(t)$. 
At times larger than the correlation time the back reaction of the density on the flow through $-\nabla p$ causes the compression to turn into rarefaction so that the time integral of $\langle w(0)w(t)\rangle$ is zero, cf. \cite{arxiv,plm}. Thus if initially $\bm x$ is in a compression region then, after time of order of the correlation time of $w(t)$ the trajectory will be typically found in a rarefaction region and vice versa.

Further details on $\sum_{i=1}^3\!\lambda_i$ and other consequences of finiteness of density are presented in Appendix \ref{a}. Of the results presented there the following is of particular interest:
\begin{eqnarray}&&\!\!\!\!\!\!\!\!\!\!\!\!\!\!\!
\langle \rho w\rangle=0,\ \ \langle \bm v\cdot\nabla\rho\rangle=0, \label{per}
\end{eqnarray}
Thus on the average the density is constant along the instantaneous streamlines of the flow. Correspondingly the density is on the average constant along a closed streamline and is locally axially symmetric around vortex filaments.

We stress that the above relations are rigorous consequences of the density finiteness. If the sum of the Lyapunov exponent is finite (and thus negative) then the density, the internal energy and the entropy diverge.
The observations of these relations however require flow resolution below the viscous scale. For instance,
the anticorrelation property of $w [t, \bm q(t, \bm x)]$ holds separately along each of the fluid trajectories along which $w$ must be sign-alternating, see Eq.~(\ref{sums}). Thus it can be missed in numerical
simulations if $\bm q(t, \bm x)$ is not well resolved below $\eta$ during the correlation time of $w$. This seems to be the reason
why non-zero sums of Lyapunov exponents were reported in simulations \cite{grauer} where the Kolmogorov scale was half the grid size and the observations of Lyapunov exponents in fact referred to Lyapunov exponents contaminated by
the inertial range.

\subsection{Supersonic inertial range and mutlifractality}

Finally we define the supersonic inertial range of scales which is of main interest in this paper. Turbulence is excited at scale $L$ which is the characteristic scale of acceleration $\bm a$ in Eq.~(\ref{basic}). It is assumed that the rms velocity at scale $L$ is larger than the speed of sound. The flow instabilities generate fluctuations (eddies) of velocity and density with scales smaller than $L$. The characteristic velocity of the eddies decreases downscales becoming of the order of the speed of sound at the sonic scale $l_s$. Turbulence below $l_s$ behaves as a low Mach number quasi-incompressible turbulence with almost uniform density in the inertial range \cite{dust}. In some cases the dissipative scale $l_d$, defined by the condition that the local (defined with velocity of eddies with this scale \cite{frisch}) Reynolds number is of order one, is larger or comparable with $l_s$ (in simulations $l_s\sim l_d$ holds often \cite{krit2007}). Thus we define $\eta=\max[l_s, l_d]$. Then the range of scales between $\eta$ and $L$ defines the supersonic inertial range. Turbulent eddies with a characteristic scale in this range are weakly influenced by viscosity and are supersonic.

Multifractality of $\rho$ and $n$ holds in the supersonic inertial range with fluctuations below $\eta$ smoothened by the effective incompressibility of the flow there or the dissipation. It will be often useful to consider the limit $\eta\to 0$ similar to the use of $\nu\to 0$ limit in the study of incompressible turbulence \cite{frisch}.

We remark that having a large supersonic inertial range, $L\gg \eta$ does not necessarily imply that ${\rm Ma}\gg 1$. Indeed if we denote the scaling exponent of the velocity in this range by $\chi$ then the sonic scale is determined by the condition $(L/l_s)^{\chi}\sim {\rm Ma}$ so that $L/l_s\sim {\rm Ma}^{1/\chi}$. For instance for Kolmogorov scaling $\chi=1/3$ we would have cubic growth of $L/l_s$ with ${\rm Ma}$. The real growth is slower because of deviations from the Kolmogorov scaling and must by roughly quadratic using $\chi$ from \cite{krit2007}. Thus the supersonic inertial range may be well-defined already at ${\rm Ma}\simeq 4-5$.
%
%
%
%

\section{Relaxation of concentration to density}\label{rel}

It might seem self-evident that all initial conditions for the continuity equation relax at large times to the same solution. Indeed, the difference of two different solutions is also a solution whose spatial integral vanishes. Positive and negative regions of an initial condition with zero spatial integral would be mixed by turbulence. Thus at large times coarse-graining over an infinitesimal scale would wash out the field contrasts producing zero.
This is quite similar to mixing by incompressible turbulence \cite{review} where the only difference is that in our case the trajectories' mixing is confined to the non-trivial support of the steady state density instead of the whole volume. Thus any two solutions of the continuity equation would agree at large times after coarse-graining.

If the above consideration using the assumption of the mixing is true then any initial (normalized) distribution of tracers relaxes after some time to a universal limiting distribution independent of the initial condition. Moreover this distribution equals the steady state fluid density. This conclusion would have far reaching theoretical and experimental consequences. Theoretically it would allow to study the fluid density using well-developed techniques for the study of passively transported concentrations \cite{review}. Experimentally it would provide the simplest way of observing the multifractal structure of the density. One could spread in
space a large number of tracer particles that obey Eq.~(\ref{Lagrangian}) and study the stationary distribution on which they settle after transients \cite{grauer}. The study of multifractals via large number of points distributed on them is standard \cite{hp}.

In this Section we undertake critical examination of the assumption that concentration and density coincide in the steady state. The reasons for questioning the equality, that is taken for granted usually see e. g. \cite{grauer}, were sketched in the Intoduction and are detailed here. The simplest argument for the equality would be uniqueness of the steady state solution that often holds due to dissipation. However the continuity
equation is not dissipative and the steady state solutions are not unique. This is well-known for incompressible non-random flows: a measure concentrated on a periodic orbit would solve the continuity equation and be different from a constant solution $n=1$, see e. g. \cite{luzzato}. Similarly for random flows a measure supported on any level set of pointwise first Lyapunov exponent is stationary \cite{yb}. Thus uniqueness cannot be used for proving the equality of the density and the concentration in the steady state. Mixing also cannot be taken for granted. The density could not mix with the tracers by creating a non-mixing set on which it would be supported due to interaction with the transporting flow, cf. the Introduction.

We start from providing an example that helps to see how the difference between passive and active scalars can arise.

\subsection{Instability as possible origin of the difference of active and passive scalars}

Difference of passive and active fields that obey identical first order in time evolution equation is a recurrent question in fluid mechanics. It was observed by Batchelor and Kraichnan already in the sixties
that vorticity in two-dimensional Navier-Stokes (NS) turbulence obeys the same equation as that of a passive scalar (here and below we consider equal diffusivity coefficients of fields) \cite{review}. Direct cascades of both quantities to smaller scales proceed similarly: fields' blobs are stretched by the large-scale flow. However vorticity is active and it rotates the blob which decelerates the stretching. This results in the steady state statistics different from that of the passive scalar \cite{fl,review}. Still the analogy is useful providing the reference from which the full solution derives \cite{kr1,kr2,kr3,batch,fl,fal}.

The key to the difference between active and passive fields is the difference of their behaviors under perturbations. For instance, one can consider the identical equations for the vorticity $\omega$ and the passive scalar
$\theta$ in incompressible two-dimensional turbulent flow $\bm u$ where the forcing $f$ and the viscosity $\nu$ coincide \cite{legras},
\begin{eqnarray}&&\!\!\!\!\!\!\!\!\!\!\!\!\!
\partial_t \omega+(\bm v\cdot \nabla)\omega=f+\nu \nabla^2\omega,\nonumber\\&&\!\!\!\!\!\!\!\!\!\!\!\!\!
\partial_t \theta+(\bm v\cdot \nabla)\theta=f+\nu \nabla^2\theta. \label{feidla}
\end{eqnarray}
Here $f$ is the curl of the force that drives $\bm u$. The difference $\phi=\omega-\theta$ obeys,
\begin{eqnarray}&&\!\!\!\!\!\!\!\!\!\!\!\!\!
\partial_t \phi+(\bm v\cdot \nabla)\phi=\nu \nabla^2\phi.\label{phi}
\end{eqnarray}
that has a possible steady state solution $\phi=0$ and $\omega=\theta$. Yet it was observed in \cite{legras} (who used a somewhat different structure of dissipative
term where instead of $\nu\nabla^2$ a combination of large-scale friction and Laplacian to power eight were used. The dissipative term however was linear so our consideration would apply.) that there is a significant difference between the fields. The forcing that was used consisted of keeping a low wavenumber component of the fields at the same constant value. However this type of forcing in fact depends on the considered field so that $f$ depends on the field that it forces. Thus the RHS of Eq.~(\ref{phi}) contains difference of forces of $\omega$ and $\theta$ that vanishes only at $\omega=\phi$. Thus despite that $\phi=0$ is a valid solution of the equations, its small perturbations result in the
appearance of force in Eq.~(\ref{phi}) that can further increase the perturbation resulting in instability. It is this instability that was observed. In the case where $f$ in Eqs.~(\ref{feidla}) are identical external forces
independent of the fields $\omega=\theta$ would hold.

Similarly in our case it is readily seen from,
\begin{eqnarray}&&\!\!\!\!\!\!\!\!\!\!\!
\partial_t \frac{n}{\rho}+\left(\bm v\cdot\nabla\right) \frac{n}{\rho}=0, \label{conservationlaw}
\end{eqnarray}
implied by Eqs.~(\ref{con}), that $n=\rho$ is a valid steady state solution. However it can get unstable if fluid particles form a configuration that preserves itself thanks to the interaction with the transporting flow. Small deviation of tracers from this configuration could grow, cf. the Introduction (in the case of continuum fields configuration is a field configuration).

\subsection{Natural measure for the tracers} \label{nar}

We consider relaxation of different initial conditions for the tracers' concentration to the same limiting distribution. This can be done by relying on the physics of the problem. The
tracers perform Brownian motion with a certain diffusion coefficient $\kappa$. Thus the full continuity equation for their concentration is $\partial_t n+\nabla\cdot(n\bm v)=\kappa \nabla^2 n$.
The presence of a finite $\kappa$ allows to overcome the non-dissipative nature of the continuity equation \cite{pesco} that obstructs the demonstration of the relaxation. Indeed, if the diffusion term is discarded then
one can introduce a conserved distance between two smooth solutions of the continuity equation. This distance
is similar to the $H-$function of Boltzmann or entropy $-H$, see its use for the relaxation of solutions of the more general Fokker-Planck equation in \cite{risken}.
We define "H-function" for two smooth positive solutions of the continuity equation $n_1(t, \bm x)$ and $n_2(t, \bm x)$ obeying $\int n_i(t, \bm x) d\bm x=1$ as,
\begin{eqnarray}&&
H(t)=\int n_1\ln \frac{n_1}{n_2}d\bm x. \label{d1}
\end{eqnarray}
We observe that $H(t)$ is a non-negative function. Using $\int n_i(\bm x, t) d\bm x=1$ we may rewrite $H$ as
\begin{eqnarray}&&
H=\int \left[n_1\ln \frac{n_1}{n_2}-n_1+n_2\right]d\bm x
\nonumber\\&&
=\int n_2 \left[
R\ln R-R+1\right]d\bm x,
\end{eqnarray}
where we introduced $R\equiv n_1/n_2$. The last term in the above equation is always not positive as it follows
from
\begin{eqnarray}&&
R\ln R-R+1=\int_1^R \ln x dx\geq 0, \label{B6}
\end{eqnarray}
that holds for any $R\geq 0$. We see that $H=0$ only at $n_1(t, \bm x)=n_2(t, \bm x)$. Thus $H(t)$ is non-negative and it vanishes only if the solutions agree pointwise providing a good definition of the distance between $n_i$.
Changing the integration variable in Eq.~(\ref{d1}) from $\bm x$ to $\bm q(t, \bm y)$ it is found that $H(t)=H(0)$. This is the consequence of mass conservation $n(t=0, \bm y)d\bm y=n(t, \bm x)d\bm x$ and conservation
of $n_1/n_2$ on $\bm q(t, \bm y)$ that follows from applying Eq.~(\ref{conservationlaw}) to $n_1/n_2$.
Thus the distance between the solutions is conserved and pointwise relaxation to $n_1=n_2$ is not possible. This problem could be dealt with by considering relaxation in the sense of measures, which is equivalent to relaxation in the coarse-grained sense. However inclusion of diffusion, that is necessarily present in applications, is simpler. It is readily seen that with the diffusion term the time derivative of $H$ is \cite{risken},
\begin{eqnarray}&&\!\!\!\!\!\!\!\!
{\dot H}=-\kappa \int n_1 \left(\nabla \ln (n_1/n_2)\right)^2 d\bm x\leq 0,
\end{eqnarray}
where the equality holds for $n_1=n_2$ only. This equation implies relaxation to $n_1=n_2$ since $H\geq 0$. Diffusion makes the equation dissipative and all initial conditions relax in the limit of large evolution times to the same smooth distribution $n^{\kappa}$ where we stress the dependence on $\kappa$. We can define the so-called  \cite{dor,sinai,sinain} natural measure $n^s$ by,
\begin{eqnarray}&&\!\!\!\!\!\!\!\!
n^s_{\epsilon}(\bm x)=\lim_{\kappa\to 0} n^{\kappa}_{\epsilon}(\bm x), \label{dfnc}
\end{eqnarray}
where we defined coarse-graining by,
\begin{eqnarray}&&\!\!\!\!\!\!\!\!\!\!\!\!\!\!
h_{\epsilon}(t, \bm x)\!=\!\frac{3}{4\pi \epsilon^3}\int_{|\bm x-\bm x'|<\epsilon}\!\! h(\bm x')d\bm x', \label{coarsed}
\end{eqnarray}
where $h(\bm x)$ is any function.
Since Eq.~(\ref{dfnc}) defines $n^s_{\epsilon}(\bm x)$ for any $\epsilon>0$ that it defines $n_s$ completely. The definition can be given an explicitly $\epsilon-$independent form by the demand that for any smooth function $f(\bm x)$ we have,
\begin{eqnarray}&&\!\!\!\!\!\!\!\!
\int f(\bm x)n^s(\bm x)d\bm x=\lim_{\kappa\to 0}\int f(\bm x) n^{\kappa}(\bm x)d\bm x. \label{sashy}
\end{eqnarray}
On scales $\epsilon$ larger than a certain diffusive scale \cite{review}, which is usually quite small, $n^{\kappa}$ coincides with $n^s$.
We observe that diffusion introduces white noise in the equation of motion of the tracer. Thus $n^{\kappa}(\bm x)$ can be considered as the probability density function (PDF) of the tracer position. The PDF's limit of $\kappa\to 0$ is not necessarily smooth so it might not be definable as an ordinary function however the limiting averages of arbitrary function of the position $f(\bm x)$ are well-defined by Eq.~(\ref{sashy}).

Thus we have proved that $n^s$ is the universal limiting distribution of tracers at scales that are not too small for diffusion to become relevant. The natural measure $n^s$ solves the ordinary continuity equation without the diffusion term (it is a weak solution). This way of constructing the natural measure $n_s$ as the zero-noise limit of the PDF of a stochastic process constitutes a rigorous mathematical approach \cite{kifer} that in our case is dictated by the physics of the problem.

It is often possible to write $n^s$ explicitly. For this it is useful to construct $n^{\kappa}$ by setting the initial condition at $t=-T$ and considering $n(\bm x)\equiv n(0, \bm x)$ in the infinite evolution time limit $T\to\infty$. We take with no loss of generality a uniform initial condition $n(-T, \bm x)=1$. We have from Eq.~(\ref{dfnc}),
\begin{eqnarray}&&\!\!\!\!\!\!\!\!
n^s_{\epsilon}(\bm x)=\lim_{\kappa\to 0}\lim_{T\to\infty} n_{\epsilon}(\bm x)=\lim_{T\to\infty}\lim_{\kappa\to 0} n_{\epsilon}(\bm x).
\end{eqnarray}
We assume here that the order of the limits can be changed which is a form of a mixing assumption. Observing that $n(\bm x)$ at $\kappa\to 0$ becomes the solution of the continuity equation given by Eq.~(\ref{solv})
we find \cite{fouxon},
\begin{eqnarray}&&\!\!\!\!\!\!\!\!\!\!\!
n_{\epsilon}(\bm x)= \lim_{T\to\infty}\left(\exp\left(-\int_{-T}^0 w(t, \bm q(t, \bm x))dt\right)\right)_{\epsilon}.\label{densrepr}
\end{eqnarray}
For other approach to this representation see \cite{ruelle}. The statistics of the natural measure can be studied using in Eq.~(\ref{densrepr}) the stationary statistics of the velocity. The rule of the thumb is that the change of the order of limits (which is the only non-rigorous assumption made in the derivation) is valid as long as the answers obtained from Eq.~(\ref{densrepr}) are finite. For instance the pair correlation function of concentration $f(\bm r)=\langle n(0)n(\bm r)\rangle$ is,
\begin{eqnarray}&&\!\!\!\!\!\!\!\!\!\!\!\!\!\!\!\!\!
f(\bm r)\!=\!\left\langle\!\exp\left(\!-\!\int_{-\infty}^0\!\!\left[w(t, \bm q(t, \bm r))\!+\!w(t, \bm q(t, 0))\right]dt\right)\!\right\rangle, \label{pair}
\end{eqnarray}
where the coarse-graining can be dropped at finite $|\bm r|$. Application of the cumulant expansion theorem to  $\langle n(0)n(\bm r)\rangle/\langle n(0)\rangle\langle n(\bm r)\rangle$ with average concentrations represented
via Eq.~(\ref{densrepr}) gives \cite{Ma,fouxon},
\begin{eqnarray}&&\!\!\!\!\!\!\!\!\!\!\!\!\!\!\!\!\!
\ln f(\bm r)=g(\bm r)+\ldots,  \label{dots}
\end{eqnarray}
where the dots stand for higher order mixed (involving both $w(t, \bm q(t, \bm r))$ and $w(t, \bm q(t, 0))$) cumulants and \cite{fouxon},
\begin{eqnarray}&&\!\!\!\!\!\!\!\!\!\!\!\!\!\!
g(\bm r)=\int_{-\infty}^0
dt dt'
\left\langle \left\langle w(t, \bm q(t, 0))w(t', \bm q(t', \bm r))
\right\rangle\right\rangle,
\label{higher}
\end{eqnarray}
where double angular brackets stand for dispersion. The representation given by Eq.~(\ref{dots}) will be useful below.

\subsection{Impact of effective diffusion on the density}

We saw above that the question of relaxation to a universal distribution that is independent of the initial condition is easy in the presence of dissipation. If we could introduce some diffusion to the continuity equation
for the density it would solve the problem of relaxation. However there is no reason to introduce this term. If it is introduced in a numerical scheme artificially then this demands a validity check. However smoothness
of the density field implies regularity of the limit of zero diffusion at finite times that is,
\begin{eqnarray}&&\!\!\!\!\!\!\!\!\!\!\!\!\!\!\!\!\!
\rho(t, \bm x)\approx \rho^{\kappa}(t, \bm x), \ \ \partial_t\rho^{\kappa}+\nabla\cdot(\rho^{\kappa}\bm v)=\kappa\nabla^2 \rho^{\kappa}.
\end{eqnarray}
Unfortunately this could not be used for proving the relaxation since the diffusive term brings dissipation and its accumulation over time can bring a finite effect in the steady state limit of infinite evolution time.
Staying in the frame of the continuity equation it is usual to use the assumption of mixing for proving the density relaxation \cite{sinai,sinain}. This can be done completely in the limit of small ${\rm Ma}$ where
the well-established mixing by incompressible turbulence allows to prove relaxation of concentration to the density.

\subsection{Relaxation of concentration to density at ${\rm Ma}\ll 1$}

It is instructive to consider the case of ${\rm Ma}\ll 1$ where there is solution for the density in terms of the flow. The solution is not unique and depends on interaction with heat \cite{bayly}, see also \cite{Lighthill,Montgomery,Matthaeus,bayly,Ristorcelli}. In the case of isothermal flow
we have $\rho(t, \bm x)=1+c{\rm Ma}^2 p(t, \bm x)$ where $c$ is a constant and $p$ is the pressure as obtained from the incompressible Navier-Stokes equations. The pressure obeys the Poisson equation $\nabla^2 p=-\nabla_ku^0_i
\nabla_iu^0_k$ where $\bm u^0$ is the incompressible turbulent flow (divergence of forcing is assumed to be zero here) and its smallest scale of variations in space \cite{frisch} is the Kolmogorov scale $\eta$. It is assumed
that $\langle p\rangle=0$. The solution makes
it explicit that the density field has a finite scale of spatial variations due to the interaction with the flow. Proportionality of the density and pressure spectra in the inertial range was confirmed for the considered regime by recent numerical simulations \cite{fractal,gotoh} for ${\rm Ma}<0.3$. The spectrum ($\sim k^{-7/3}$ where $k$ is the wavenumber) indicates that the density is a large-scale non-fractal field with positive scaling exponents in the inertial range. The non-fractal nature of the density fluctuations at low Mach numbers is supported also by other scenarios that hold for ${\rm Ma}\ll 1$, having $k^{-5/3}$ spectrum \cite{bayly,nu}, see also \cite{dj,eyink,jd}.

It is readily seen that the assumption that $n/\rho=const$ pointwise is self-contradictory. Passive concentration transported by low Mach number
flow must approximately obey transport by incompressible turbulence. However that creates finite contrasts over infinitesimal scales associated with indefinite growth of $\nabla n$. This is in contradiction with
finite $\nabla \rho$.

In fact, the relaxation of $n$ to $\rho$ holds in the sense of measures that is,
\begin{eqnarray}&&\!\!\!\!\!\!\!\!\!\!\!
\lim_{t\to\infty}\frac{\int f(\bm x) n(t, \bm x) d\bm x}{\int f(\bm x) \rho(t, \bm x) d\bm x}=1,
\end{eqnarray}
for any smooth function $f$. This equation is equivalent to the statement that the concentration $n_{\epsilon}(t, \bm x)$ coarse-grained over scale $\epsilon$ relaxes to $\rho(t, \bm x)$ at large $t$. This is indeed so as we demonstrate.
We assume that the fluid flow is in the steady state and consider the evolution of concentration of tracers
injected at some $t<0$ with initial distribution $n(t, \bm x)=n_0(\bm x)$. We study the limit of infinite evolution time by considering the concentration at time zero $n(\bm x)$ at $t\to-\infty$. We have from Eq.~(\ref{conservationlaw}),
\begin{eqnarray}&&\!\!\!\!\!\!\!\!\!\!\!
n(\bm x)=\rho(\bm x)\frac{n_0(\bm q(t, \bm x))}{\rho(t, \bm q(t, \bm x))}.
\end{eqnarray}
With no loss of generality we can consider the initial condition $n_0=1$, cf. subsection \ref{nar}. Coarse-graining over $\epsilon\ll\eta$ gives,
\begin{eqnarray}&&\!\!\!\!\!\!\!\!\!\!\!
\frac{n_{\epsilon}(\bm x)}{\rho(\bm x)}=\int_{|\bm x'-\bm x|<\epsilon} \frac{d\bm x'}{(4\pi \epsilon^3/3)\rho(t, \bm q(t, \bm x'))}.\label{cg}
\end{eqnarray}
Relaxation of $n_{\epsilon}(\bm x)$ to $\rho(\bm x)$ means that the RHS of Eq.~(\ref{cg}) relaxes to one.
We find using the form of the density at ${\rm Ma}\ll 1$,
\begin{eqnarray}&&\!\!\!\!\!\!\!\!\!\!\!
\int_{|\bm x'\!-\!\bm x|<\epsilon}\!\! \frac{d\bm x'}{\rho(t, \bm q(t, \bm x'))}
\!\approx \!\!\int_{|\bm x'\!-\!\bm x|<\epsilon}\!\!\! (1\!-\!c{\rm Ma}^2 p(t, \bm q(t, \bm x')))d\bm x'.\nonumber
\end{eqnarray}
We observe that we can use in the integral in the considered order in ${\rm Ma}$ the trajectories of incompressible flow. Then the mixing property of the incompressible turbulence implies that at large times $\int_{|\bm x'\!-\!\bm x|<\epsilon} p(t, \bm q(t, \bm x'))d\bm x'$ is proportional to the spatial average of pressure (see the proof in the next subsection where $\rho=1$ must be used in Eq.~(\ref{mixing1})). That spatial average is zero by the made assumption. Therefore we find $n_{\epsilon}(\bm x)=\rho(\bm x)$. Finally the proof that any smooth initial condition on the density with $\langle n\rangle=1$ will also relax to $\rho$ can be constructed using
combination of incompressible mixing and small ${\rm Ma}$ expansion.

The extension of the above proof to the case of finite Mach number is non-evident and requires another approach.

\subsection{Mixing}

The property that is needed for proving relaxation of concentration to density is mixing \cite{sinai,sinain}. The assumption of mixing tells that different time correlation functions defined with the help of $\rho$ in the limit of large times reduce to the product of averages. In our case the averaging measure is time-dependent, being stationary only statistically, which demands a slight modification in the form of the mixing assumption. We assume that for any smooth functions $f(\bm x)$ and $g(\bm x)$,
\begin{eqnarray}&&\!\!\!\!\!\!\!\!\!\!
\lim_{t\to\infty}\frac{ \int f(\bm q(t, \bm x)) g(\bm x) \rho(\bm x) d\bm x}{\int f(\bm x)\rho(t, \bm x)d\bm x\int g(\bm x) \rho(\bm x) d\bm x}
=1,\label{mixing1}
\end{eqnarray}
where the integral in the numerator of the LHS defines the correlation function $\langle f(t)g(0)\rangle$. For time-independent flows this reduces to the usual form \cite{sinai,sinain} on using $\rho(t, \bm x)=\rho(\bm x)$. Our Eq.~(\ref{mixing1}) incorporates that in the decomposition $\langle f(t)g(0)\rangle\approx \langle f(t)\rangle\langle g(0)\rangle$, that holds at large times, we must
use $\rho(t, \bm x)$ for averaging $f$ and not $\rho(t=0, \bm x)$. We observe that Eq.~(\ref{mixing1}), is a bilinear relation. Its elementary form is obtained by taking for $f$ and $g$ the indicators $\chi_{\epsilon}(\bm x-\bm x_1)$ and $\chi_{\epsilon}(\bm x-\bm x_2)$ for some $\bm x_1$ and $\bm x_2$.
Here $\chi_{\epsilon}(\bm x)$ equals one for $|\bm x|<\epsilon$ and zero otherwise. We have from Eq.~(\ref{mixing1}),
\begin{eqnarray}&&\!\!\!\!\!\!\!\!\!\!
\lim_{t\to\infty}\frac{\int_{|\bm x-\bm x_2|\!<\!\epsilon}\!\chi_{\epsilon}(\bm q(t, \bm x)-\bm x_1) \rho(\bm x) d\bm x}{\int_{|\bm x-\bm x_1|\!<\!\epsilon} \rho(t, \bm x) d\bm x\int_{|\bm x-\bm x_2|\!<\!\epsilon} \rho(\bm x) d\bm x}\!=\!1.
\end{eqnarray}
This becomes for $\epsilon\ll \eta$,
\begin{eqnarray}&&\!\!\!\!\!\!\!\!\!\!
\lim_{t\to\infty}\frac{\int_{|\bm x-\bm x_2|<\epsilon} \chi_{\epsilon}(\bm q(t, \bm x)-\bm x_1)   d\bm x}{(4\pi\epsilon^3/3)^2\rho(t, \bm x_1)}=1.
\end{eqnarray}
We observe that,
\begin{eqnarray}&&\!\!\!\!\!\!\!\!\!\!
\frac{ \int_{|\bm x-\bm x_2|<\epsilon} \chi_{\epsilon}(\bm q(t, \bm x)-\bm x_1)   d\bm x}{(4\pi\epsilon^3/3)^2}=n_{\epsilon}(t, \bm x_1),
\end{eqnarray}
where $n(t, \bm x)$ is (normalized) solution of the continuity equation obeying the initial condition $n(t=0, \bm x)=\chi_{\epsilon}(\bm x-\bm x_2)/(4\pi\epsilon^3/3)$. We obtain combining the last equations,
\begin{eqnarray}&&\!\!\!\!\!\!\!\!\!\!
\lim_{t\to\infty}\frac{n_{\epsilon}(t, \bm x_1)}{\rho(t, \bm x_1)}=1.
\end{eqnarray}
Finally linearity of the continuity equation and arbitrariness of $\bm x_2$ and $\epsilon$ imply that coarse-grained concentration relaxes to the density in the limit of large evolution time for arbitrary initial condition.

We can get insight into the nature of the mixing assumption given by Eq.~(\ref{mixing1}) by using for $f$ and $g$ the indicators $\chi_A$ and $\chi_B$ of some volumes $A$ and $B$ in space \cite{sinai}. We find,
\begin{eqnarray}&&\!\!\!\!\!\!\!\!\!\!
\lim_{t\to\infty}\frac{ \int_B \chi_A(\bm q(t, \bm x)) \rho(\bm x) d\bm x}{\int_A \rho(t, \bm x)d\bm x\int_B\rho(\bm x) d\bm x}
=1.
\end{eqnarray}
The assumption in this form tells that asymptotically at large times the mass fraction of points of $B$ that are found inside $A$ equals the mass fraction of $A$ in the whole volume.
In other words, the mass of $B$ is redistributed over the volume uniformly with respect to $\rho$. Mixing can also be given probabilistic interpretation: given the particle is initially in $B$ the probability of finding it in $A$ is the probability of $A$. Thus asymptotically at large times the fluid particle distributes in space independently of where it was initially, see details in \cite{luzzato}. This is not self-evident though. However large the time is, there is memory of the initial condition in Eq.~(\ref{solv}): the trajectory $\bm q(t, \bm x)$, the divergence on it and $\rho(\bm x)$ must combine in smooth and finite field $\rho(t, \bm x)$. For instance
in the previously considered case of ${\rm Ma}\ll 1$ the density evolution is non-generic: it preserves smoothness in contrast with the concentration evolution that does not.

The memory effect can be illustrated by considering the evolution in the steady state,
\begin{eqnarray}&&\!\!\!\!\!\!\!\!\!\!\!\!\!
\rho(\bm x)\!=\! \rho(-T, \bm q(-T, \bm x))\exp\left(\!-\!\int_{-T}^0\!\! w(t, \bm q(t, \bm x))dt\right), \label{denmsu}
\end{eqnarray}
that differs from Eq.~(\ref{densrepr}) by the $\rho(-T, \bm q(-T, \bm x))$ prefactor. If the density forms a special configuration that preserves itself thanks to coupling with the flow then $\rho(-T, \bm x)$ differs from a generic function that relaxes to the natural measure. It depends on past interactions of the density and the flow, and is correlated with the instantaneous flow. Its evolution is tuned with the velocity and $\bm q(t, \bm x)$ depends on $\rho(-T, \bm x)$.
Thus Eq.~(\ref{denmsu}) would not describe the relaxation to the natural measure as it would for a generic (normalized) prefactor. This fits the general knowledge that there are stationary invariant measures different from the natural measure. These are atypical and evolve from initial conditions of measure zero (in functional space) \cite{grass}. In contrast with the concentration obeying Eq.~(\ref{densrepr}), the fluid density cannot be written as a functional of the stationary velocity because it back-reacts on that velocity.

The (possible) difference can be illustrated further by trying to apply for the fluid density the same procedure that we used for deriving the representation for the steady state concentration given by Eq.~(\ref{densrepr}). We start with unit initial condition on the density $\rho(-T)=1$. This initial condition was used for the build up of stationary turbulence in the simulations of \cite{fedor}. In that approach the solution in Eq.~(\ref{densrepr}) holds for $\rho(\bm x)$ however $\bm w(t, \bm x)$ in the integrand is not stationary. Indeed, the transient period of relaxation to the steady state at times close to $-T$ is never forgotten. In this sense the important and profound distinction between the tracers' concentration and the fluid density, is that initial conditions are forgotten for the former but not for the latter, which is correlated with the flow velocity, cf. \cite{grass}.

Further study of concentration relaxation to the density and the validity of the mixing assumption must probably be done numerically and experimentally. Our study below does not depend on the validity of $n=\rho$ in the steady state, however if the equality holds then our results for $n$ can be transferred to $\rho$.

%
%

\section{Multifractal growth of volumes in the inertial range}
\label{forma}

In this Section we describe the behavior of volumes of tracers with size within the supersonic inertial range. We demonstrate that at large times the volumes obey a power-law dependence on the time that is associated with formation of a multifractal structure in space. We demonstrate that the volumes decay at large times to zero with probability one. This does not contradict the conservation of the total volume since the decay is strongly non-uniform in space (intermittency). At any, however large but finite, time there are volumes that expanded much providing for the constant total volume. Thus the concept of the sum of Lyapunov exponents can be generalized to the inertial range in logarithmic time variable. The decay implies formation of asymptotically singular (multifractal) concentration. The consideration parallels
the consideration of similar facts in smooth flows \cite{fb,review} from which review we start. The considerations of this Section are mostly phenomenological as in the Richardson law or many other considerations of turbulence. Precise calculations can be done in the frame of the Kraichnan model, see \cite{review} and below.


\subsection{Introduction: volumes in smooth flows}

We consider behavior of tracers in an unconstrained, generic, compressible flow. This flow has a negative sum of Lyapunov exponents, see above. Thus here and only here we do not consider the degenerate flow with zero sum of Lyapunov exponents considered in the rest of the paper. The tracers are attracted to regions with negative divergence. This is because the flux of the particles to any region is proportional to the integral over the region's surface $\int \bm v\cdot d\bm S$ that equals the volume integral $\int\nabla \cdot\bm v d\bm x$. Thus there is an effective attraction of particles
that move to the same regions of the flow. This attraction causes shrinking of volumes of particles to zero and formation of multifractal. The conclusion is known for transport of tracers by smooth flows where it can be seen as a form of the second law of thermodynamics \cite{arxiv,ff}.

The evolution of infinitesimal volume $V_s(t, \bm x)$ of tracers initially located near $\bm x$ is determined by the Jacobian of the Lagrangian mapping defined in Eq.~(\ref{Lagrangian}),
\begin{eqnarray}&&\!\!\!\!\!\!\!\!\!\!\!\!\!\!
\ln V_s(t, \bm x)\!\propto \!\ln \det \frac{\partial \bm q(t, \bm x)}{\partial x}\!=\!\!\int_{0}^t \!\!w(t', \bm q(t', \bm x))dt',  \label{defs1}
\end{eqnarray}
where the subscript stands for smooth. At times much larger than the correlation time of $w(t, \bm q(t, \bm x))$ the RHS is roughly a sum of a large number of independent random variables. Its cumulants grow linearly with time and we find using the cumulant expansion theorem,
\begin{eqnarray}&&\!\!\!\!\!\!\!\!\!\!\!
\langle V^k_s(t)\rangle\sim \exp(\gamma(k)t),\label{defs2}
\end{eqnarray}
that holds for space-averages at large times \cite{fouxon, review}. H\"{o}lder's inequality implies that $\gamma(k)$ is a convex function. Besides the trivial zero at $k=0$ this function has also a zero at $k=1$ because this moment gives the conserved total volume of the flow \cite{arxiv}. Thus if $\gamma(k)$ is not degenerate we have $\gamma(k)<0$ for $0<k<1$ and $\gamma(k)>0$ otherwise. In fact, it is readily seen from Eqs.~(\ref{defs1})-(\ref{defs2}) that $\gamma'(0)$ is the non-positive sum of Lyapunov exponents $\sum_{i=1}^3\lambda_i$ considered in Eqs.~(\ref{sums})-(\ref{sums01}). This is in agreement with $\gamma(0)=0$ and $\gamma(k)<0$ (where strict inequality holds in the non-degenerate case) for $0<k<1$.

Writing the moments of $V_s$ with the help of the PDF $P(\sigma_s, t)$ of $\sigma_s=\ln V_s(t)/t$,
\begin{eqnarray}&&\!\!\!\!\!\!\!\!\!\!\!
\langle V^k_s(t)\rangle\equiv \int P(\sigma_s, t) \exp(k t\sigma_s) d\sigma_s\sim \exp(\gamma(k)t),\label{pdfcon}
\end{eqnarray}
provides an equation for $P(\sigma_s, t)$. Its solution is given by \cite{grb},
\begin{eqnarray}&&\!\!\!\!\!\!\!\!\!\!\!
P\sim \exp\left(-tS\left(\sigma_s 
\right)\right),\ \ \max_x[kx-S\left(x
\right)]=\gamma(k), \label{pdfs}
\end{eqnarray}
where $S(x)$ is called large deviations or rate function. This function is the Legendre transform of the convex function $\gamma(k)$ and thus also convex. It is readily seen by considering $\gamma(0)$ and $\gamma'(0)$ via $S(x)$ that the rate function is non-negative and has a unique minimum of zero taken at $x=\gamma'(0)$. We obtained these properties directly from the behavior of the moments given by Eq.~(\ref{pdfcon}) which is useful for the study of the inertial range below. However these conclusions could also be obtained directly from Eq.~(\ref{defs1}), see \cite{fb,review}.

It is seen from Eq.~(\ref{pdfs}) using the properties of $S(x)$ that the limiting distribution $P(\sigma_s, t\to \infty)$ is $\delta\left(\sigma_s - \sum_{i=1}^3\lambda_i\right)$. We conclude that,
\begin{eqnarray}&&\!\!\!\!\!\!\!\!\!\!\!
\lim_{t\to \infty}\frac{\ln V_s(t, \bm x)}{t}=\gamma'(0)=\sum_{i=1}^3\lambda_i<0,
\end{eqnarray}
with probability one (that is the limit exists and is independent of $\bm x$ for almost every $\bm x$), cf. Eq.~(\ref{sums}). In application to real volumes with small but finite size the limit above can be used only as long as the largest linear size of $V_s$ is smaller than the smoothness scale of the flow (that is the scale below which velocity difference is well described by Taylor series).

We reach the conclusion that in the non-degenerate case of $\sum_{i=1}^3\lambda_i<0$ infinitesimal volumes decay to zero with probability one, cf. the discussion of Eqs.~(\ref{defs1})-(\ref{defs2}). This is how one sees that at large times the particles accumulate on a set with zero volume (the strange attractor) in the case of smooth dynamics. It is impressive that this conclusion is so general. It uses smoothness, finite correlation time (that underlies the exponential behavior of the moments in Eq.~(\ref{defs2})), the assumption that $\gamma(k)$ is not identically zero and volume conservation $\gamma(1)=0$. In fact its generality is of a  similar status as the second law of thermodynamics since $\gamma'(0)\leq 0$ can be seen as a form of that law \cite{ruelle,ff}.

\subsection{Self-similar flow in the supersonic inertial range}

Remarkably there is a counterpart of the above consideration in the supersonic inertial range provided that the velocity is self-similar. Self-similarity holds above a critical Mach number ${\rm Ma}_s$ where the subscript stands for scaling. This number is defined by the condition that the scaling of the solenoidal, $\bm v_s(\bm x, t)$, and potential, $\bm v_p(\bm x, t)$, components of the velocity is approximately equal at ${\rm Ma}>{\rm Ma}_s$. Thus for instance for a second order structure function of the components we have,
\begin{eqnarray}&&\!\!\!\!\!\!\!\!\!\!\!
\left\langle \left(\bm v_s(\bm x)-\bm v_s(0)\right)^2\right \rangle \sim x^{\chi_2},\nonumber\\&&\!\!\!\!\!\!\!\!\!\!\! \left\langle \left(\bm v_p(\bm x)-\bm v_p(0)\right)^2\right \rangle \sim x^{\chi_2+\Delta_2},
\end{eqnarray}
where $\Delta_2\ll \chi_2$. At small Mach numbers $\Delta_2$ is comparable with $\chi_2$ however as the Mach number increases the gap between the components' scalings closes. The scaling exponents of $\bm v _s$ and $\bm v_p$ differ by less than ten per cent already for ${\rm Ma}$ of order one \cite{krit2007}. Apparently the scalings differ at all ${\rm Ma}$ (so $\Delta_2$ is never zero) however the difference decreases with ${\rm Ma}$ becoming negligible at ${\rm Ma}>{\rm Ma}_s$.

The critical number ${\rm Ma}_s$ is not defined sharply since it depends on the desired accuracy. For our purposes the criterion is that at the considered scale $L/l$ raised in the power of characteristic difference of scalings $\Delta_2$ is approximately one. If this condition holds then the pair-correlation function of the concentration of tracers approximately obeys a power-law, see below.

We consider the behavior of a volume $V_l(t, \bm x)$ of tracers which occupy at $t=0$ a ball of radius $l$ centered at $\bm x$. It is assumed that $l$ belongs to the supersonic inertial range, $\eta\ll l\ll L$.
The volume of the image $V_l(t)$ considered as a function of time obeys,
\begin{eqnarray}&&\!\!\!\!\!\!\!\!\!
\dot V_l\!=\int_{\partial V_l} \bm v\cdot d\bm S\!=\!\int_{V_l} \!\!\!w(t, \bm x) d\bm x,\label{fl}
\end{eqnarray}
where $\partial V_l$ is the boundary of the volume and $d\bm S$ is surface element vector. We consider the behavior of $V_l(t)$ at ${\rm Ma}>{\rm Ma}_s$ where the velocity in the supersonic inertial range
obeys scaling $r^{\sigma}$, where $0<\sigma<1$. As will be discussed at length below the flow scaling implies that at large times the moments obey a power-law dependence on time,
\begin{eqnarray}&&\!\!\!\!\!\!\!\!\!\!\!
\langle V^k_l(t)\rangle\sim   t^{\gamma_1(k)}, \label{scalingvol}
\end{eqnarray}
where $\gamma_1(k)$ are independent of $l$ (here and below we do not write the dimensional factors that are irrelevant for the result). We find similarly to the smooth case that $\gamma_1(k)$ in Eq.~(\ref{scalingvol}) is a convex function, that has two zeros at $k=0$ and $k=1$. This function is negative for $0<k<1$ and positive otherwise. The function is non-linear so that the growth of the volumes is not self-similar despite the self-similarity of the transporting flow (we neglect intermittency for the moment). The solved case of transport by incompressible self-similar flow reviewed in \cite{review} demonstrates that statistics of tracers is not self-similar despite the flow self-similarity. We describe the origin of the breakdown of the self-similarity here.

The Richardson law for the inertial range growth of separation between two tracers in incompressible turbulence \cite{review} states that $R^2(t)-R^2(0)\sim t^3$. Similar law holds for compressible turbulence. Using the consideration identical with that of \cite{review} for incompressible turbulent flow with exponent $\sigma$, we find,
\begin{eqnarray}&&\!\!\!\!\!\!\!\!\!\!\!\!\!\!
R^{1-\sigma}(t)-R^{1-\sigma}(0)\sim t. \label{scals}
\end{eqnarray}
The separation is independent of $R(0)$ at large $t$ where we can drop $R^{1-\sigma}(0)$ in the LHS. Disregarding intermittency the evolution of the PDF of $R(t)$ must be self-similar at those times \cite{review}.
We observe that tracers with infinitesimally small initial separation $R(0)$ separate by a finite distance in a finite time which is the so-called explosive separation. This type of separation is typical in rough flows \cite{phase}. It causes the volume to develop a complex multifractal form spread over a large range of spatial scales in a finite time. It is at these times that Eq.~(\ref{scalingvol}) holds when the statistics of the shape of the volume is time-independent in complete similarity with the incompressible case \cite{review}. The volume at a later time is obtained by rescaling all lengths of the volume at an earlier time with a magnifying factor. The fine details of the volume change however are statistically stationary. The growth can be described by using the following order of magnitude estimate for the integral in Eq.~(\ref{fl}),
\begin{eqnarray}&&\!\!\!\!\!\!\!\!\!
\dot V_l\!=\int_{\partial V_l} \bm v\cdot d\bm S\!\sim |\delta v(l(t))| A(l(t)),\label{fl1}
\end{eqnarray}
where by assumption we do not distinguish between the velocity and its potential component in scaling estimates. Here $l(t)$ is the typical linear size of the volume that could be defined as the gyration radius. After transients that size obeys $l(t)\sim t^{1/(1-\sigma)}$, see Eq.~(\ref{scals}) so that $|\delta v(l(t))|\sim l^{\sigma}(t)\sim t^{\sigma/(1-\sigma)}$.
The growth is not self-similar because of the multifractality of the volume's shape that causes non-trivial dependence of moments of $A(l)$ on $l$. We can write $A(l)\sim l^{\delta}$ where $\delta$ fluctuates due to multi-fractality and typically obeys $2<\delta<3$. We find $A(l(t))\sim t^{\delta/(1-\sigma)}$ which use in Eq.~(\ref{fl1}) gives that,
\begin{eqnarray}&&\!\!\!\!\!\!\!\!\!
\frac{dV_l}{dt}\!\sim t^{(\sigma+\delta)/(1-\sigma)};\ \ V_l(t)\sim t^{(1+\delta)/(1-\sigma)}.\label{volumesfg}
\end{eqnarray}
Thus the growth of the volume depends on both the scaling of velocity, described by the constant $\sigma$, as well as on the strength of the volume multifractality, described by the fluctuating exponent $\delta$. The volume grows faster the higher the fractality or $\delta$ is, because stronger fractality leaves the volume with more area to grow through. Similarly the growth is faster for rougher flow (smaller $\sigma$). Non-linearity of $\gamma_1(k)$ results from averaging powers of $V_l(t)$ in Eq.~(\ref{volumesfg}) over $\delta$. Similar considerations hold for time-reversed motion of the tracers so we can introduce
the scaling functions $\gamma_i(k)$,
\begin{eqnarray}&&\!\!\!\!\!\!\!\!\!\!\!\!\!\!\!\!\!
\left\langle V^{k}(t)\right\rangle\sim t^{\gamma_1(k)},\ \ t>0,\ \  \left\langle V^{k}(t)\right\rangle\sim |t|^{\gamma_2(k)},\ \ t<0, \label{vgrw}
\end{eqnarray}
cf. Eq.~(\ref{scals1}). The functions $\gamma_1(k)$ and $\gamma_2(k)$ differ because the flow statistics is not time-reversible \cite{review}.

We consider how the flow intermittency changes the above consideration. The velocity scaling exponent $\sigma$ is no longer constant. It changes in space, see e. g. the multifractal model \cite{frisch}. This implies the breakdown of self-similarity for the growth of $R(t)$. That can be described by the non-trivial scaling exponent function ${\tilde \gamma}(k)$ that describes the growth of the moments at large times,
\begin{eqnarray}&&\!\!\!\!\!\!\!\!\!\!\!\!\!\!
\left\langle R^{k}(t)\right\rangle\sim t^{{\tilde \gamma}(k)}, \label{scals1}
\end{eqnarray}
where at negligible intermittency (that depends on the Reynolds number) ${\tilde \gamma}(k)=k/(1-\sigma)$, see \cite{lanot} and references therein for observations of non-linearity. The intermittency causes fluctuations of $\sigma$ in $V_l(t)$ in Eq.~(\ref{volumesfg}) and also changes the statistics of fluctuations of $\delta$. All these fluctuations enter the final form of $\gamma_i(k)$ in Eqs.~(\ref{vgrw}).

We observe that Eq.~(\ref{scalingvol}) is Eq.~(\ref{defs2}) in logarithmic time variable. Thus we can simply use $\ln t$ instead of $t$ in the expressions for smooth case for the study of the PDF. We introduce the random variable,
\begin{eqnarray}&&\!\!\!\!\!\!\!\!\!\!\!
x_l(t)=\frac{\ln V_l(t)}{\ln t},\ \ V_l(t)=t^{x_l(t)}.
\end{eqnarray}
The variable $x_l$ describes the $V_l$ scaling with time and its dependence on the realization. In the phenomenological approach of Eq.~(\ref{volumesfg}) this variable is $(1+\delta)/(1-\sigma)$ corrected by
the $l-$dependent prefactor that is irrelevant asymptotically. The PDF $P(x_l, t)$ of $x_l(t)$ obeys,
\begin{eqnarray}&&\!\!\!\!\!\!\!\!\!\!\!
\langle V^k_l(t)\rangle\equiv \int P(x_l, t) t^{k x_l} dx_l\sim   t^{\gamma_1(k)},
\end{eqnarray}
cf. Eq.~(\ref{pdfcon}).
This equation occurs in the theory of fractal dimensions, see \cite{grb} and below. Its solution is,
\begin{eqnarray}&&\!\!\!\!\!\!\!\!\!\!\!
P(x_l, t)\sim t^{-S_1\left(x_l\right)}, \label{pdf}
\end{eqnarray}
where $S_1$ is the Legendre tranform of $\gamma_1$. This formula is what we would find by using in the formulas of the smooth case $\ln t$ instead of $t$. Similarly to the smooth case $S_1(x)$ has a unique maximum of zero at $x=\gamma_1'(0)$. Thus $\gamma_1'(0)$ can be considered as the counterpart of the sum of Lyapunov exponents in the inertial range (in contrast with the sum, it is dimensionless). In the limit of $t\to\infty$ the PDF becomes $\delta\left(x_l - \gamma_1'(0)\right)$. We reach the conclusion that volumes in the inertial range decay with probability one,
\begin{eqnarray}&&\!\!\!\!\!\!\!\!\!\!\!
\lim_{t\to \infty}\frac{\ln V_l(t, \bm x)}{\ln t}=\gamma_1'(0)<0.
\end{eqnarray}
This conclusion is of the same level of generality as for the smooth flows and constitutes one of the central results of this Section. Volumes of tracers decay at ${\rm Ma}>{\rm Ma}_s$ with probability one (which does not contradict conservation of $\langle V_l\rangle$ or growth of $\langle V_l^2\rangle$, cf. similar fact for the smooth flow). The volumes decay implies that the natural measure is singular as in non-degenerate smooth case.
Conversely, if ${\rm Ma}<{\rm Ma}_s$ then scalings of the velocity components differ and correlations functions of the concentration of tracers must not obey scaling as necessary for multifractality. We conclude that the transition of tracers to multifractality occurs at ${\rm Ma}_s$. This conclusion will be also obtained differently later.

We remark that the described behavior of $V_l$ can be used as long as the relevant linear dimensions belong to the supersonic inertial range. Our consideration describes volumes of tracers and
does not imply that fluid volumes would also decay at ${\rm Ma}>{\rm Ma}_s$ (of course the fluid fills the whole volume of the flow and the volumes' decay here is only the asymptotic property of the inertial range).
This is because of the volumes' back reaction on the transporting flow: the volumes resist compression. Should this be the case, the transition to multifractality would occur for the density at ${\rm Ma}_{cr}>{\rm Ma}_s$. It would happen on further increase of ${\rm Ma}$ since increased compressibility brings increased effective attraction between the fluid particles.

\subsection{Cascade picture of volume growth}

Here we give a more detailed, cascade picture of growth of volumes. Similarly to the traditional use of the cascade pictures in turbulence \cite{frisch} we consider the volume evolution as a sequence of steps. At each step the characteristic linear size $l$ of the volume changes by a factor of order one. This change is determined by turbulent eddies with size of order $l$. The locality of interactions \cite{frisch} implies that changes at different steps are approximately independent. In contrast with the usual cascade models, this one allows an almost rigorous derivation.

We rewrite Eq.~(\ref{fl}), which is valid at both positive and negative $t$, in the form
\begin{eqnarray}&&\!\!\!\!\!\!\!\!\!
\frac{d \ln V_l}{dt}\!=\!\int_{V_l(t)} \!\!\!w(t, \bm x) \frac{d\bm x}{V_l(t)}, \label{log0}
\end{eqnarray}
that demonstrates that the logarithmic rate of growth of the volume is given by the velocity divergence coarse-grained over that volume. We consider the long-time asymptotic regime where the statistics of the shape of $V_l(t)$ is time-independent, see the discussion after Eq.~(\ref{scalingvol}). The coarse-grained divergence has a non-trivial scaling in $l(t)$ with realization-dependent scaling exponent. We assume that the time of variations of the coarse-grained divergence is determined uniquely by $l(t)$ and denote this time by $\tau_{l(t)}$. This time is also the correlation time of the divergence. Assuming that the difference of the scaling exponents of the solenoidal and potential components can be neglected we find that $\tau_l$ scales inversely proportionally to $\int w(t, \bm x)d\bm x/V_l(t)$. This time-scale is imposed by the convective term of the Navier-Stokes equation, $\partial_t\bm v\sim (\bm v\cdot\nabla)\bm v$, in the same way as the time-scale $l/\delta v_l$ of eddies with scale $l$ and characteristic velocty difference $l/\delta v_l$, see \cite{frisch}. Thus the dimensionless random variable,
\begin{eqnarray}&&\!\!\!\!\!\!\!\!\!
\kappa_l\!=\!\tau_l\int_{V_l} \!\!\!w(\bm x) \frac{d\bm x}{V_l}, \label{kappa}
\end{eqnarray}
has distribution which is independent of $l$ and $V_l$ (and time). This key observation does not neglect intermittency because it includes the possibility of fluctuations of the scaling exponents of the flow divergence coarse-grained over $V_l(t)$. The time $\tau_l$ scales in $l$ so that characteristic linear size of the volume $l(t)$ increases within time $\tau_{l(t)}$ by an $l-$independent fluctuating factor $p>1$ which is of order one. We find by integration of Eq.~(\ref{log0}),
\begin{eqnarray}&&\!\!\!\!\!\!\!\!\!
\ln\left(\frac{V_l(t)}{V_l(0)}\right)\!=\!\int_0^t dt' \int_{V_l(t')} \!\!\!w(t', \bm x) \frac{d\bm x}{V_l(t')}\approx \sum_{i=1}^{N(t)}\kappa_i, \label{kaps}
\end{eqnarray}
where $N(t)$ is determined by the product of independent scaling factors $p_i$ where $p_i$ is the scale increase factor at the $i-$th step of the cascade,
\begin{eqnarray}&&\!\!\!\!\!\!\!\!\!\!\!\!\!\!\!
l\left\langle\prod_{i=1}^{N(t)} p_i \right \rangle\! =\! \left\langle l(t)\right \rangle\propto t^{1/(1-\sigma)},\ \ N(t)\!=\!\log_{\langle p\rangle}\frac{\left\langle l(t)\right \rangle}{l}.\label{kaps11}
\end{eqnarray}
Thus $N(t)$ has logarithmic behavior in time $t$. We observe that since intermittency is not disregarded then the velocity scaling exponent $\sigma$ is not defined uniquely. It is the law $\left\langle l(t)\right \rangle\propto t^{1/(1-\sigma)}$ that provides the unique definition in our consideration. The random variables $\kappa_i$ are independent identically distributed random variables so that,
\begin{eqnarray}&&\!\!\!\!\!\!\!\!\!\!\!\!\!\!
\left\langle \frac{V_l^k(t)}{V_l^k(0)}\right \rangle\!\sim\! \left\langle \exp\left(k\kappa\right)\right \rangle^{N(t)}\!\sim\! \left(\langle p\rangle^{N(t)}\right)^{\log_{\langle p\rangle} \left\langle \exp\left(k\kappa\right)\right \rangle}.
\end{eqnarray}
We conclude that,
\begin{eqnarray}&&\!\!\!\!\!\!\!\!\!
\left\langle V_l^k(t)\right \rangle\!\sim\! V_l^k(0)\left(\frac{\left\langle l(t)\right \rangle}{l}\right)^{\log_{\langle p\rangle} \left\langle \exp\left(k\kappa\right)\right \rangle},    \label{volgr}
\end{eqnarray}
which yields $\gamma_1(k)$ in Eq.~(\ref{vgrw}) as,
\begin{eqnarray}&&\!\!\!\!\!\!\!\!\!
\left\langle V_l^k(t)\right \rangle\!\sim\! V_l^k(0)t^{\gamma_1(k)},\ \
\gamma_1(k)= \frac{\log_{\langle p\rangle} \left\langle \exp\left(k\kappa\right)\right \rangle}{1-\sigma}. \label{fina}
\end{eqnarray}
This derivation provides a "microscopic" view of what forms $\gamma_1(k)$. The formula becomes transparent if we neglect fluctuations of $p$ and $\kappa$ using $\langle p\rangle$ instead of $p$ and some characteristic constant value $\kappa$. We would have then, using that in one step of the cascade the volume increases by factor of $\exp(\kappa)$ that,
\begin{eqnarray}&&\!\!\!\!\!\!\!\!\!
l\langle p\rangle^{N(t)}=\langle l(t)\rangle,\ \ V_l(t)\sim V_l(0) \exp(N(t)\kappa),
\end{eqnarray}
which is equivalent to Eq.~(\ref{volgr}).

The power-law for the volume growth given by Eq.~(\ref{fina}) is similar to the Richardson law however there is a significant difference. The volume is proportional to the initial volume and in the limit of zero $V_l(0)$ we find that $\left\langle V_l^k(t)\right \rangle$ also tends to zero. Thus, in contrast with Richardson law, the volume growth is not explosive. This could seem contradicting the observations of growing volume of initial "points" (small balls with radius in the inertial range) in incompressible turbulence \cite{review,phase}. However that growth is due to finite resolution scale: the infinitely resolved volumes are conserved by incompressibility. The proportionality of $V_l(t)$ to $V_l(0)$ seems to be necessary for consistency of formulas for concentration, see Eq.~(\ref{conc}) and the next Section.

We observe that the power-law behavior originates in the $l-$independence of the statistics of $\kappa_l$ in Eq.~(\ref{kappa}). If the scalings of the solenoidal and potential components do not agree then, since $\tau_l$ is determined by the solenoidal component of the flow, $\kappa_l$ is $l-$dependent. The volume growth does not obey then a power law. This introduces the condition ${\rm Ma}>{\rm Ma}_s$ in Eq.~(\ref{scalingvol}).

\section{Evolution to natural measure} \label{phen}
In this Section we consider the problem of describing the evolution of initial conditions to the natural measure. Here this process is described in terms of growth of the moments of coarse-grained concentration. This evolution can occur for instance when there is an externally caused influx of dust particles into a molecular cloud which is in a regime of stationary compressible turbulence. The influx prepares an initial condition for the continuity equation. We assume diluteness so the evolution of concentration of the injected dust particles is independent of other dust that may already be present in the cloud.


We study relaxation of smooth initial conditions to the singular natural measure by setting the initial condition in the remote past, at a negative time $t$, and studying the evolution of concentration at time zero $n(\bm x| t)$ at increasing $|t|$ (thus the second argument of $n(\bm x| t)$ is the time of setting the initial condition. The dependence on this time is studied when the moment of observation, taken to be zero, is fixed). The relaxation of $n(\bm x| t)$ to the stationary measure is characterized by growth of moments of coarse-grained concentration $n_l(\bm x| t)$ where the coarse-graining scale $l$ belongs to the supersonic inertial range, see the definition of notations in Eq.~(\ref{coarsed}). The steady-state moments of $n_l(t, \bm x)$ contain a large parameter $L/l$ and are much larger than the initial moments. We model the initial condition as a uniform distribution that we normalize by $n(t)=1$. In numerical simulations this setup can be realized by generating a stationary compressible turbulence with simulations of the Navier-Stokes equations starting from some negative time
so that the steady state is reached by the time $t$. Then tracers are distributed uniformly over the volume of the flow at time $t$ and their spatial distribution at time zero $n(\bm x| t)$ is studied. In other words we study the time zero concentration of tracers that are at time $t$ distributed uniformly over a stationary turbulent flow that by itself exists from very large negative times.

The moments $\left\langle n_l^k(\bm x| t)\right\rangle$ grow starting from $\left\langle n_l^k(t, \bm x)\right\rangle=1$ and ending with the steady state values at $|t|\to\infty$. We observe that mass conservation means that the mass $4\pi l^3n_l(\bm x| t)/3$ equals $V_l(t)$. Indeed, the trajectories of all tracers located inside $V_l(t)$ converge inside the ball of radius $l$ at time zero and the total mass of these tracers is $V_l(t)$ because of $n(t)=1$. We find,
\begin{eqnarray}&&\!\!\!\!\!\!\!\!\!\!\!\!\!\!
\left\langle n_l^k(\bm x| t)\right\rangle=\frac{\left\langle V_l^k(t)\right\rangle}{(4\pi l^3/3)^k}\sim |t|^{\gamma_2(k)}, \label{conc}
\end{eqnarray}
where we used Eq.~(\ref{vgrw}).
This law holds at large $|t|$ as long as the typical linear size of $V_l(t)$ (similar to $l(t)$ in the consideration above) remains much smaller than $L$. This law could be more accessible for experimental tests than directly Eq.~(\ref{vgrw}).

We designate the (negative) time when the relevant size of $V_l(t)$ becomes comparable with $L$ by $t^*$. Since distances' evolution obeys a power law this time scales with $L$.
At $t<t^*$ the volume would not grow much more because it will become larger than the correlation length of the velocity divergence, cf. \cite{arxiv}.
This time depends on the order of the moment so we introduce $t^*(k)$ as the time beyond which the moment of order $k$ stops growing. We find that in the steady state,
\begin{eqnarray}&&\!\!\!\!\!\!\!\!\!\!\!\!\!\!
\left\langle n_l^k(\bm x| t)\right\rangle\sim |t^*(k)|^{\gamma_2(k)}\sim \left(\frac{L}{l}\right)^{{\tilde \zeta}(k)}, \label{conc11}
\end{eqnarray}
where ${\tilde \zeta}(k)$ includes both $\gamma_2(k)$ and the $k-$dependence of $t^*(k)$. We introduced $l$ for dimensional reasons as the only scale from which a dimensionless quantity can be formed with $L$. Similar consideration for the smooth case was performed in \cite{arxiv}.

\section{Hentschel-Procaccia and R\'{e}nyi dimensions}\label{s1}

In this Section we review the definitions of the generalized fractal dimensions and provide the multifractal formalism. This is done for filling the gap in the literature on compressible turbulence and demonstrating that many facts that hold for multifractal attractors of smooth chaotic systems can be transferred without change to multifractals formed by  non-differentiable rough flow in the inertial range. For instance for fractals formed by a smooth flow the local dimension is constant almost everywhere, that is except for points that carry zero total mass \cite{yb}. This is true also in our case. Other reason is introducing the less used R\'{e}nyi dimensions as a tool for studying the density. These overcome many problems in measurement.

When density and concentration fields are considered in the supersonic inertial range (that is when these fields are coarse-grained over a scale in the range) they manifest multifractality above ${\rm Ma}_{cr}$ and ${\rm Ma}_s$, respectively. For simpler notation we will consider the density and concentration fields in the formal limit of $\eta\to 0$, having in mind the inertial range at the intermediate range between $\eta$ and $L$. Then these smooth and differentiable fields become singular with support on a multifractal set in space. This set has zero volume so the probability of hitting it with a point is zero. The values of the density or the concentration are therefore either zero or infinity so that the fields somewhat resemble $\delta-$functions: they vanish almost everywhere but have finite integral. However the set on which the fields are infinite is much more complex than a point. For studying the distribution quantitatively one must therefore consider a finite physical quantity. This is provided by the total fluid mass in a ball of radius $l$,
\begin{eqnarray}&& \!\!\!\!\!\!\!\!\!\!\!\!\!\!
m_l(t, \bm x)=\int_{|\bm x'-\bm x|\leq l}\rho(t, \bm x')d\bm x'=\frac{4\pi l^3 \rho_l(t, \bm x)}{3}, \label{coarse}
\end{eqnarray}
which is considered as a function of center' position $\bm x$, and defines the coarse-grained density over scale $l$, $\rho_l(t, \bm x)$. Our considerations here are formulated in terms of the fluid density $\rho$ but apply equally well to the tracers' concentration $n$.  For smooth distributions the knowledge of the limiting behavior of $m_l(\bm x)$ at small $l$ is equivalent to the density. However, for non-smooth distributions there is no well-defined density (the limit is either zero or infinite) and we operate directly with $m_l(\bm x)$. The multifractality property is the statement of the existence of the limits,
\begin{eqnarray}&& \!\!\!\!\!\!\!\!\!\!\!\!\!\!
\lim_{l\to 0}\lim_{\eta\to 0} \frac{\ln m_l(t, \bm x)}{\ln (l/L)}\!=\!d(t, \bm x),\ \ m_l(t, \bm x)\!\sim\! \left(\frac{l}{L}\right)^{d(t, \bm x)},\label{inf1}
\end{eqnarray}
for $\bm x$ belonging to the multifractal, see e. g. \cite{yb,Harte}. Here for transparency we wrote explicitly the limit of zero $\eta$. This limit is taken before the limit of zero $l$ and the order of limits is significant. If we took first $l$ to zero and then $\eta$ to zero we would find $d(t, \bm x)=3$ for all spatial points $\bm x$ because the density is a smooth field at finite $\eta$. For $\bm x$ outside the multifractal the mass $m_l(t, \bm x)$ is zero in the limit of $l\to 0$ taken after $\eta\to 0$. (There are points of the multifractal for which the limit in Eq.~(\ref{inf1}) does not exist, however these points give zero contribution to the relevant moments of mass, see \cite{yb} and below). The corresponding limit field and the multifractal set for the concentration may be different but the line of consideration here and below is identical with the density. In the consideration below we assume that the limit of zero $\eta$ is taken so all distributions are multifractal, singular ones.

We observe that for one-dimensional curve, two-dimensional surface and three-dimensional continuum, $m_l$ has linear, quadratic and cubic dependencies on $l$, respectively. Thus $d(t, \bm x)$ defines a local dimension of the multifractal. The field of dimensions $d(t, \bm x)$ is statistically stationary. For homogeneous fractals this field is constant on the fractal but not for multifractals. However, it is true that $d(\bm x)$ is "almost" constant \cite{yb}. We transfer the proof from smooth systems for demonstrating this. The fluctuations of $d(\bm x)$ in space are studied by considering $d_l(\bm x)=\ln m_l(\bm x)/\ln (l/L)$ at arbitrarily small but
finite $l/L$. The single-point probability density function (PDF) $P(d_l)$ of $d_l(\bm x)$ is defined by,
\begin{eqnarray}&& \!\!\!\!\!\!\!\!\!\!\!\!\!\!
P(d_l)\!=\!\!\int\!\! \delta\left(\frac{\ln m_l(\bm x)}{\ln (l/L)}\!-\!d_l\right)\rho(\bm x) d\bm x. \label{defin}
\end{eqnarray}
The presence of $\rho(\bm x)$ in the averaging guarantees that only points $\bm x$ in the multifractal count so that $m_l(\bm x)\neq 0$. The PDF obeys,
\begin{eqnarray}&& \!\!\!\!\!\!\!\!\!\!\!\!\!\!
\int m_l^k(\bm x) \rho(\bm x) d\bm x=\int \left(\frac{l}{L}\right)^{kd}P(d_l) dd_l\sim\left(\frac{l}{L}\right)^{\xi(k)},\label{scaling1}
\end{eqnarray}
where $\xi(k)$ is the scaling exponent of the $k-$th moment. The H\"{o}lder's inequality $\left\langle m_l^{(1-\alpha)x+\alpha y}\rho\right\rangle \leq \left\langle m_l^{x}\rho\right\rangle^{1-\alpha}\left\langle m_l^{y}\rho\right\rangle^{\alpha}$ (recall that the angular brackets stand for spatial averaging) implies that $\xi(k)$ considered as a function of $k$ is a concave function. The scaling of the moments given by Eq.~(\ref{scaling1}) implies that $P(d_l)$ obeys the asymptotic form \cite{yb,grb,hals,krzysztof},
\begin{eqnarray}&& \!\!\!\!\!\!\!\!\!\!\!\!\!\!
P(d_l)\sim \left(\frac{L}{l}\right)^{S(d_l)}, \label{de}
\end{eqnarray}
where $S(d_l)$ is the rate function. This is similar to that introduced in the time domain in the previous Section and it is considered similarly. Carrying out now the integral in Eq.~(\ref{scaling1}) by the saddle point method we obtain that $\xi(k)=min_{d_l}\left[kd_l-S(d_l)\right]$. We find that $S(d_l)$ is concave, non-positive and has a unique maximum of zero. We denote the value for which that maximum is attained by $D(1)$. The PDF $P(d_l)$ becomes $\delta\left(d_l-D(1)\right)$ in the limit $l\to 0$. We conclude that $d(t, \bm x)=D(1)$ for all the points $\bm x$ on the multifractal except those that do not contribute to $P(d_l)$ in Eq.~(\ref{defin}) at $l\to 0$ and thus have zero total mass. The points $\bm x$ for which $d(t, \bm x)\neq D(1)$ are thus very "dilute" in space. They form zero mass set inside the zero volume multifractal. Still they are not negligible because they are dense in the multifractal. Thus however small ball of radius $l$ is considered, we still cannot set $m_l(\bm x)\sim l^{D(1)}$ for $\bm x$ obeying $d(t, \bm x)=D(1)$. This is because the ball contains points with $d(t, \bm x)\neq D(1)$ due to which $m_l(\bm x)\sim l^{D(1)}$ fails, see \cite{yb} and below.

We stress that Eq.~(\ref{de}) relies on the scaling of the moments only. Thus though it is derived originally for smooth chaotic systems \cite{yb,grb,hals,krzysztof} it holds in our case of rough velocity as well. It proves that the property that $d(\bm x)=D(1)$ holds for all $\bm x$ but those who have zero mass characterises also the considered rough velocity case. The function $S(d)$ represents the rate of decrease with $l$ of the fraction of the multifractal on which $d_l(\bm x)\neq D(1)$.

The multifractal nature of the distribution can be described by using the Hentschel-Procaccia (HP) \cite{hp,yb} spectrum of dimensions defined as,
\begin{eqnarray}&& \!\!\!\!\!\!\!\!\!\!\!\!\!\!
D(k)=\lim_{l\to 0}\frac{\ln \int m_l^{k-1}(\bm x) \rho(\bm x) d\bm x}{(k-1)\ln(l/L)}=\frac{\xi(k-1)}{k-1}. \label{hp}
\end{eqnarray}
The dimension $D(1)$ is ill-defined and the limit of $D(q)$ at $q \to 1$ in non-generic cases may depend on whether it is taken from the right or from the left. In our case the limits agree and the limiting value is found by L'Hospital's rule,
\begin{eqnarray}&& \!\!\!\!\!\!\!\!\!\!\!\!\!\!
D(1)=\lim_{l\to 0}\frac{\int \ln m_l(\bm x) \rho(\bm x) d\bm x}{\ln(l/L)}. \label{infim}
\end{eqnarray}
The $D(1)$ dimension is called information or entropy dimension since it measures information in the scatter of the multifractal in space, see below. The dimension is unique in that the logarithm of the mass is taken before the integration while it is the other way around in the definition of the $D(k)$ for $k\neq 1$ in Eq.~(\ref{hp}). Interchanging the order of the limit and the integral and observing that points with zero mass do not contribute to the density-weighted integral we see that $D(1)$ is that unique fractal dimension that holds for almost all the points of the multifractal, as is inferred from Eq.~(\ref{inf1}) and from the discussion that follows it.

Using Jensen's inequality it is seen that $D(k)$ is a non-increasing function of $k$. The value of $D(k)$ at zero is the box counting dimension of the multifractal \cite{hp,dimens,serguei}. The limiting value of $D(k)$ at $k=\infty$  in our case of random flow must be zero because of the existence of optimal fluctuation, see \cite{arxiv,krzysztof}, below and cf. \cite{hp}. We observe that $D(k)$ can be written as,
\begin{eqnarray}&& \!\!\!\!\!\!\!\!\!\!\!\!\!\!
D(k)=3-\lim_{l\to 0}\frac{\ln \left\langle \rho_l^{k-1} \rho\right\rangle}{(k-1)\ln(L/l)}, \label{sp}
\end{eqnarray}
where the coarse grained density $\rho_l$ is defined in Eq.~(\ref{coarse}). This form explicitly expresses the dimension deficit which is the difference between the space dimension three and $D(k)$.

We comment on the relevance of points $\bm x$ for which $d(t, \bm x)\neq D(1)$ in the calculation of $D(k)$ in Eq.~(\ref{hp}). These points have zero mass and thus can be taken out of the domain of integration. However, as told previously, this does not mean that these points are irrelevant and we can put $m_l(\bm x)\sim l^{D(1)}$ in the integral. This is because $m_l(\bm x)$ can be much larger or smaller than $l^{D(1)}$ however small $l$ is. This is due to persistent presence of points with $d(t, \bm x)\neq D(1)$ inside the ball of radius $l$. This phenomenon comprises the strong intermittency of multifractal statistics.

{\it Difficulty in experimental studies of dimensions and the R\'{e}nyi dimensions.}---Finding the moments of mass in Eq.~(\ref{scaling1}) from numerical simulations or experiments can be not straightforward. The usual procedure \cite{hp} starts from a large number $N$ of points with coordinates $\bm x_i$ on the multifractal. One has using discrete approximation $\rho(\bm x)=\sum_{i=1}^N \delta(\bm x-\bm x_i)/N$ that,
\begin{eqnarray}&& \!\!\!\!\!\!\!\!\!\!\!\!\!\!
\int m_l^k(\bm x) \rho(\bm x) d\bm x\approx \frac{\sum_{i=1}^N m_l^k(\bm x_i)}{N},
\end{eqnarray}
which becomes exact in the continuum limit of $N\to\infty$. However getting particles on the multifractal support of the fluid density can be non-trivial. The procedure of seeding the flow with a large number of tracer particles and studying their distribution is valid only provided that $n=\rho$ holds in the steady state which demands a separate proof.

Thus, in order to facilitate the calculations of the fractal dimensions independent of $n=\rho$ from numerical or experimental data, we introduce a different set of dimensions \cite{renyi} which seems to be more suitable for working with the density, the continuous R\'{e}nyi dimensions ${\tilde D}(k)$,
\begin{eqnarray}&& \!\!\!\!\!\!\!\!\!\!\!\!\!\!
{\tilde D}(k)=\lim_{l\to 0}\frac{\ln \int m_l^{k-1}(\bm x) \rho_l(\bm x) d\bm x}{(k-1)\ln(l/L)}\nonumber\\&&\!\!\!\!\!\!\!\!\!\!\!\!\!\!
=\lim_{l\to 0}\frac{\ln \int m_l^{k}(\bm x) d\bm x}{(k-1)\ln(l/L)}-\frac{3}{k-1}.
\end{eqnarray}
The difference between this definition and Eq.~(\ref{hp}) is that it uses for averaging of $m_l^{k-1}(\bm x)$ the coarse-grained density $\rho_l(\bm x)$ rather than the fine density $\rho(\bm x)$ (our definition uses balls rather than cubes which could make a difference \cite{renyi,dimens} but in our case seems irrelevant). In contrast with the HP dimensions, the R\'{e}nyi dimensions are ill-defined for $k<0$ since any hole with finite volume gives non-integrable $m_l^k$ for small $l$. However for $k>0$ the dimensions are well-defined and coincide with the HP dimensions as we demonstrate below. Thus ${\tilde D}(k)$ give a way of finding $D(k)$ overcoming the difficulties to the difference of the density and the natural measure.

We observe that $\int m_l^{k}(\bm x) d\bm x$ considered at fixed $l$ and $k\to 0$ is roughly the volume of points $\bm x$ for which $m_l\neq 0$. Correspondingly $\int m_l^{k-1}(\bm x) \rho_l(\bm x) d\bm x$ is that volume divided by $4\pi l^3/3$ which is the box counting dimension ${\rm dim}_{box}({\rm supp}\  \rho)$ of the support of $\rho$. Indeed, despite that interchange of the limits of zero $l$ and $k$ is not valid it will be seen below that,
\begin{eqnarray}&& \!\!\!\!\!\!\!\!\!\!\!\!\!\!
\lim_{k\to 0^+}{\tilde D}(k)={\rm dim}_{box}({\rm supp}\  \rho), \label{box}
\end{eqnarray}
(it is recalled that $\rho$ in this equation is the multifractal singular distribution obtained in the limit of zero $\eta$).

Assuming it exists, the dimension ${\tilde D}(1)$ yields the limiting value of the Gibbs entropy $-\left\langle \rho_l\ln \rho_l\right\rangle$ derived from the coarse-grained density $\rho_l$,
\begin{eqnarray}&& \!\!\!\!\!\!\!\!\!\!\!\!\!\!
{\tilde D}(1)\!=\!\lim_{l\to 0}\frac{\int \ln m_l(\bm x) \rho_l(\bm x) d\bm x}{\ln(l/L)}\!=\!3\!-\!\lim_{l\to 0} \frac{\left\langle \rho_l\ln \rho_l\right\rangle}{\ln (L/l)}.\label{ntr}
\end{eqnarray}
Thus $D(1)$ which equals ${\tilde D}(1)$, see below, derives from the entropy which is the reason why it is called the information dimension.

To gain further insight into the nature of the fractal dimensions, we notice that:
\begin{eqnarray}&& \!\!\!\!\!\!\!\!\!\!\!\!\!\!
\int m_l^k(\bm x) \rho_l(\bm x) d\bm x\sim\left(\frac{l}{L}\right)^{k{\tilde D}(k+1)}. \label{sm1}
\end{eqnarray}
The R\'{e}nyi dimensions may therefore be written as:
\begin{eqnarray}&&\!\!\!\!\!\!\!
{\tilde D}(k)=3-\lim_{l\to 0} \frac{\ln \left\langle \rho_l^k\right\rangle}{(k-1)\ln (L/l)}, \label{dimensions}
\end{eqnarray}
so that ${\tilde D}(k)$ describes the spatial statistics of the coarse-grained density, rather than the statistics obtained by density-weighted averaging as in Eq.~(\ref{sp}). The last term in Eq.~(\ref{dimensions}) is minus the dimension deficit. For smooth distributions $\rho_l$ is independent of $l$ at small $l$ and ${\tilde D}(k)=3$. We also observe that,
\begin{eqnarray}&&\!\!\!\!\!\!\!
\left\langle \rho_l^k\right\rangle=c_l \left(\frac{L}{l}\right)^{\zeta(k)},\ \ {\tilde D}(k)=3-\frac{\zeta(k)}{k-1}, \label{fc}
\end{eqnarray}
where $c_l$ are constants of order one. When these formulae are applied to the concentration they provide a way for finding fractal dimensions from the cascade model, cf. Eq.~(\ref{conc11}) and see below.

The R\'{e}nyi dimensions ${\tilde D}(k)$ lend themselves to easier calculations from numerical or experimental data. We demonstrate that in the range of their definition, $k>0$, they equal the HP dimensions, $D(k)={\tilde D}(k)$. The equality is readily demonstrated for positive integer $k$ greater than one. Indeed, in that case $D(k)$ describes the correlations of the positions of $k$ particles \cite{hp}. Writing $m_l^k(\bm x)=\int_{|\bm x_i-\bm x|<l} \prod_{i=1}^k \rho(\bm x_i) d\bm x_i$ we have:
\begin{eqnarray}&& \!\!\!\!\!\!\!\!\!\!\!\!\!\!
\int m_l^k(\bm x) \rho(\bm x) d\bm x\!=\!\int_{r_i<l}\!\! \left\langle \rho(0)\rho(\bm r_1)\ldots \rho(\bm r_k)\right\rangle\prod_{i=1}^k d\bm r_i, \label{cor1}
\end{eqnarray}
where the correlation function $\left\langle \rho(0)\rho(\bm r_1)\ldots \rho(\bm r_k)\right\rangle$ describes the probability of simultaneously finding $k$ particles at distances $\bm r_i$ from the origin given that there is a particle at the origin (here particle is the fluid particle). Similarly,
\begin{eqnarray}&& \!\!\!\!\!\!\!\!\!\!\!\!\!\!
\int\!\!  m_l^k(\bm x) \rho_l(\bm x) d\bm x\!\label{cor12}\\&&\!\!\!\!\!\!\!\!\!\!\!\!\!\!
=\!\frac{3}{4\pi l^3}\int_{r_i<l}\!\! \left\langle \rho(0)\rho(\bm r_2-\bm r_1)\ldots \rho(\bm r_{k+1}-\bm r_1)\right\rangle\prod_{i=1}^{k+1} d\bm r_i.\nonumber
\end{eqnarray}
These representations demonstrate therefore the identical scaling of $\left\langle m_l^k \rho \right \rangle$ and $\left\langle m_l^k \rho_l \right \rangle$ in $l$. Thus for all integers larger than one we have $D(k)={\tilde D}(k)$. Furthermore, it was proved in \cite{renyi} that $D(1)={\tilde D}(1)$. This equality is often taken for granted without proof \cite{hp}. We conclude that $D(k)={\tilde D}(k)$ holds for all positive integers. It was proved in \cite{renyi} that the equality can be continued for all $k>0$. Heuristic proof is obtained by observing that since the scale of spatial variations of $m_l(\bm x)$ is of order $l$ then at $l_0\ll l$ we have approximate equality,
\begin{eqnarray}&& \!\!\!\!\!\!\!\!\!\!\!\!\!\!
\int m_l^k(\bm x) \rho(\bm x) d\bm x\approx \int m_l^k(\bm x) \rho_{l_0}(\bm x) d\bm x. \label{equality}
\end{eqnarray}
As there is no pathology \cite{dimens}, we can continue this equality asymptotically to $l_0\sim l$ finding $D(k)={\tilde D}(k)$.

Other advantage of the use of the R\'{e}nyi dimensions is that these are directly addressed by the more intuitive Frisch-Parisi multifractal formalism \cite{frisch,frischparisi} which could be the starting point of the study of multifractality, cf. \cite{mandel0}. The multifractal is considered as a union of fractals formed by the level sets of $d(t, \bm x)$. We assign to the set of points with $d(t, \bm x)=d$ the Hausdorff dimension $S(d)+d$ where $S(d)$ for consistency with $D(k)={\tilde D}(k)$ must be the same function that appears in Eq.~(\ref{de}), see \cite{grb,yb} and below. The probability that a ball with radius $l$ randomly placed in the domain of the flow intersects a fractal set with dimension $S(d)+d$ behaves as
$l^{3-d-S(d)}$. This gives \cite{frisch,frischparisi},
\begin{eqnarray}&& \!\!\!\!\!\!\!\!\!\!\!\!\!\!\!\!
\int\!\! m_l^{k+1}(\bm x) d\bm x\!\sim \!\int_{d_{min}}^{d_{max}} \!\!\!\!\!\l^{kd+3-S(d)}dd\!\sim\! l^{3+\min_d[kd-S(d)]},
\end{eqnarray}
where $(d_{min}, d_{max})$ is the domain of variation of $d(t, \bm x)$. Comparing this with $\int\!\! m_l^{k+1}(\bm x) d\bm x\sim l^{3+k{\tilde D}(k+1)}$, see Eq.~(\ref{sm1}), we find that $k{\tilde D}(k+1)$ and $S(d)$ form Legendre transform pair. Since $kD(k+1)$ and $S(d)$ also form Legendre transform pair, see Eqs.~(\ref{scaling1})-(\ref{hp}) then we have $D(k)={\tilde D}(k)$ confirming the self-consistency of the consideration. The multifractal formalism described above lies at the origin of the multifractal model of turbulence \cite{frisch,frischparisi}. In fact, our singularities of mass in the inertial range are more similar to the singularities of turbulent velocity difference than to singularities of mass of attractors of smooth systems.

We comment that the scaling of mass and its moments is only approximate. This is due to finite difference of scalings of solenoidal and potential components of velocity discussed previously. When the above definitions are used for concentration of tracers, the simplest way of measuring the dimensions is by working with a large number of discrete particles obeying Eq.~(\ref{Lagrangian}). It must then be ensured that the number of particles in the considered small volumes is large. Otherwise discreteness of the particles causes deviations from the continuum behavior which we consider here. The impact of discreteness was obtained at small compressibility in \cite{itzhak}.

\section{Fractal dimensions of isothermal turbulence}\label{f}

In the case of isothermal turbulence numerical simulations performed at zero dissipative coefficients demonstrated that single-point spatial statistics of the density is log-normal, see e. g. \cite{krit2007,fractal}. The single-point density in these simulations is asymptotically the density in the inertial range coarse-grained over the resolution scale $l_0$ that is $\rho_{l_0}$
(the viscous scale and viscosity are set to zero). We take it as observed fact that $\rho_l$ in isothermal turbulence is log-normal at $l>\eta$ and derive from this the fractal dimensions.

The log-normality holds at intermediate Mach numbers and deviations from it were observed at higher Mach numbers \cite{fractal}. We demonstrate that deviations from lognormality must hold for the tail of the distribution for avoiding contradictory results concerning the fractal dimensions. These deviations are the consequence of the fact that the maximal value of $\rho_l$ cannot exceed the total mass of the system divided by $4\pi l^3/3$. Despite the triviality of this observation it has far-reaching consequences. The PDF of $\rho_l$, in contrast with the single-point PDF of the density $\rho$, has compact support given by the interval $(0, 3/(4\pi l^3))$. Thus the tail of the PDF is not log-normal. This corresponds to very special behavior of $D(k)$ at large $k$: we have $\langle m_l^k\rangle\sim l^{\delta_0}$ where $\delta_0$ is a $k-$independent constant. This type of behavior is caused by the optimal fluctuation where the maximal possible mass of order one is compressed in the ball of radius $l$, see \cite{arxiv,krzysztof,fz}. It must be clarified that $D(k)$ derive from the behavior of $m_l$, see the previous Section, and for them log-normality cannot hold for all $k$, see below. In contrast, the single-point PDF of the density can be log-normal with no contradiction. Finally the PDF of finitely resolved density
is considered at the PDF of $\rho_l$ with $l$ equal to the scale of spatial resolution.

We demonstrate that for lognormal statistics all the fractal dimensions ${\tilde D}(k)$ depend only on one parameter: the density spectrum decay exponent $\delta$. For that purpose we first notice that the inverse Fourier transform of the spectrum $E(k)\propto k^{\delta-1}$ results in the following pair-correlation function at separation in the inertial range,
\begin{eqnarray}&&\!\!\!\!\!\!\!\!\!\!\!\!
\langle \rho(0)\rho(\bm r)\rangle=\int \exp\left(i\bm k\cdot \bm r\right) E(k) \frac{d\bm k}{4\pi k^2}\approx c \left(\frac{L}{r}\right)^{\delta},
\end{eqnarray}
where $c$ is a constant of order one. We therefore find from Eqs.~(\ref{cor1})-(\ref{cor12}) for $k=1$ that $D(2)={\tilde D}(2)=3-\delta$. Since $D(2)$ is derived from pair correlations then it is called correlation dimension. Next, lognormality of $\rho_l$ is Gaussianity of $s_l\equiv \ln \rho_l$ so that the probability density function (PDF) $P_l(s)$ of $s_l$ obeys,
\begin{eqnarray}&&\!\!\!\!\!\!\!\!\!\!\!\!\!\!
P_l(s)=\langle \delta(s_l-s)\rangle=\frac{1}{\sqrt{2\pi \sigma_l^2}}\exp\left(-\frac{(s-\langle s_l\rangle)^2}{2\sigma_l^2}\right),\label{log}
\end{eqnarray}s
where we designated the dispersion of $s_l$ by $\sigma_l^2$. We have,
\begin{eqnarray}&&\!\!\!\!\!\!\!
\langle \rho_l^k\rangle=\left\langle \exp\left(ks_l\right)\right\rangle=\exp\left(k\langle s_l\rangle+\frac{k^2 \sigma_l^2}{2}\right),
\end{eqnarray}
where we used the Gaussian averaging formula for the average of the exponent of the Gaussian variable $ks_l$. Since $\langle \rho_l\rangle=1$ then setting in the above formula $k=1$ we find the identity $\langle s_l\rangle+\sigma_l^2/2=0$ which in turn yields:
\begin{eqnarray}&&\!\!\!\!\!\!\!\!\!\!\!\!\!\!\!\!\!
\langle \rho_l^k\rangle\!=\!\exp\left(\frac{k(k\!-\!1) \sigma_l^2}{2}\right). \label{ln}
\end{eqnarray}
Employing now the last relationship in Eq.~(\ref{dimensions}) results in the following expression for the R\'{e}nyi dimensions for lognormal statistics:
\begin{equation}
{\tilde D}(k)\!=\!3\!-\!\lim_{l\to 0} \frac{k\sigma_l^2}{2\ln(L/l)}.
\label{ln1}
\end{equation}
Thus in the lognormal case the spectrum of codimensions $3-{\tilde D}(k)$ is a linear function of $k$, see a similar observation for inertial particles below the viscous scale in \cite{fouxon}. Setting finally $k=2$ we find,
\begin{eqnarray}&&\!\!\!\!\!\!\!
\delta=3-{\tilde D}(2)=\lim_{l\to 0} \frac{\sigma_l^2}{\ln(L/l)},\ \ \sigma_l^2\sim \delta \ln \left(\frac{L}{l}\right).\label{kl}
\end{eqnarray}
This implies
\begin{eqnarray}&&\!\!\!\!\!\!\!\!\!\!\!\!\!\!\!\!\!
\langle \rho_l^k\rangle=\left(\frac{L}{l}\right)^{k(k-1) \delta/2}. \label{densitysc}
\end{eqnarray}
We conclude that fractality, which implies that the correlation dimension $D(2)$ is strictly smaller than $3$ (recall that $D(2)$ is not larger than the information ($D(1)$) and box counting ($D(0)$) dimensions by Jensen's inequality) then $\delta=3-D(2)$ must be a positive number when the distribution is multifractal.

It is of interest to note the logarithmic dependence of the dispersion $\sigma_l^2$ on the coarse-graining scale. That dispersion has been intensively studied in terms of the single-point density which asymptotically gives rise to $\sigma_{l_0}^2$, see e. g. \cite{krit2007,fractal}. However, these studies did not consider the dependence of the dispersion on the resolution's scale, with exception of \cite{fractal}. Indeed, the logarithmic dependence of $\sigma_{l_0}^2$ on $l_0$ is quite slow and it was not detected. In the passage from $256^3$ grid to $1024^3$ considered in \cite{fractal} the dispersion changes only by about twenty per cent which is the probable reason for why the dependence was not observed previously. We stress that these properties of $\sigma_{l_0}^2$ apply only in the case where the density is multifractal which is the regime of supercritical Mach numbers larger than ${\rm Ma}_{cr}\sim 7$, see below. Thus there is no contradiction with previous studies that examined $\sigma_{l_0}^2$ at smaller ${\rm Ma}$.

Returning now to Eqs.~(\ref{ln})-(\ref{kl}), the fractal dimensions may be conveniently expressed in the following way:
\begin{eqnarray}&&\!\!\!\!\!\!\!
D(k)={\tilde D}(k)=3- \frac{k\delta }{2}, \ \ k>0.  \label{fd}
\end{eqnarray}
We consider the implication of the above relationship on the HP and the R\'{e}nyi dimensions. We observe that lognormality of $\rho_l$ with respect to the spatial averaging implies also lognormality with respect to the mass-weighted averaging weighted by the coarse-grained density $\rho_l$. Indeed, using Eq.~(\ref{log}) we find \cite{hopkins2013},
\begin{eqnarray}&&\!\!\!\!\!\!\!
{\tilde P}_{l}(s)\equiv\int \delta(\ln \rho_l(\bm x)-s)\rho_l(\bm x) d\bm x=\exp(s)P_l(s)\nonumber\\&&\!\!\!\!\!\!\! =
\frac{1}{\sqrt{2\pi \sigma_l^2}}\exp\left(-\frac{(s+\langle s_l\rangle)^2}{2\sigma_l^2}\right),
\end{eqnarray}
where we used $\langle s_l\rangle=-\sigma_l^2/2$. Using the definition of $D(k)$ in Eq.~(\ref{sp}) we have from Eq.~(\ref{fd}),
\begin{eqnarray}&& \!\!\!\!\!\!\!\!\!\!\!\!\!\!
\left\langle \rho_l^{k-1} \rho\right\rangle\sim \left(\frac{L}{l}\right)^{k(k-1)\delta/2}.
\end{eqnarray}
This implies that,
\begin{eqnarray}&&\!\!\!\!\!\!\!\!\!\!\!\!\!\!\!\!\!
\int\!\! \delta(\ln \rho_l(\bm x)\!-\!s)\rho(\bm x) d\bm x\!\sim\! \frac{1}{\sqrt{2\pi \sigma_l^2}}\exp\left(\!-\frac{(s\!+\!\langle s_l\rangle)^2}{2\sigma_l^2}\right).\label{asymp}
\end{eqnarray}
Thus in accord with Eq.~(\ref{equality}) at $l_0=l$ we find that $D(k)={\tilde D}(k)$ implies that,
\begin{eqnarray}&&\!\!\!\!\!\!\!
\int \delta(\ln \rho_l(\bm x)-s)\rho(\bm x) d\bm x\sim \int \delta(\ln \rho_l(\bm x)-s)\rho_l(\bm x) d\bm x.\nonumber
\end{eqnarray}
This consideration raises the question whether the asymptotic equality in Eq.~(\ref{asymp}) can be replaced by approximate equality, that is if $\rho_l$ is lognormal not only with respect to the spatial average with weight $d\bm x$ but also with respect to the mass-weighted average with weight $\rho(\bm x)d\bm x$. This question is relevant because log-normality in cascade model would hold for mass-weighted and not space-weighted average. This question is left for future work.

\subsection{Correcting phenomenology of compressible turbulence}

A main problem of the currently existing phenomenology of supersonic turbulence is that scaling of the density is described with only one scaling exponent, cf. the Introduction. Thus the density is assumed to be fractal and not multifractal. This causes problems in the agreement with the data as inspection of \cite{krit2007} reveals. The authors considered the fit of the phenomenology with the observed scalings of the first and second order structure functions. They saw that the density scaling exponent that must be constant in the phenomenological theory differs for these orders by forty per cent. If we consider also the data brought for the third order structure function we find the variation of already about one hundred per cent. Thus the correction of the phenomenology for multifractality seems necessary. Here we demonstrate difficulties in this correction.

We use consideration similar to \cite{krit2007}. It was assumed in accord with the observations that $\rho^{1/3}\bm v$ has behavior similar to the velocity of incompressible turbulence,
\begin{eqnarray}&&\!\!\!\!\!\!\!
\left\langle |\rho^{1/3}(\bm r)\bm u(\bm r)-\rho^{1/3}(0)\bm u(0)|^p\right\rangle\sim  r^{p/3},
\end{eqnarray}
where the intermitency corrections are assumed to be small. The issue is how given the above formula and the scaling of the density described by Eq.~(\ref{densitysc}) we can infer the scaling exponents of the velocity. The straightforward approach of considering
\begin{eqnarray}&&\!\!\!\!\!\!\!\!\!\!\!\!\!\!\!
\left\langle |\bm u(\bm r\!+\!\bm l)\!-\!\bm u(\bm r)|^p\right\rangle\!\sim \! \frac{\left\langle |\rho^{1/3}(\bm r)\bm u(\bm r)\!-\!\rho^{1/3}(0)\bm u(0)|^p\right\rangle}{\left\langle\rho_l^{p/3}\right\rangle},
\end{eqnarray}
is invalid. Indeed this would produce Kolmogorov scaling of the third order structure function on setting $p=3$. That scaling is at significant variance with the observations \cite{krit2007}. We leave the question of how the phenomenology must be corrected for future work.

\subsection{Breakdown of lognormal approximation}

The linear dependence of the fractal dimensions on the order, given by Eq.~(\ref{fd}), gives wrong prediction of negative $D(k)$ at large $k$. The contradiction with the demand that $D(k)\geq 0$ is caused by the use of lognormal distribution beyond the domain of its applicability. High order moments of the mass are determined by the optimal fluctuation where the flow produces as large mass $m_l$ in a ball of radius $l$ as possible, which is mass of order one, see \cite{arxiv,krzysztof,fz}. This can be seen by considering which spatial regions determine the space average of $\left\langle m_l^k\right\rangle$ at $k$ as large as we wish. We observe that there are rare regions in space where $m_l$ reaches the maximal possible value. These regions are formed when the mass is compressed from all sides from the correlation length $L$ inside $l$ giving $m_l\sim 1$ (the total mass of the system is one and we assume here that $L$ is comparable with the system size. If the system size is much larger than $L$ the conclusion does not change). In these regions the density is increased by factor of $(L/l)^3$ in comparison with the average. When we consider larger $k$ the contribution of these regions in $\left\langle m_l^k\right\rangle$ becomes more and more pronounced until these regions become the regions that define the moment.
A straightforward way to see this is to write 
\begin{eqnarray}&&\!\!\!\!\!\!\!
\left\langle m_l^k\right\rangle=\int_0^1 m_l^k P(m_l)dm_l,
\end{eqnarray}
where $P(m_l)$ is the PDF of $m_l$ and we stress that $m_l$ has a limited domain of variation. This way of writing makes it obvious that as $k$ increases the integral becomes determined by the upper limit of integration. 
The probability $\sim P(m_l=1)$ of the optimal fluctuation where mass of order one is compressed into the ball of radius $l$ is not describable by the lognormal distribution. It corresponds to large deviations that form the right tail of the PDF of $\ln \rho_l$. This breaks the lognormality assumption even in the simpler case of concentration of tracers transported by smooth weakly compressible flow \cite{fz,krzysztof}. The inconsistency of lognormal approximation to multifractality for high-order moments was stressed in \cite{mandel}.

Other consequence of Eq.~(\ref{fd}) is the equality of the box counting dimension $D(0)$ to the full spatial dimension three. The equality of spatial and box counting dimensions was observed previously \cite{dor,tasaki,fouxon,fz}. It is plausible that $D(0)=3$ holds also for other non-isothermal types of compressible turbulence, see \cite{fz}.

We conclude that $D(k)$ at large $k$ are determined by the non-lognormal right tail of the PDF of $\ln \rho_l$, respectively. The actual range of $k$ where Eq.~(\ref{fd}) is valid must be determined numerically. It is plausible that this range includes $k=1$ and $k=2$. Indeed, $D(1)$ is determined by typical events (see the previous Section) that determine the peak of the lognormal distribution. The mass moment that determines $D(2)$ is determined by the right portion of the distribution of $s_l$ which is well-described by lognormal distribution at relevant Mach numbers \cite{krit2007}. We conclude that the information dimension obeys $D(1)=3-\delta/2$.

We remark that the last formula must be taken with reservation.  It is found that $k$, starting from which $\langle m_l^k\rangle$ and $D(k)$ are determined by the optimal fluctuation, does not need to be large due to intermittency. For the solution of a two-dimensional smooth system (which does not differ in the considered aspect much from our rough case) presented in \cite{krzysztof} the correlation dimension $D(2)$ can be determined by the optimal fluctuation. Then the log-normal approximation for $D(k)$ given by Eq.~(\ref{fd}) cannot be used for connecting $D(1)$ and $D(2)$ and $D(1)=3-\delta/2$ does not hold. It is thus highly relevant for the future (numerical) studies to be able to address the $k-$dependence of $D(k)$ and test Eq.~(\ref{fd}).

\subsection{Consistency with literature}

Our derivation in this Section provides a rigorous connection between the spectrum of the density and multifractality. We compare this result with literature in the field.

The density spectrum $E(k)$ in the supersonic inertial range of wavenumbers obeys a power-law $k^{\delta-1}$ which is cut off at $\eta^{-1}$. The exponent $\delta$ is an increasing function of ${\rm Ma}$ \cite{kimryu,fractal}.  It crosses zero from below \cite{krit2007,fractal} at a critical Mach number ${\rm Ma}_{cr}\sim 6-8$. The crossing produces a qualitative change of behavior of the average squared density $\int_0^{\infty} E(k)dk$. At ${\rm Ma}<{\rm Ma}_{cr}$ the integral $\int_0^{\infty} E(k)dk$ is determined by wavenumbers of the order of $L^{-1}$ staying finite in the limit $\eta\to 0$. In contrast, at ${\rm Ma}>{\rm Ma}_{cr}$ the integral is determined by wavenumbers of order $\eta^{-1}\gg L^{-1}$ and the average squared density is infinite at $\eta\to 0$. This gives description of the transition to multifractality.
The density at ${\rm Ma}<{\rm Ma}_{cr}$ is a large-scale non-fractal field whose pair correlation function is constant in the inertial range. Second moment of density difference at points separated by a distance $r$ in the inertial range scales as $r^{|\delta|}$. In contrast, density at ${\rm Ma}>{\rm Ma}_{cr}$ is a small-scale field which pair-correlation function proportional to $r^{-\delta}$ grows downscales until the cut off at $\eta$. Density dispersion is determined by spatial fluctuations with scale $\eta$ and diverges in the limit $\eta\to 0$. The statistics is multifractal in the inertial range above $\eta$ with $3-\delta$ called the correlation dimension.

The current description of this transition to multifractality in the literature is misleading.

It was observed in \cite{krit2007} that the information dimension $D(1)$ is smaller than three at $\delta<0$. Since the correlation dimension must be smaller than the information dimension, see \cite{hp} and Sec. \ref{s1}, then this observation implies $D(2)<3$ at $\delta<0$. This is in contradiction with the paragraph above that gives $D(2)=\min[3, 3-\delta]$. This contradiction and the paragraph above hold irrespective of the validity of lognormality. However since lognormality is obeyed in the considered simulations then there is also contradiction with the formula $D(1)=3-\delta/2$. One reason for the discrepancy could be the incomplete resolution of $\delta$: it was found to be resolution-dependent where the sign of $\delta$ depends on the resolution. The other reason seems to be the consequence of non-rigorous procedure used in determining $D(1)$ that employed the following definition: 
\begin{eqnarray}&&\!\!\!\!\!\!\!
D(1)={\tilde D}(1)=\lim_{l\to 0}\frac{\int \ln m_l(\bm x) \rho_{l_0}(\bm x) d\bm x}{\ln(l/L)},
\end{eqnarray}
where $l_0$ is the resolution scale. The procedure for calculating the integral in the above definition used only points $\bm x$ where the density $\rho_{l_0}(\bm x) $ is higher or equal to half the maximal density in the considered snapshot (with further time averaging). These points were identified as belonging to the multifractal. The rationale for this procedure is not obvious.

It was proposed in \cite{fractal} that transition to fractality occurs when $\delta$ crosses minus one and not zero. However the $k^{-2}$ spectrum corresponding to $\delta=-1$ describes linear scaling of the squared density difference with separation. This can be the case of a passive scalar in incompressible flow \cite{review} or Burgers turbulence \cite{fb} both of which are not fractal. However as we stressed, for $0>\delta>-1$ the average squared density $\int E(k)dk$ is finite at $\eta\to 0$ and this cannot be so for a singular fractal distribution. Moreover this reference proposed that ``the"  fractal dimension, which is probably the information dimension, is $5/2-\delta/2$ that differs from Eq.~(\ref{fd}). The proposed heuristic derivation does not consider the fluctuations of the scaling exponents of the mass $m_l$ with $l$ in space. Consequently there are differences in the prediction for the limiting information dimension at ${\rm Ma}=\infty$. In that limit $E(k)$ becomes a constant, corresponding to $\delta=1$. The prediction for the limiting dimension from Eq.~(\ref{fd}) would be
$5/2$ and not $2$ proposed in \cite{fractal}. However our prediction for the limiting dimension must be taken with a reservation. It is probable from the indication of lognormality breakdown at higher ${\rm Ma}$ in \cite{fractal} that at ${\rm Ma}\to\infty$ lognormality breaks down and a reconsideration of Eq.~(\ref{fd}) is needed.

The considerations descibed above apply only in the case where the density is multifractal which is the regime of supercritical Mach numbers larger than ${\rm Ma}_{cr}\sim 7$, see below. We see that all $D(k)$ describable by Eq.~(\ref{fd}) are uniformly smaller than three at $\delta>0$. This is not necessarily so: in fact, moments of mass of different orders may transit to multifractal behavior at different critical Mach numbers. This could be the case of $D(k)$ with large $k-$s that are not describable by the lognormal distribution. For instance dimensions of moderate orders could still be three when the high $k$ moments of the mass are already determined by the optimal fluctuation producing $D(k)<3$. Thus the critical Mach number for which $D(k)$ becomes smaller than three could depend on $k$. The study of these questions is beyond our scope here.

\section{Cascade and transition to multifractality}\label{cas}

In this Section we apply the cascade model developed in the previous Section to the steady state statistics.

\subsection{Cascade model for correlation functions}

We consider the representation of the pair-correlation function given by Eq.~(\ref{pair}). We demonstrate that for $f(r)$ to obey a power law, which is the case of the multifractal phase of the concentration, the scalings of the compressible and solenoidal components of velocity must coincide. This is in accord with previous considerations.

We designate the turnover time of eddies of size $r$ by $t_r$. We observe that the trajectories $\bm q(t, \bm r))$ and $\bm q(t, 0)$ in Eq.~(\ref{pair}) diverge backward in time by distance $L$ in characteristic time $t_L$. Beyond this time $w(t, \bm q(t, \bm r))$ and $w(t, \bm q(t, 0))$ become decorrelated so the contribution of times $t<-t_L$ is negligible. As a result the lower integral limit in Eq.~(\ref{pair}) may be replaced by $-t_L$.
Then we represent the exponent as sum of contributions of time intervals during which the scale depletes by a factor of $e$ (or any other factor of order one),
\begin{eqnarray}&&\!\!\!\!\!\!\!\!
f(\bm r)\!\approx \!\left\langle\!\exp\left(\!-\!\sum_{i=1}^{N+1} \!\!\int_{t_{i-1}}^{t_i}\!\!\!\!\left[w(t, \bm q(t, \bm r))\!+\!w(t, \bm q(t, 0))\right]dt\right)\!\right\rangle,\nonumber
\end{eqnarray}
where $N=\ln (L/r)$ and $t_i=-t_{L/e^i}$. We set $t_{N+1}=0$. The contribution of the $i-$th interval is determined by eddies of size $Le^{-(i-1)}$ and can be considered independent of the other intervals because of the approximate independence of eddies with very different size (locality of interactions) \cite{frisch}. We find,
\begin{eqnarray}&&\!\!\!\!\!\!\!\!
f(\bm r)\!\sim \!\prod_{i=1}^{N+1} \!\! \left\langle\!\exp\left(\!-\! \!\!\int_{t_{i-1}}^{t_i}\!\!\!\!\left[w(t, \bm q(t, \bm r))\!+\!w(t, \bm q(t, 0))\right]dt\right)\!\right\rangle,\nonumber
\end{eqnarray}
This formula manifests the cascade like formation of pair correlations \cite{frisch}. It is readily seen now that we need to assume that all the terms in the exponents are $i$-independent in order to obtain a scaling law for the pair correlation. This independence indeed occurs only if the scalings of the compressible and solenoidal components are the same: both $t_i^{-1}$ as well as the divergence in that time interval are then estimated as $\delta v_r/r$ where $\delta v_r$ is the characteristic velocity difference at scale $r$, cf. the previous Section. Designating therefore the positive average of the $i-$th term by $\exp(\beta)$ we find ($N+1\approx N$),
\begin{eqnarray}&&\!\!\!\!\!\!\!\!\!\!\!\!\!\!\!\!\!
f(r)\sim \exp(N\beta)=\left(\frac{L}{r}\right)^{\beta}.
\end{eqnarray}
Thus the correlation dimension of the concentration, namely, $3-\beta$ obeys $\beta=\ln \left\langle  \exp(2\kappa)\right\rangle$ where the random variable ${\tilde \kappa}$ is the product of $t_r$ and the fluctuating divergence $w_r$ at scale $r$ which is assumed to have $r-$independent PDF. This variable is equivalent of $\kappa_l$ in Eq.~(\ref{kappa}).

Finding $f(r=l)$ is equivalent to the knowledge of the dispersion of the coarse-grained concentration  $n_l$, see Eq.~(\ref{cor12}) at $k=1$. Thus the above consideration establishes the cascade model representation of $\langle n_l^2\rangle$. Similar considerations hold for the moments of $n_r$ with higher integer order. The moment of order $k$ is found by integration of the $k-$point correlation function $f_k=\langle n(\bm r_1)n(\bm r_2)\ldots n(\bm r_k)\rangle$ that obeys,
\begin{eqnarray}&&\!\!\!\!\!\!\!\!\!\!\!\!\!\!\!\!\!
f_k\!=\!\left\langle\!\exp\left(\!-\!\sum_{i=1}^k\int_{-\infty}^0\!\! w(t, \bm q(t, \bm r_i))dt\right)\!\right\rangle. \nonumber
\end{eqnarray}
generalizing Eq.~(\ref{pair}). We find that similarly to $f(\bm r)$ we can asymptotically cut the integrals at the time $-t^k_L$ at which the distances $|\bm q(t, \bm r_i)-\bm q(t, \bm r_l)|$ become equal to $L$. This brings the corresponding cascade representation of the $k-$th moment of $n_l$.

\subsection{Cascade model of natural measure}

We construct the cascade model representation for $n_l(\bm x)$ which contains more information than only the integer moments considered above. Our starting point is the representation $n_l\sim V_l(t^*)/(4\pi l^3/3)$ introduced at the end of Sec.~\ref{phen}. Here the asymptotic equality must be understood in the sense that the scaling of both sizes of the equation in $l$ is identical. This representation assumes that the growth
of the volume $V_l(t)$ backward in time can be characterized by a single length scale $l(t)$ giving the overall size of the volume. It is assumed that the growth effectively stops when $l(t)$ is of order $L$ so there is no secondary growth after $l(t)$ growing backward in time exceeds $L$. This reasonable assumption agrees with the consideration of integer moments of $n_l(\bm x)$ above and can be further argued for by using the Green function representation of $n(t, \bm x)$. Then the cascade model for creation of fluctuations of $n_l(\bm x)$ is obtained from the cascade model of the volume growth introduced at the end of Sec.~\ref{phen}. We find using Eq.~(\ref{kaps}),
\begin{eqnarray}&&\!\!\!\!\!\!\!\!\!\!\!\!\!\!
q_l=\ln n_l\sim \ln \frac{V_l(-t_*)}{4\pi l^3/3}=\sum_{i=1}^{N_l} \kappa_i,\label{logcascade}
\end{eqnarray}
where the number of cascade steps $N_l$ is determined by setting $l(t)=L$ in Eq.~(\ref{kaps11}).
In accord with the discussion at the end of Sec.~\ref{phen} we can disregard the fluctuations of $p$ which gives $l \left\langle p\right\rangle^{N_l}=L$. Thus $q_l$ is a sum of a large number $N_l=\log_{\langle p\rangle} (L/l)$ of identically distributed independent random variables. We stress \cite{frisch} that this does not necessarily imply lognormality but only a large deviations form, namely, $P_l(q)\sim \exp\left(-N_lH(q)\right)$ of the probability density function (PDF) $P_l(q)$ of $q_l$. The non-negative convex large deviations function $H(q)$, similar to the entropy of statistical physics, has a unique minimum of zero at $q=\langle q_l\rangle$. Since $N_l\gg 1$ then $P_l(q)$ is sharply peaked at $q=\langle q_l\rangle$. The moments of $q_l$ of not high order are determined by the region near the peak and can be obtained by using quadratic expansion of $H(q)$ near the minimum \cite{frisch}. This reproduces the central limit theorem. In contrast, the high-order moments are determined by the tail of the distribution and cannot be described by the Gaussian peak. Correspondingly the moments of the coarse-grained concentration $n_l=\exp(q_l)$ are not determined by the minimum because of exponentiation that includes the high-order moments. Thus $n_l=\exp(q_l)$ cannot be described by lognormal distribution \cite{frisch}. We rather have from Eq.~(\ref{logcascade})
\begin{eqnarray}&&\!\!\!\!\!\!\!\!\!\!\!\!\!\!
\langle n_l^k\rangle= \langle \exp(k \kappa)\rangle^{\log_{\langle p\rangle} \left(L/l\right)}=\left(\frac{L}{l}\right)^{\log_{\langle p\rangle} \langle \exp(k \kappa)\rangle}. \label{scalingexponents}
\end{eqnarray}
The scaling exponents of $\langle n_l^k\rangle$ depend on $k$ non-linearly which is the benchmark of the multifractal behavior. The exponent $\log_{\langle p\rangle} \langle \exp(k \kappa)\rangle$ divided by $k-1$ gives the dimension deficit $3-D(k)$, see Eq.~(\ref{dimensions}). The cascade model tells that multifractality arises because of fluctuations of concentration increase factor in one step $\exp(\kappa)$.

Construction of the cascade model above used evolution of trajectories backwards in time since it is this evolution that determines the concentration \cite{arxiv}. The more traditional forward in time form is obtained by time reversal of the previous consideration. Fluctuations of the concentration at scale $l$ are formed by compression of a blob of tracers whose initial size is of the order of the integral scale $L$ of the turbulence. Initial concentration inside the blob is roughly the average concentration $\langle n\rangle$. Transport of the particles leads to the fragmentation of the blob by a sequence of steps each of which decreases the characteristic length by a factor of $p>1$. Different steps are determined by eddies with significantly different scales and can be considered independent. The continuity equation implies that the increase factor of concentration in one step is $\exp\left(\tau_l w_l\right)=\exp(\kappa_l)$ where $w_l$ is the coarse-grained divergence. The rest of the consideration is straightforward.

The above conclusions rely on the assumption that the concentration's reaction on the flow is negligible. That assumption, of course, does not hold for the fluid density, which back reacts on the fluid velocity in a significant manner. Still the cascade model developed here for the passive concentration can be to some extent transferrable to the active fluid density in the isothermal case. The reason that this is possible despite the density's back reaction on the transporting velocity is that the force per unit mass exerted by the fluid pressure is proportional to gradient of the density logarithm, which is independent of the magnitude of the density, cf. \cite{nordlund99,bsk}. We observe that this view of formation of inhomogeneities is quite different from the model of superposition of many shocks used for the density \cite{Passot1998,nordlund99,bsk}. This demands further studies. Though the consideration of the last two Sections is quite qualitative, it provides a more consistent derivation of the cascade representation than used usually in turbulence \cite{frisch}. In what follows, these considerations are confirmed quantitatively.


\section{Multifractality of tracers: Markov property and zero modes}\label{mod}

Unless the equality of density and concentration, if true, is proved we have more knowledge of the statistics of active fluid density than of passive concentration in the multifractal phase. The density in isothermal turbulence is lognormal and no fact of similar simplicity holds for the concentration. The lognormality of the statistics of the concentration corresponds to neglecting higher than quadratic terms in the cumulant expansion of $N-$point correlation function, see Eq.~(\ref{dots}) and \cite{fouxon}, that would not hold usually. In this Section we study pair correlations of tracers and present theoretical reasons for the breakdown of lognormality.

\subsection{Pair correlations}

We study the pair correlation function $f(\bm r)=\langle n(0)n(\bm r)\rangle $ that gives the concentration spectrum $E_c(k)$,
\begin{eqnarray}&&
E_c(k)=4\pi k^2 \int \exp\left(-i\bm k\cdot\bm r\right)f(\bm r) d\bm r.
\end{eqnarray}
If anisotropy is relevant then averaging over directions of $\bm k$ must be introduced in the RHS. The pair-correlation function $f(\bm r)$ equals $\langle n\rangle^2$ times the radial distribution function (RDF) $g(\bm r)$, see Appendix \ref{rdf}. The RDF gives the probability of having two tracer particles separated by $\bm r$ in the steady state,
\begin{eqnarray}&&\!\!\!\!\!\!\!\!\!\!\!\!\!\!
g(\bm r)=\lim_{t\to\infty} P(\bm r, \bm r', t). \label{pdf}
\end{eqnarray}
Here $P(\bm r, \bm r', t)$ is the PDF of the distance between two tracers transported by turbulence given that the initial distance is $\bm r'$
\begin{eqnarray}&&\!\!\!\!\!\!\!\!\!\!\!\!\!\!
P(\bm r, \bm r', t)=\left\langle \delta(\bm q(t, \bm x_1+\bm r')-\bm q(t, \bm x_1)-\bm r)\right\rangle, \label{propog}
\end{eqnarray}
where the average is independent of $\bm x_1$ by spatial homogeneity.
Here and below the angular brackets stand for averaging over the statistics of the flow and at $t=0$:
\begin{equation}
P(\bm r, \bm r', 0)=\delta(\bm r'-\bm r). \label{cond}
\end{equation}
The equivalence of time averaging and averaging over realizations of the flow is demonstrated in Appendix \ref{rdf} where further details on definitions of $f(\bm r)$ and $g(\bm r)$ can be found. Below we do not distinguish $g(\bm r)$ and $f(\bm r)$ working in units with $\langle n\rangle=1$.

We consider the time evolution of the distance between two tracers in Eq.~(\ref{propog}). Typically the particles initially separate by distance of order of the size of the vessel which can be $L$ or larger. Then the particles perform occasional rare excursions to distances in the inertial range. The accumulation of statistics of these excursions forms $f(\bm r)$. This way of obtaining the RDF is inconvenient for the study because it involves besides the transport in the inertial range also the transport on the scale of the whole vessel. Below we describe how the properties of $f(\bm r)$ can by studied using only the inertial range statistics.


We derive local stationarity condition on $f(r)$ in the supersonic inertial range $r\ll L$. We observe that the Chapman-Kolmogorov equation is approximately true,
\begin{eqnarray}&&
P(\bm r, \bm r', t_1+t)\approx \int P(\bm r, \bm r'', t)P(\bm r'', \bm r', t_1) d\bm r'',\nonumber\\&& t_r\ll t \lesssim t_L, \ \ t_1\to \infty, \label{ma}
\end{eqnarray}
where $t_r$ is the turnover time of eddies with size $r$. For proving this we write,
\begin{eqnarray}&&\!\!\!\!\!\!\!\!\!\!\!\!\!\!
P(\bm r, \bm r', t_1\!+\!t)\!=\!\left\langle  \delta(\bm q(t_1\!+\!t, \bm x_1\!+\!\bm r')\!-\!\bm q(t_1\!+\!t, \bm x_1)\!-\!\bm r)\right\rangle
\nonumber\\&&\!\!\!\!\!\!\!\!\!\!\!\!\!\!
=\int \left\langle \delta(\bm q(t_1+t, \bm x_1+\bm r')-\bm q(t_1+t, \bm x_1)-\bm r)
\right.\nonumber\\&&\!\!\!\!\!\!\!\!\!\!\!\!\!\!\left.
\delta(\bm q(t_1, \bm x_1+\bm r')-\bm q(t_1, \bm x_1)-\bm r'')\right\rangle d\bm  r''.  \label{kl0}
\end{eqnarray}
We consider increasing $t$ at fixed large $t_1$ (we'll take $t_1\to\infty$ eventually). For flow fluctuations that created $\bm r(t_1+t)=\bm r$ the value of $r(t_1)$ increases with $t$. Indeed, the particles are most of the time separated by the distance of order one (size of the vessel) and we consider events where at some time before $t_1+t$ the particles started to approach each other for reaching the distance $r\ll L$ at the moment of observation. The decrease of distance from $\sim L$ to $r$ occurs by a cascade of (on average) contraction events. Qualitatively the distance decreases from $L$ to $L/2 $ due to transport by eddy of size $L$, then from $L/2$ to $L/4$ by transport by eddy of size $L/2$ and then this process continues until $r\ll L$ is reached. The eddies at the different steps of the cascade are independent because of locality of interactions. Thus if we take $t$ much larger than the eddy turnover time of eddies of scale $r$ then we can assume approximate independence of the degrees of freedom of the flow that form $\bm r(t_1+t)$ and $\bm r(t)$: the dependence comes only through eddies of size of order $r(t)$ that are correlated with $r(t)$ and determine the first step of the cascade process. Thus $\bm r(t)$ on large time-scales is approximately Markovian and we find Eq.~(\ref{ma}) by performing independent averaging of the $\delta-$functions on the RHS of Eq.~(\ref{kl0}). We find the stationarity condition,
\begin{eqnarray}&&\!\!\!\!\!\!\!\!\!\!\!\!\!\!
f(\bm r)\approx \int P(\bm r, \bm r', t)f(\bm r') d\bm r', \ \ t_r\ll t\ll t_L, \label{opa}
\end{eqnarray}
where we used Eq.~(\ref{pdf}). This condition was found previously in \cite{phase} for the white-noise model (considered below) where the Markov property holds exactly due to zero correlation time. This condition holds also in the smooth chaotic systems for $r$ in the smoothness (viscous) range of the flow \cite{do,krzysztof}.

Keeping in mind the condition given by Eq.~(\ref{cond}), the stationarity condition given by Eq.~(\ref{opa}) has a power law solution $f(r)\propto r^{-\beta}$,
\begin{eqnarray}&&\!\!\!\!\!\!\!\!\!\!\!\!\!\!
\int P(\bm r, \bm r', t)r'^{-\beta} d\bm r'=r^{-\beta}, \label{cons}
\end{eqnarray}
that holds at not too large times so that the characteristic value of $r'$ that determines the integral belongs to the inertial range, see Eq.~(\ref{opa}). For incompressible flow the PDF $P(\bm r, \bm r', t)$ is normalized not only with respect to $\bm r$ but also with respect to $\bm r'$ (the operator is Hermitian) so that uniform distribution with $\beta=0$ is a solution. For compressible flow $\int P(\bm r, \bm r', t) d\bm r'$ differs from one and $\beta\neq 0$. Assuming now isotropy of small-scale turbulence that holds when the inertial range is appropriately large, averaging of Eq.~(\ref{opa}) over the directions of $\bm r$ yields,
\begin{eqnarray}&&\!\!\!\!\!\!\!\!\!\!\!\!\!\!
f(r)\approx 4\pi \int_0^{\infty} {\tilde P}(r, r', t)f(r') r'^2dr',\label{sym}
\end{eqnarray}
where we denoted the angle-averaged $P(\bm r, \bm r', t)$ by ${\tilde P}(r, r', t)$, using the fact that it is independent of the direction of $\bm r'$ due to isotropy \cite{phase}. This condition is compatible with the power-law solution $f(r)\propto r^{-\beta}$ provided that ${\tilde P}(r, r', t)$ has a proper scaling dependence on its arguments. This scaling dependence holds only when the scalings of the compressible and solenoidal components of the flow coincide, see the previous Section and study of the model below, that is for ${\rm Ma}>{\rm Ma}_s$. We thus have,
\begin{eqnarray}&&\!\!\!\!\!\!\!\!\!\!\!\!\!\!
r^{-\beta}\approx 4\pi \int_0^{\infty} {\tilde P}(r, r', t) r'^{2-\beta}dr'.
\end{eqnarray}
This generalizes the condition on the scaling exponent of the pair-correlation function for smooth chaotic systems \cite{do} to our case of non-smooth rough velocity in the inertial range. We briefly sketch the derivation of the condition in the smooth case.

We start with Eq.~(\ref{sym}) that holds also in the smooth case. In this case smoothness implies that $\bm r(t)=W(t)\bm r'$ where $W(t)$ is the Jacobi matrix \cite{review}. We find that ${\tilde P}={\tilde P}(r, r', t)$ obeys,
\begin{eqnarray}&&\!\!\!\!\!\!\!\!\!\!\!\!\!\!
 {\tilde P}\!=\!\int\!\! \frac{d{\hat r}}{4\pi}\left\langle\delta\left(r{\hat r}\!-\!W(t)\bm r'\right)\right\rangle\!=\!\left\langle \frac{\delta(r|W(t){\hat r'}|^{-1}\!-\!r')}{4\pi  r'^2|W(t){\hat r'}|^3}\right\rangle.\nonumber
\end{eqnarray}
We find that $f(r)\sim r^{-\beta}$ solves Eq.~(\ref{sym}) provided the $\beta-3$th moment of the distance between the particles $\left\langle r^{\beta-3}(t) \right \rangle$ is conserved. This condition was derived in \cite{do}, the simplicity of finding the correlation dimension $3-\beta$ in comparison with other fractal dimensions was stressed in \cite{gp}. In the inertial range we cannot make similar angle averaging that would allow rewriting Eq.~(\ref{cons}) in terms of conserved moment of $r(t)$. In fact, we demonstrate that in our case $\left\langle r^{\beta-3}(t) \right \rangle$ is not conserved - it is rather divergent.
We observe that the characteristic time of reaching $r$ from initial distance $r'\gg r$ is independent of $r$. This is because of acceleration of the cascade: the total duration of the cascade process is of order of duration of the first step when the distance changes from $r'$ to $r'/2$ that is the turnover time of eddies with size $r'$. Hence $r'$ determining the integral in Eq.~(\ref{sym}) is independent of $r$: it is determined by the condition $t_{r'}\sim t$. Taking $t\sim t_L$ we find,
\begin{eqnarray}&&\!\!\!\!\!\!\!\!\!\!\!\!\!\!
f(r)\sim  P(r, L, t_L)L^3, \label{l}
\end{eqnarray}
where we used that concentration at scale $L$ decorrelates so that $f(L)$ is of order of squared mean concentration which is one in our normalization. For self-similar flow with identical (in reality close) scaling of compressible and solenoidal components  ${\tilde P}(r, r', t)$ has power-law behavior ${\tilde P}(r, r', t)\sim r^{-\beta}$ at small $r$, see concrete calculation for the model below. We find from Eq.~(\ref{l}),
\begin{eqnarray}&&\!\!\!\!\!\!\!\!\!\!\!\!\!\!
f(r)=\left(\frac{{\tilde L}}{r}\right)^{\beta},\ \ r\ll L, \label{par}
\end{eqnarray}
where ${\tilde L}\sim L$ so that the matching condition at $r\sim L$ holds. In contrast, if the flow is not self-similar then there is no power-law behavior and thus no multifractality. This supports that the transition to multifractality happens at the critical Mach number ${\rm Ma}_s$ where the difference of scalings of the velocity components becomes negligible. The value of ${\rm Ma}_s$ defined in this way depends on the needed resolution of the exponents and the resulting power-law of $f(r)$.

We conclude that the correlation codimension $3-\beta$ can be obtained as small first argument asymptotic behavior of ${\tilde P}(r, r', t)$,
\begin{eqnarray}&&\!\!\!\!\!\!\!\!\!\!\!\!\!\!
f(r)\!\propto \! r^{-\beta},\ \ {\tilde P}(r, r', t) \!\propto\! r^{-\beta}; \ \ r, r' \!\ll \!L,\ \ t\!\ll\! t_L.\label{predictions}
\end{eqnarray}
This behavior is independent of $r'$ and holds also in the limit $r'\to 0$, see the concrete calculation for the white-noise model in the next Section. The limit $r'\to 0$ is finite due to the explosive separation of trajectories in the inertial range, see below and \cite{phase}. Thus ${\tilde P}(r, t)={\tilde P}(r, 0, t)$ is the minimal object from which we can infer $\beta$. For self-similar velocities ${\tilde P}(r, t)$ has self-similar evolution with scaling variable determined by the counterpart of the Richardson law for compressible turbulence.

It must be stressed that the non-triviality of Eq.~(\ref{predictions}) is that we use here the asymptotic PDF in the inertial range that depends on time $t$. The equation itself is also true for long-time limit $P(\bm r, \bm r', t\to\infty)$ however it is trivial there, see Eq.~(\ref{pdf}). We observe also that since ${\tilde P}(r, r', t) \sim r^{-\beta}$ at small $r$ then $\langle r^{\beta-3}\rangle$ is the moment with largest order that diverges: moments of order larger that $\beta-3$ are finite and moments with smaller order diverge at zero argument. Very similar statement holds for correlation dimension of attractors of smooth systems \cite{yb}.  The same moment determines the correlation codimension in smooth chaotic flow where $\langle r^{\beta-3}(t)\rangle=r^{\beta-3}(t=0)$ is the unique non-trivial conserved moment of the inter-particle distance \cite{do}.

\subsection{Anomalous scaling of higher order correlations}

We observe from Eq.~(\ref{opa}) that the pair correlation function is invariant under the action of operator with kernel $P(\bm r, \bm r', t)$. Thus it is similar to the so-called zero mode \cite{phase}. Zero modes
are statistically conserved functions of the evolving spatial configuration of $n$ particles. They can have non-trivial scaling exponents, the fact that provided the key to the understanding of anomalous scaling of the passive scalar in incompressible turbulence, see \cite{shr,gr,kr,kr11} and the review \cite{review}. In the case of compressible turbulence the zero modes are the reason for strong breakdown of lognormality in the miltifractal regime, cf. \cite{krzysztof}. Lognormality property $\left\langle n(\bm r_1)n(\bm r_2)\ldots n(\bm r_n)\right\rangle=\prod_{i>k}\left\langle n(\bm r_i)n(\bm r_k)\right\rangle$ entails the normal scaling of the $n-$th order correlation function given by $n/2$ times the scaling exponent of $\left\langle n(0)n(\bm r)\right\rangle$. We say that the lognormality is broken weakly if the equality $\left\langle n(\bm r_1)n(\bm r_2)\ldots n(\bm r_n)\right\rangle=\prod_{i>k}\left\langle n(\bm r_i)n(\bm r_k)\right\rangle$ breaks down but the normal scaling of the $n-$th order correlation function still holds. Strong breakdown occurs when the scalings do not agree, the case referred to in \cite{review} as anomalous scaling.

The study of anomalous scaling involves considering the stationarity condition on the higher-order correlation function $f(\bm r_1,\ldots, \bm r_n)=\left\langle n(0)n(\bm r_1)\ldots n(\bm r_n)\right\rangle$. The derivation proceeds similarly to the pair correlation.
We find that the $(n+1)-$point correlation function is determined by the joint PDF of the distances between $n+1$ particles, that is the transition probability  $P(\bm R, \bm R', t)$. Here we introduced $(\bm R)=(\bm r_1, \bm r_2,\ldots, \bm r_n)$ where $\bm r_i$ is the distance from $i-$th particle to ``zeroth" particle, cf. \cite{phase}. We find,
\begin{eqnarray}&&\!\!\!\!\!\!\!\!\!\!\!\!\!\!
f(\bm R)\approx \int P(\bm R, \bm R'', t)f(\bm R'') d\bm R'', \ \ t_r\ll t\ll t_L, \label{higc}
\end{eqnarray}
which is direct generalization of Eq.~(\ref{opa}). Thus $\left\langle n(0)n(\bm r_1)\ldots n(\bm r_n)\right\rangle$ is the zero mode of the operator of Lagrangian evolution of distances between the particles $P(\bm R, \bm R'', t)$, cf. \cite{phase}. Inspection of the zero mode mechanism of anomalous scaling of passive scalar in incompressible turbulence \cite{review} reveals that normal scaling is highly implausible. The problem of actual computation of the exponents is formulated for a model in the next Section.



\section{Using the Markov property: the Kraichnan model}\label{Kraic}

In this Section we introduce the Kraichnan model of the statistics of the flow velocity relying on the Markov property derived in the previous Section. The purpose of this model is to facilitate the investigation of the dependence of the concentration statistics on the Mach number, both as determined by the ratio of the magnitude of the compressible and solenoidal components and by the difference of the scaling exponents of the components. Moreover the model can be used for the consistent study of anomalous scaling.

We observe that the Chapman-Kolmogorov equation given by Eq.~(\ref{ma}) can also be written for finite $t_1$ since all the used considerations apply. This time must be large so that the decoupling in the product holds. However we pick it not so large that $P(\bm r'', \bm r', t_1)$ in  Eq.~(\ref{ma}) can still be considered as the inertial range quantity. The general solution of the Chapman-Kolmogorov equation is,
\begin{eqnarray}&&\!\!\!\!\!\!\!\!\!\!\!\!\!\!\!\!\!
P(\bm r, \bm r', t)=\exp(t{\hat L})(\bm r, \bm r'),\ \ \partial_t P={\hat L}P, \ \ t\gg t_{r'},
\end{eqnarray}
where ${\hat L}$ is a time-independent linear operator. This operator describes long-time transport of pairs of particles and it 
depends on the properties of turbulence non-locally both in space and time. The direct study of ${\hat L}$ is hardly possible. 
For approximations we can consider this operator as a series in derivative operators. By the Pawula theorem, for not developing negative transition probabilities $P(\bm r, \bm r', t)$, approximations of this series must either stop at the second derivative terms or contain an infinite number of terms \cite{risken}. We will consider the most general second order approximation consistent with the spatial homogeneity and isotropy. It will be demonstrated in the next subsection that this approximation is a rigorous way of introducing eddy diffusivity. The approximation is provided by the famous Kraichnan model that helped the breakthrough in the undertsanding of transport by incompressible turbulence, see \cite{review} and references therein.
We use the formulation of \cite{arxiv}. The flow $\bm u$ is assumed to be a Gaussian random field with zero mean and pair-correlation function, 
\begin{eqnarray}&&\!\!\!\!\!\!\!\!\!\!\!\!\!\!\!\!\!\!\!
\langle u_i(t_1, \bm x_1)u_k(t_2, \bm x_2)\rangle=\delta(t_2-t_1)\left[V_0\delta_{ik}-K_{ik}(\bm r)\right],\label{mdl}
\end{eqnarray}
where $\bm r=\bm x_2-\bm x_1$. The most general form of $K_{ik}$ (obeying $K_{ik}(r=0)=0$) which is consistent with isotropy is ,
\begin{eqnarray}&&\!\!\!\!\!\!\!\!\!\!\!
2K_{ik}=\left[\frac{(r^4u)'}{r^3}-c\right]r^2\delta_{ik}-\left[\frac{(r^2u)'}{r}-c\right]r_ir_k, \label{tensor}
\end{eqnarray}
where $u$ and $v$ are certain scalar functions of $r$. Here the used white noise in time structure of the statistics is fixed uniquely by the Markov property. The Gaussianity is not a necessary assumption as long as the study is confined to the pair correlations. Indeed, the increments of white noise over small but finite time intervals are Gaussian by the central limit theorem, cf. with the Gaussianity of Langevin forces in the theory of Brownian motion and see the Kramers-Moyal coefficients in \cite{risken}. The only assumption that is introduced by the model is that in this model ${\hat L}(\bm r, \bm r')=\nabla_i\nabla_k K_{ik}(\bm r)\delta(\bm r-\bm r')$ where the operators act to the right, see \cite{arxiv}. Thus the operator ${\hat L}$ is the most general differential operator of the second order which is consistent with conservation of probability and isotropy (there could be also another term of $\nabla_i {\tilde u}_i(\bm r) \delta(\bm r-\bm r')$ however this term would correspond to a mean flow). 

The tensor $K_{ik}(\bm r)$ or equivalently the functions $u(r)$ and $c(r)$ must be picked for best fitting of the data. For instance the arguments of the previous Section demonstrate that at small $r$ the NS turbulence corresponding to the multifractal phase of the tracers gives,
\begin{eqnarray}&&\!\!\!\!\!\!\!\!\!\!\!\!\!\!\!\!\!
P(\bm r, 0, t)\sim \frac{const}{r^{\beta}}, \ \ r\to 0. 
\end{eqnarray}
This behavior is reproduced by the model described by Eq.~(\ref{mdl}) with certain values of the constant in the numerator and the exponent $\beta$ in the denominator derived from $u(r)$ and $c(r)$, see below. These values must be gauged so that the above behavior is reproduced. More information on how the functions $u(r)$ and $c(r)$ are fixed can be found in Appendix \ref{krm} where the model is introduced in detail. It is demonstrated there that $u(r)$ is a linear combination of the inverse Fourier transforms $f(r)$ and $h(r)$ of effective solenoidal and potential spectral functions. The functions $f(k)$ and $h(k)$ are not the spectra of the solenoidal and potential components of turbulence and their scalings differ from those of these spectra. They are similar to spectrum in frequency-wavenumber domain evaluated at zero frequency, see details in the Appendix.  
On long time scales the model reproduces the pair dispersion in the NS flow at least qualitatively. Finally the function $c(r)$ is non-zero if and only if the flow has finite compressibility. The correlation function of the velocity divergence may therefore be calculated in terms of the $c(r)$, and is given by:
\begin{eqnarray}&&\!\!\!\!\!\!\!\!\!\!\!\!\!\!\!
\langle w(t, \bm x)w(t', \bm x')\rangle=\delta(t'-t)\left(3c(r)+rc'(r)\right),\label{comprs}
\end{eqnarray}
The functions $c$ and $r$ have regular Taylor expansion in the viscous range. 
The sum of the Lyapunov exponents is given by Eq.~(\ref{va}) which gives $\sum_{i=1}^3\lambda_i=-(1/2)\int \langle w(0)w(t)\rangle dt=-3c(0)/2$, see \cite{arxiv} for details. Thus we must set $c(0)=0$. However this would produce zero single-point fluctuations of $w(t, \bm x)$, see Eq.~(\ref{comprs}). The reason is that $\delta-$function correlation cannot describe the zero-correlation time limit in the viscous range: vanishing of $\int_{-\infty}^0 \langle w(0)w(t)\rangle dt$ and non-vanishing of  $\int_{-\infty}^0 \langle w(0)w(t)\rangle t dt$ considered at the end of Appendix \ref{a} imply that $\langle w(0)w(t)\rangle$ has the behavior of $\delta'(t)$ not of $\delta(t)$. Nevertheless as long as the model is used in the inertial range only, it produces physically reasonable results. For $r$ in the inertial range the $\delta(t)$ behavior is reasonable because $\int \langle w_r(0)w_r(t)\rangle dt$ is non-zero where $w_r$ is coarse-grained over scale $r$. There are anti-correlations of $w_r$ however they are not that restrictive. Moreover we demonstrate that at least for the pair correlation of the concentration the results are reasonable everywhere. This can be understood by considering the model as a result of a suitable limit process with infinitesimal $c(0)$ that describe small inertia of tracers. We use below $c(0)=0$.

\subsection{Yaglom-type relation and eddy diffusivity assumption} \label{eddy}

The Kraichnan model provides a consistent way for resolving the ambiguity in the eddy diffusivity approximation that occurs due to compressibility. We start from deriving an exact relation for the tracers concentration in a Navier-Stokes (NS) turbulence. This is the counterpart of Yaglom's relation for scalar turbulence \cite{moninyaglom,review}.
Starting with the stationarity condition $\partial_t \langle n(\bm x_1)n(\bm x_2)\rangle=0$, after moving the time derivative under the average and using Eq.~(\ref{con}) we find,
\begin{eqnarray}&&\!\!\!\!\!\!\!\!\!\!\!
\nabla\cdot\langle\left(\bm v(\bm r)-\bm v(0)\right)n(0)n(\bm r)\rangle=0,\label{dived}
\end{eqnarray}
where $\bm r=\bm x_2-\bm x_1$ and we used the statistics homogeneity. Using isotropy and regularity at zero we find,
\begin{eqnarray}&&\!\!\!\!\!\!\!\!\!\!\!
\langle\left(\bm v(\bm r)-\bm v(0)\right)n(0)n(\bm r)\rangle=0.\label{yaglom}
\end{eqnarray}
For Yaglom's relation the RHS in Eq.~(\ref{yaglom}) is a finite constant and the scaling of the scalar in the inertial range is found from that of the velocity by power counting. In our case, however, the scaling of $\langle n(0)n(\bm r)\rangle$ is determined not by the absolute scaling of the velocity but rather by the relative scalings and magnitudes of the compressible and solenoidal components, see below.

We cannot decouple the velocity and the concentration in Eq.~(\ref{yaglom}) however it is plausible that similarly to the passive scalar case \cite{moninyaglom} we can use the eddy diffusivity approximation, at least qualitatively,
\begin{eqnarray}&&\!\!\!\!\!\!\!\!\!\!\!\!\!\!\!\!\!\!\!
\langle\left(\bm v(\bm r)\!-\!\bm v(0)\right)n(0)n(\bm r)\rangle
\!=\!-\!\nabla_k\left(K_{ik}(\bm r)\langle n(0)n(\bm r)\rangle\right), \label{turb}
\end{eqnarray}
where $K_{ik}(\bm r)$ is the eddy diffusivity tensor that is taken for fitting the data. In the Kraichnan model Eq.~(\ref{turb}) is exact with $K_{ik}$ given by Eqs.~(\ref{mdl})-(\ref{tensor}) as can be seen from the equation on the pair correlation function \cite{arxiv}. We thus have from Eqs.~(\ref{yaglom})-(\ref{turb}),
\begin{eqnarray}&&\!\!\!\!\!\!\!\!\!\!\!
K_{ik}\nabla_k\ln \langle n(0)n(\bm r)\rangle=-\nabla_k K_{ik}.\label{drd}
\end{eqnarray}
This equation becomes a first order ODE for $\langle n(0)n(\bm r)\rangle$ upon the use of isotropy. Its solution is given by:
\begin{eqnarray}&&\!\!\!\!\!\!\!\!\!\!\!
\langle n(0)n(\bm r)\rangle=\exp\left[\int_r^{\infty}\frac{c(r')dr'}{r'u(r')}\right].\label{solut}
\end{eqnarray}
This solution was presented in \cite{arxiv} where the flow with non-zero sum of the Lyapunov exponents and divergent $\langle n^2\rangle$ was considered. In our case $c(0)=0$ guarantees the regular behavior of $\langle n(0)n(\bm r)\rangle$ at the origin where the pair-correlation has a finite maximum. The behavior in the inertial range depends on the magnitudes of scaling exponents of the compressible component $c(r)$ and solenoidal component $u(r)$. For ${\rm Ma}<{\rm Ma}_s$ the scaling exponent of $c(r)$ in the inertial range is smaller than that of the solenoidal component. For instance in the pseudo-sound regime at small Mach numbers the spectrum of the compressible component is proportional to $k^{-3}$ which decays much faster than the almost Kolmogorov spectrum of the solenoidal component \cite{gotoh}. In this case and also below the sonic scale at ${\rm Ma}>{\rm Ma}_s$ we find from Eq.~(\ref{solut}) that $\langle n(0)n(\bm r)\rangle$ is smoothly dependent on $r$. There is no divergent power-law dependence that characterises multifractal distributions. In contrast, above the sonic scale at ${\rm Ma}>{\rm Ma}_s$ the scalings of $c(r)$ and $u(r)$ can be approximated as identical. Then the ratio $c(r)/u(r)$ is given by the constant $\beta$ in the inertial range where the notation is used for consistency with the results of the previous Sections. We therefore find,
\begin{eqnarray}&&\!\!\!\!\!\!\!\!\!\!\!
\langle n(0)n(\bm r)\rangle=\left(\frac{{\tilde L}}{r}\right)^{\beta}, \ \ \frac{c(r)}{u(r)}\approx \beta,\ \ \eta\ll r\ll L,\label{prc}
\end{eqnarray}
where ${\tilde L}$ is a cut-off scale of the order of the integral scale $L$. Eq.~(\ref{par}) is thus recovered. It is remarkable that the common scaling exponent of $c(r)$ and $u(r)$ drops from $\langle n(0)n(\bm r)\rangle$. In contrast the relative magnitude $\beta$ of the compressible and solenoidal components determines the scaling. Qualitative reasons for this are provided by the cascade model presented in Section \ref{cas}.

\subsection{Supercritical transport in Kraichnan model} \label{supercritical}

We formulate the Kraichnan model of transport in the supersonic inertial range at ${\rm Ma}>{\rm Ma}_s$. In that regime, the solenoidal and potential components of the velocity are characterised by the same scaling exponent. We therefore introduce in the inertial range $u(r)=c_0'r^{\xi-2}$ and $c(r)=c_0 r^{\xi-2}$  where $\xi$, $c_0'$ and $c_0$ are constants \cite{phase},
\begin{eqnarray}&&\!\!\!\!\!\!\!\!\!\!\!\!\!\!
2K_{ik}\!=\!\left(c_0'(2\!+\!\xi)\!-\!c_0\right)r^{\xi}\delta_{ik}\!+\!\left[\frac{c_0}{\xi}\!-\!c_0'\right]\xi r^{\xi-2}r_ir_k.\label{oir}
\end{eqnarray}
The scaling exponent $\xi$ is chosen so that the condition of modelling the NS turbulence given by Eq.~(\ref{scaling}) in Appendix \ref{krm} holds. Thus if the scaling exponent of the NS velocity is $\alpha$ (so that the spectrum decays as $k^{-1-2\alpha}$) then $\xi=1+\alpha$, see the consideration after Eq.~(\ref{77}). The Kolmogorov value $\alpha=1/3$ corresponds therefore to $\xi=4/3$ and not $\xi=2/3$, see \cite{review}. Besides $\xi$ the model is characterized by another dimensionless parameter, $\beta=c_0/c_0'$, see Eq.~(\ref{prc}). Here the overall magnitude of $K$ determines a dimensionless time-scale $\tau=c_0't$ that has no qualitative relevance and drops from the steady state averages. Therefore, without loss of generality, we set below $c_0'=1$. Instead of $\beta$ we can use the compressibility degree ${\cal P}$ that is defined as the ratio of $\left\langle(\nabla\cdot\bm u)^2\right\rangle$ and $\left\langle(\nabla\bm u)^2\right\rangle$ and is given by \cite{phase},
\begin{eqnarray}&&\!\!\!\!\!\!\!\!\!\!\!
{\cal P}=\frac{\beta}{\xi\left[3+\xi-\beta\right]}, \ \ 0\leq {\cal P}\leq 1. \label{compel}
\end{eqnarray}
Thus the model is determined by ${\cal P}$ and $\xi$ both of which can be modeled as monotonously growing functions of ${\rm Ma}$. The compressibility ${\cal P}$ grows from a certain finite value at ${\rm Ma}_s$ to $1$ at infinite Mach number. The exponent $\xi$ grows from some value larger than $4/3$ at ${\rm Ma}_s$ up to a value $\xi_{\infty}$ at  ${\rm Ma}=\infty$. The value of $\xi_{\infty}=3/2$ that corresponds to $k^{-2}$ spectrum of the Burgers equation is a reasonable and widely accepted conjecture \cite{bold}.

\subsection{Clumping transition} \label{lumpin}

The clumping transition has been invoked in \cite{phase} as an important mechanism for the transition to multifractality. Within that scenario, increasing the Mach number leads to a transition from a finite number of encounters between a pair of particles to particles sticking to each other with probability one. Here we briefly consider that transition and demonstrate that it is not likely to occur in NS compressible turbulence. In order to do that we revert to Eq.~(\ref{compel}) that yields:
\begin{eqnarray}&&\!\!\!\!\!\!\!\!\!\!\!
\beta=\frac{{\cal P}\xi\left[3+\xi\right]}{1+{\cal P}\xi}.
\end{eqnarray}
As $\beta$ represents the difference between the space dimension and the correlation dimension, it is necessarily a growing function of ${\rm Ma}$. That difference between the space and the correlation dimensions, called dimension deficit, grows linearly with the compressibility degree ${\cal P}$ at small compressibility (this range is purely theoretical since there are no cases of weakly compressible flows with identical scalings of solenoidal and potential components known to us. This situation would be highly interesting, see Appendix \ref{krms}). For ${\cal P}>3/\xi^2$ the value of $\beta$ becomes larger than three, which corresponds to negative correlation dimension. For such values of $\beta$, namely bigger than $3$, the integral of the correlation function given by Eq.~(\ref{prc}) diverges at zero thus yielding an (unphysical) infinite mass in an arbitrarily small ball.  Thus the expression for the correlation function breaks down for ${\cal P}$ above $3/\xi^2$. This signifies the clumping transition as two tracers glue to each other at large times with probability one. The PDF $P(\bm r, \bm r_0, t)$ in this case is a sum of a regular term and a $\delta(\bm r)$ term whose amplitude grows from zero at $t=0$ to one at $t=\infty$, see \cite{phase}. Correspondingly the steady state correlation function is $\delta(\bm r)$ in this case.

We saw previously that for compressible turbulence the value of $\xi$ is bounded from above by $3/2$. The resulting value of $3/\xi^2$ is larger than one so that the range ${\cal P}>3/\xi^2$ is unphysical. Thus assuming that the Kraichnan model provides a realistic description of the NS compressible turbulence leads to the conclusion that no clumping transition occurs in that turbulence. We hypothesize that this conclusion is true however complete settling of this issue requires further studies. We finally remark that in contrast to the clumping transition, the infinite recurrence transition that is discussed in the next two subsections may occur in NS turbulence.

\subsection{PDF of the distance and pair correlations} \label{lumpi}

We confirm Eq.~(\ref{predictions}) by direct calculation. In Kraichnan model $P(\bm r, \bm r', t)$ obeys the Fokker-Planck equation \cite{phase,review},
\begin{eqnarray}&&\!\!\!\!\!\!\!\!\!\!\!\!\!\!
\partial_{|\tau|} P=\nabla_i\nabla_l\left(K_{il}(\bm r) P\right),\ \ P(\tau=0)=\delta(\bm r-\bm r'), \label{hrm}
\end{eqnarray}
where the evolution can be considered both for positive and negative dimensionless time $\tau$. This equation necessarily has the same form as the evolution equation of $\langle n(0)n(\bm r)\rangle$ provided in \cite{arxiv}, cf. Eq.~(\ref{opa}). The turbulent diffusion operator $\nabla_i\nabla_l K_{il}$ describes a power-law growth of $r(|\tau|)$ in the inertial range,
\begin{eqnarray}&&\!\!\!\!\!\!\!\!\!\!\!
\frac{d\langle r^k(|\tau|)\rangle}{d|\tau|}=\frac{d}{d|\tau|}\int r^k P(\bm r, \bm r', |\tau|)d\bm r=\int r^k d\bm r\nonumber\\&&\!\!\!\!\!\!\!\!\!\!\!\nabla_i\nabla_l \left(K_{il}(\bm r) P\right)
=k\left(k+1+\xi-\beta\right)\langle r^{k+\xi-2}\rangle,
\end{eqnarray}
where we integrated by parts and employed Eqs. (79) and (80) as well as the relationship:
\begin{equation}
K_{il}\nabla_i\nabla_l r^k=kr^{k+\xi-2} \left(k+1+\xi-\beta\right)
\end{equation}
that holds in the inertial range and is obtained by direct calculation using Eq.~(\ref{oir}). We find setting $k=2-\xi$,
\begin{eqnarray}&&\!\!\!\!\!\!\!\!\!\!\!
\langle r^{2-\xi}(\tau)\rangle=r^{2-\xi}(0)+(2-\xi)\left(3-\beta\right)|\tau|,\label{sm}
\end{eqnarray}
which is the form that the Richardson law, given by Eq.~(\ref{scals}), takes in the Kraichnan model. As discussed after Eq.~(\ref{scals}), the separation $r(\tau)$ is independent of the initial condition at times larger than $r^{2-\xi}(0)$. There the power-law growth holds $\langle r^{2-\xi}(\tau)\rangle\approx (2-\xi)\left(3-\beta\right)|\tau|$. Such a type of separation of trajectories that remains finite in the limit of zero $r(0)$ is called explosive \cite{phase} in order to distinguish them from the more usual chaotic separation where $r(t)=0$ at $r(0)=0$. The explosive separation is a characteristic property of the inertial range separation by non-differentiable rough velocity where roughness causes non-uniqueness of the trajectories.

Returning to Equation (\ref{hrm}), it has a self-similar solution,
\begin{eqnarray}&&\!\!\!\!\!\!\!\!\!\!\!\!\!\!\!
f_s(\bm r, \tau)\!=\!\frac{|\tau|^{b_0-1}(2\!-\!\xi)^{2b_0-1} }{4\pi r^{\beta} \Gamma(1\!-\!b_0)}\exp\left(-\frac{r^{2-\xi}}{(2\!-\!\xi)^2|\tau|}\right) 
,\label{sep}
\end{eqnarray}
where $b_0=(\beta-\xi-1)/(2-\xi)$ and we normalized the solution so that $\int f(\bm r)d\bm r=1$. This solution coincides with the finite limit $P(\bm r, r'\to 0, \tau)$, see \cite{phase} and Appendix \ref{krms}. The solution is quite similar in properties to the fundamental solution of the ordinary diffusion equation. In fact considering the dependence of the model on $\xi$ as a parameter that varies in the maximal allowed \cite{review} range $0\leq 0<2$ we have that at $\xi=0$ and $\beta=0$ turbulent diffusion reduces to the ordinary diffusion. The above formula reduces then to the Green function of diffusion equation. At finite $\xi$ and $\beta$, as in ordinary diffusion, $f_s(r, \tau)$ describes the long-time asymptotic form of $P(\bm r, \bm r', \tau)$. Thus at large times the evolution of separation is self-similar giving $\langle r^n(|t|)\rangle\propto |t|^{n/(2-\xi)}$. The counterpart of the Richardson law is then $\langle r^2(|t|)\rangle\propto |t|^{2/(2-\xi)}$.

We see from Eq.~(\ref{sep}) that the scaling exponent of the pair-correlation function of the concentration determines the scaling of $P(r, r'\to 0, \tau)$ at small $r$ in accord with Eq.~(\ref{predictions}). The confirmation in the case of $r'\neq 0$ is more complex. It is provided in Appendix \ref{krms} along with formula for the moments of the distance between two tracers.

\subsection{Infinite recurrence transition}

We observe that at small compressibility ${\cal P}$ the parameter $b$ is negative and the integral of $P(r, r_0\to 0, t)$ over $t$ converges at large times. Thus the particles collide reaching $r(t)=0$ at most a finite number of times, cf. similar considerations for usual diffusion. However as the compressibility increases, a transition occurs at $\beta=1+\xi$ or ${\cal P}=(1+\xi)/(2\xi)$ and $b$ becomes positive for larger compressibilities. The time integral diverges at large times so that particles will collide infinite number of times with probability one. This trapping effect of compressibility discovered in \cite{phase} makes the behavior of pairs of tracers qualitatively different from that in the incompressible flow.

In contrast with the clumping transition, for the infinite recurrence transition the Kraichnan model indicates that this transition can occur in the NS turbulence. Indeed, for $4/3\leq \xi\leq 3/2$ the value of ${\cal P}=(1+\xi)/(2\xi)$ is below one. Deciding whether the transition does occur requires numerical studies.

\subsection{Anomalous scaling and zero modes}\label{inetr}

Finally we comment on the scaling of higher order correlation functions of the concentration that determine the fractal dimensions of positive integer order. These functions obey in the Kraichnan model a closed PDE: they are zero modes of the operator $\sum_{nl}\nabla_{r_n^i}\nabla_{r_l^k} K_{ik}(\bm r_n-\bm r_l)$ where $\bm r_n$ are the points in the correlation function \cite{review,krzysztof}. Zero modes provide known way of producing anomalous scaling exponents, in our case non-trivial dependence of $D(k)$ on $k$, cf. the previous Section.

We examine the validity of this consideration outside the Kraichnan model. We observe that similarly to our study of the pair correlations we find that $P(\bm R, \bm R', t)$ in Eq.~(\ref{higc}) is given by $\exp(t{\hat L}_n)(\bm R, \bm R')$ with certain linear operator ${\hat L}_n$. In the Kraichnan model ${\hat L}_n$ reduces to linear combination of the pair-dispersion operators ${\hat L}$. This reduction would not hold for propagators $P(\bm R, \bm R', t)$ of the NS flow. However the reduction introduced by the Kraichnan model is similar to neglecting the intermittency of the flow and it is valid qualitatively. The reason is that the intermittency of the statistics of the transported quantity, tracers' concentration in our case, is much stronger than the intermittency of the transporting velocity. For instance the difference of the scaling exponent of the fourth order correlation function of the concentration and twice the scaling exponent of the pair correlation is finite even when the transporting velocity is self-similar: the concentration is intermittent even when the flow is not. This was found in the case of a passive scalar transported by incompressible turbulence where the scalar is not self-similar despite that the transporting flow is a self-similar Gaussian flow with little structure \cite{review}. The reason for this phenomenon, called anomalous scaling, is the zero modes described in the previous Section. These modes define the correlation functions and have intrinsically anomalous scaling independently of whether the velocity scaling is anomalous. The situation seems similar in our case also though the detailed calculations are outside the scope of this paper. It must be kept in mind though that for other questions the neglect of intermittency could be not valid. For instance pair dispersion in the Kraichnan model is self-similar however intermittency of turbulence would cause breakdown of this self-similarity.

\subsection{Transition to multifractality in NS turbulence}

We consider Eq.~(\ref{solut}) at any ${\rm Ma}$, not necessarily in the multifractal phase.
We find using the formula for $c/ru$ derived in Appendix \ref{krm},
\begin{eqnarray}&&\!\!\!\!\!\!\!\!\!\!\!\!\!
\ln \langle n(0)n(\bm r)\rangle\!=\!\int_{r/{\tilde L}}^{1}\!\! \frac{(b\!+\!1)(b\!+\!7)(a\!+\!7)\Gamma' r'^{\Delta-1}dr'}{8(b\!+\!7)\!+\!2(b\!+\!3)(a\!+\!7)\Gamma' r'^{\Delta}},\label{frm}
\end{eqnarray}
where $\Delta=(b-a)/2$ is half the difference of decay exponents $b$ and $a$ of the spectra of potential and solenoidal components of velocity, respectively, and $\Gamma'$ is the ratio of the magnitudes of potential and solenoidal components, see details in Appendix \ref{krm}. The integration variable here is the ratio of the scale and the upper cutoff scale ${\tilde L}$ which is determined by the breakdown of scaling of $c$ and $u$ and is of order of the integral scale $L$. This scale is ${\tilde L}$ in Eq.~(\ref{prc}).

Integration of Eq.~(\ref{frm}) gives,
\begin{eqnarray}&&\!\!\!\!\!\!\!\!\!\!\!\!\!
\langle n(0)n(\bm r)\rangle\!=\!\left(\frac{\sigma_0+1}{\sigma_0+ (r/{\tilde L})^{\Delta}}\right)^{(b+1)(b+7)/(2(b+3)\Delta)}. \label{concise}
\end{eqnarray}
We introduced the dimensionless quantity $\sigma_0=4(b+7)/\left((b+3)(a+7)\Gamma'\right)$. We have $\sigma_0\sim 1/\Gamma'$ for physically relevant values of $a$ and $b$ so $\sigma_0$ characterizes the ratio of the magnitudes of solenoidal and potential components.

\begin{figure}
  \centerline{\includegraphics[width=8cm, height=6cm]{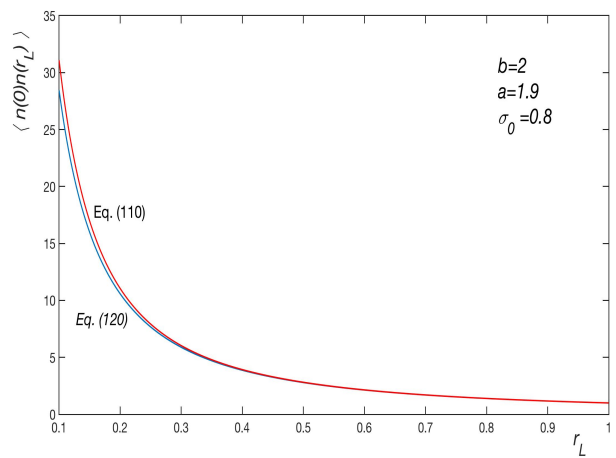}}
  \caption{The tracers pair correlation function in the high Mach number regime. $r_L =r/{\tilde L}$}
\label{fig:paircorrelation1}
\end{figure}

\begin{figure}
  \centerline{\includegraphics[width=8cm, height=6cm]{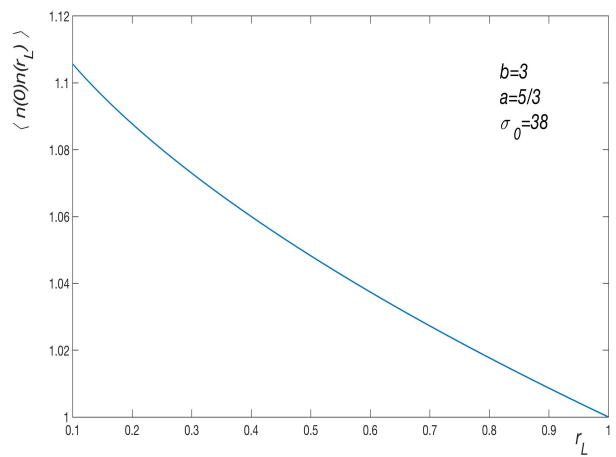}}
  \caption{The tracers pair correlation function in the low Mach number regime as predicted by Eq.~(\ref{concise}). $r_L =r/{\tilde L}$}
\label{fig:paircorrelation2}
\end{figure}

The above equation is a concise prediction of the model that holds at any ${\rm Ma}$. In the limit of small Mach numbers we have $\sigma_0\gg 1$ and the pair-correlation function is nearly constant at $r\ll L$. At ${\rm Ma}$ and $\Delta$ of order one where $\sigma_0\sim 1$ the pair-correlation function has some changes in the inertial range. The power law becomes valid as the Mach number increases and $\Delta$ becomes much smaller than one. For the study of this limit it is useful to rewrite the pair correlation function as,
\begin{eqnarray}&&\!\!\!\!\!\!\!\!\!\!\!\!\!
\ln \langle n(0)n(\bm r)\rangle\!=\!\frac{(b\!+\!1)(b\!+\!7) }{2(b\!+\!3)\Delta}
\\&&\!\!\!\!\!\!\!\!\!\!\!\!\!
\ln\left(\frac{2(b\!+\!3)(a\!+\!7)\Gamma'(r/{\tilde L})^{\Delta}\left( ({\tilde L}/r)^{\Delta}-1\right)}{8(b\!+\!7)\!+\!2(b\!+\!3)(a\!+\!7)\Gamma' (r/{\tilde L})^{\Delta}}+1\right),\nonumber
\end{eqnarray}
This can be approximated for $\Delta\ll 1$ as,
\begin{eqnarray}&&\!\!\!\!\!\!\!\!\!\!\!
\langle n(0)n(\bm r)\rangle\approx \exp\left[\frac{\beta}{\Delta}\left(\left(\frac{{\tilde L}}{r}\right)^{\Delta}-1\right)\right],
\end{eqnarray}
with
\begin{eqnarray}&&\!\!\!\!\!\!\!\!\!\!\!\!\!
\beta\!=\!\frac{(a\!+\!7)(b\!+\!1)(b\!+\!7)\Gamma'}{8(b\!+\!7)\!+\!2(b\!+\!3)(a\!+\!7)\Gamma' }.
\end{eqnarray}
We see that Eq.~(\ref{prc}) is a good approximation under the condition $\Delta \ln ({\tilde L}/r)\ll 1$. These formulas can be used for fitting the parameters of the model with the help of future numerical data. Fig. \ref{fig:paircorrelation1} depicts the tracers pair correlation function as obtained from the Kraichnan model, i.e. Eq.~(\ref{concise})
(blue line) as compared to the approximating power low given by Eq.~(\ref{prc}) (red line). It should be mentioned that Fig. \ref{fig:paircorrelation1} represents a valid picture only down to the sonic length. As commented above, below that scale the pair correlation function flattens significantly.
The pair correlation function according to Eq.~(\ref{concise}) in the small Mach number regime is shown in Fig. \ref{fig:paircorrelation2}. As expected, in that regime the concentration fluctuations are small and close to a constant value over the inertial range.

Thus we confirmed again that multifractality is approximate. This is because the density fluctuations are created by the potential component of the velocity but particles' separation is determined by the complete velocity. Strictly speaking, the multifractality holds only when the scalings of the velocity components are identical.

\section{Acceleration of formation of planetesimals}\label{acceleration}

Transition to multifractality implies strong increase of the pair correlation function of concentration at small distances. The flow transport brings the tracers close much more often than below the transition. This implies a strong increase in the collision rate of particles above the transition as we demonstrate in this Section. The rate increases by a large parameter over a short interval of Mach numbers where the pair correlation transitions from the stretched exponential to the power-law form, see Eq.~(\ref{solut}). The physical system that we use for consideration is formation of planetesimals.

We consider a model where formation of planetesimals occurs due to coalescence of particles of dust. The particles are transported by the compressible turbulent flow of a gas that is characterised by large Mach and Reynolds numbers. The particles are assumed to have negligible inertia and move as tracers.

The solution is assumed to be dilute. Thus we can neglect the particles' back reaction on the flow and consider only binary collisions. Two dust particles of radii $a_1$ and $a_2$ collide when their centers are at a distance $a_1+a_2$ from each other. Thus $a_1+a_2$ is the effective interaction radius of the particles. We assume that the collision leads to coalescence with probability $P_{12}$ that depends on $a_i$. This probability characterizes short-range interactions and is the counterpart of the collision efficiency in the similar problem of coalescence of droplets in rain formation \cite{PruppacherKlett}. The rate of coalescence per unit volume $\gamma_{12}$ of dust particles of radii $a_1$ and $a_2$ is given by \cite{st,PruppacherKlett},
\begin{eqnarray}&&\!\!\!\!\!\!\!\!\!\!\!\!\!
\gamma_{12}\!=\!-P_{12}\int_{w_r<0, r=a_1+a_2}  dS \left\langle  n_2(0) n_1(\bm r)w_r(\bm r)\right\rangle,\label{rate}
\end{eqnarray}
where $n_i$ is the concentration of the particles with radius $a_i$ and $w_r(\bm r)=v_r(\bm r)-v_r(0)$ is the radial component of the velocity difference of the colliding particles.  The integral above is over that part of the surface of the ball of radius $a_1+a_2$ on which $w_r<0$. This condition ensures that the particles approach and not separate. The angular brackets stand for the spatial averaging. We can simplify Eq.~(\ref{rate}) employing the continuity equation,
\begin{eqnarray}&&\!\!\!\!\!\!\!\!\!\!\!\!\!
\partial_tn_i+\nabla\cdot(n_i\bm v)=0,
\end{eqnarray}
where $i=1, 2$. Using the same steps taken for deriving Eq.~(\ref{dived}) we find,
\begin{eqnarray}&&\!\!\!\!\!\!\!\!\!\!\!\!\!
\nabla \cdot \langle n_2(0) n_1(\bm r) \left(\bm v_1(\bm r)-\bm v_2(0)\right)\rangle=0.
\end{eqnarray}
Finally, we find by integrating over the volume of the ball with radius $a_1+a_2$ and using the divergence theorem, the following constraint:
\begin{eqnarray}&&\!\!\!\!\!\!\!\!\!\!\!\!\!
\int  \langle n_2(0) n_1(\bm r)w_r\rangle dS=0.
\end{eqnarray}
Thus we can write Eq.~(\ref{rate}) in the form,
\begin{eqnarray}&&\!\!\!\!\!\!\!\!\!\!\!\!\!
\gamma=\frac{P_{12}}{2}\int_{r=a_1+a_2}  dS \langle n_2(0) n_1(\bm r)\left| w_r\right|\rangle ,\label{rat}
\end{eqnarray}
The derivation of similar identity in incompressible isotropic case was done in \cite{st}. This way of rewriting $\gamma_{12}$ is useful because averaging conditioned on sign of $w_r$ is more difficult.

We observe from the cascade picture of the formation of fluctuations of the concentration that for $r$ in the inertial range $\langle n_2(0) n_1(\bm r)\rangle$ is determined by many steps of the cascade and only the last step is correlated with $w_r$. Thus, neglecting one step in comparison with many, we can perform independent averaging, $\left\langle  n_2(0) n_1(\bm r)\left| w_r\right|\right\rangle=S(r)\left\langle n_2(0) n_1(\bm r)\right\rangle$, where $S(r)= \left\langle  \left| w_r(r)\right|\right\rangle$. Invoking isotropy, we find:
\begin{eqnarray}&&\!\!\!\!\!\!\!\!\!\!\!\!\!\!\!\!
\gamma_{12}\!=\!2\pi P_{12} (a_1\!+\!a_2)^2  S(a_1\!+\!a_2)\left\langle n_2(0) n_1(a_1\!+\!a_2)\right\rangle.
\end{eqnarray}
The structure function $S(r)$ changes smoothly at the transition to multifractality (at the transition the scaling exponents of solenoidal and potential components of the velocity become similar which does not bring a strong change of $S(r)$). Thus we can concentrate on the change of $\gamma_{12}$ due to $\left\langle n_2(0) n_1(a_1+a_2)\right\rangle$. In the multifractal phase this is given simply by,
\begin{eqnarray}&&\!\!\!\!\!\!\!\!\!\!\!\!\!\!\!
\left\langle n_2(0) n_1(a_1+a_2)\right\rangle=\langle n_1\rangle\langle n_2\rangle \left(\frac{L}{a_1+a_2}\right)^{3-D_t(2)},
\end{eqnarray}
where $D_t(2)$ is the correlation dimension of the multifractal formed by the tracers. We remark that particles with different $a_i$ are identically moving tracers so the probability of finding another particle at fixed distance $r$ from a given particle is independent of the particle radii $a_i$. The above formula assumes that $a_1+a_2$ belongs in the supersonic inertial range that is larger than the sonic scale $l_s$ but smaller than $L$. In this case at the transition $\gamma_{12}$ is increased by the factor of $(L/(a_1+a_2))^{3-D_t(2)}$.

\subsection{Pair correlations and collisions below the sonic scale}

It is probable that in many applications the case of colliding particles with $a_1+a_2<l_s$ would be relevant. The description of the collision rate in this case requires the knowledge of the pair-correlation function of the concentration below $\eta$. Turbulence below $\eta$ is either dissipative or it has a small Mach number. In both cases the fluctuations of the density stop increasing below $\eta$ resulting in the flattening of the correlation functions at $r<\eta$,
\begin{eqnarray}&&\!\!\!\!\!\!\!\!\!\!\!\!\!
\left\langle n(0) n(r)\right\rangle\sim \langle n\rangle^2\left(\frac{L}{\eta}\right)^{3-D_t(2)},\nonumber \\&&\!\!\!\!\!\!\!\!\!\!\!\!\!
\left\langle \rho(0) \rho(r)\right\rangle\sim \left(\frac{L}{\eta}\right)^{3-D(2)}.
\end{eqnarray}
This can be confirmed for the Kraichnan model using Eq.~(\ref{solut}). Correspondingly we find for the correlation function in the rate of collisions,
\begin{eqnarray}&&\!\!\!\!\!\!\!\!\!\!\!\!\!
\left\langle n_2(0) n_1(a_1+a_2)\right\rangle\sim \langle n_1\rangle\langle n_2\rangle \left(\frac{L}{\eta}\right)^{3-D_t(2)}.
\end{eqnarray}
Thus the smaller the interaction radius of the colliding particles, the larger the increase of the collision kernel is up to $a_1+a_2=\eta$. The increase factor for smaller $a_1+a_2$ is size-independent.

Finally we remark the total collision rate is given by summing over the rates of collisions of particles with all radii combinations $a_1$ and $a_2$.

\section{Conclusions and open questions}

The main goal of the current work was to shed light on the complex behaviour of compressible turbulent flow and in particular on the mass density of the fluid and the concentration of passive tracer particles. As a
result, the following issues and fundamental problems have been addressed and solved.

We raised the fundamental question of possible difference of the active fluid density and passive tracer concentration in the steady state.
We demonstrated that despite that the fluid density and the concentration of tracers obey the same continuity equation the proof of their equality is elusive. There is no uniqueness of the steady state solution and mixing of the fluid density may fail due to the possibility of self-organization of the fluid density in a state that is stabilized by the interaction with the flow. The same state could be unstable for the concentration. We reminded of other
cases in the fluid mechanics where the instability causes the active and passive scalars to differ in the steady state. We demonstrate that the current numerical data, though inconclusive, seems to indicate that the fields differ. We provided a number of testable predictions that will help to resolve the question unequivocally. These are rigorous and semi-rigorous results on the density and the concentration that are valuable independently of the fields' equality.

We traced the origin of the multifractality of the concentration to the increasing self-similarity of the flow at increasing Mach number due to the approach of the scalings of the solenoidal and potential components of the flow. It was demonstrated that the concentration's transition to multifractality is smooth. Detailed description of the turbulent transport was provided with the help of the Kraichnan model of turbulence whose parameters were related to the actual properties of the NS turbulence.

We provided formula for the fractal dimensions of the density of isothermal compressible turbulence. We demonstrated that the high-order dimensions are determined by rare events that correspond to the tails of the density distribution and are not describable by the lognormal distribution.

Multifractality is a form of clustering which appears at increasing compressibility of the flow due to the increasing tendency of the Lagrangian trajectories to cluster in regions with negative divergence. This tendency would lead to exactly multifractal distribution of tracers if the scalings of the solenoidal and potential components of the velocity were identical. However the scalings approach a common value only asymptotically with increasing the Mach number. The difference $2\Delta$ of the decay exponents of the spectra of potential and solenoidal  components is of order one at ${\rm Ma}<1$. It decreases as ${\rm Ma}$ increases: for instance the difference is about five per cent at ${\rm Ma}=6$ in the case of \cite{krit2007}. We derived the pair-correlation function of concentration for the general case $\Delta\neq 0$ and demonstrated that the power-law scaling is a good approximation at scale $r$ if $\Delta\ln (L/r)\ll 1$. Correspondingly the mutlifractality is a good approximation in the whole of the supersonic inertial range provided that $\Delta\ln (L/l_s)\ll 1$. Since $L/l_s$ is a power of ${\rm Ma}$ then, neglecting weak logarithmic dependence on the Mach number, we find that the concentration of tracers is multifractal under the condition $\Delta \ll 1$. This is only an order of magnitude condition so the critical Mach number ${\rm Ma}_s$ at which the transition occurs depends on the desired accuracy.

We conclude from the above that though it is not clear whether there is a sharply defined critical Mach number beyond which the scalings agree and multifractality holds exactly (probably not) however in practice we can use $\Delta\ll 1$ as the transition criterion. At $\Delta\sim 1$ the influence of the potential component on the small-scale evolution of the concentration is negligible and the fluctuations of the concentration below $L$ are small or of order one. At $\Delta\ll 1$ the dust particles concentrate on a constantly evolving, statistically stationary, multifractal seen as clusters and filaments in space, cf. \cite{dust}.

The considerations above assume that the potential and solenoidal components of the flow obey a power-law scaling. The scaling has been observed, see e. g. \cite{krit2007}. However careful investigation of the provided
data demonstrates that the power law could only be an approximation to a more complex dependence. If that is the case then our theory in terms of $\Delta$ also becomes an approximation. However the results of the Kraichnan model for the pair-correlation function of tracers do not assume the scaling of velocity components and could be used for the study of more complex dependencies. These generalizations might be needed in the future should it be found that the power law scaling of the flow components is not a good approximation.    

We remark that the singularity of the natural measure (the steady state concentration) of a self-similar compressible flow is a universal property independent of the details of the flow. This universality is well known for chaotic flows below the scale of smoothness (which is the case of trivial self-similarity determined by the linear scaling of the velocity difference). In that case the singularity of the natural measure is a consequence of the exponential dependence of small particles' volumes on time. That dependence and conservation of the total volume of the flow imply that the sum of the Lyapunov exponents is non-negative. Thus for non-degenerate flows with a strictly negative sum the steady state concentration, described by the sum, is singular. Similar universality holds for the natural measure of a rough self-similar flow (the flow is rough
if the scaling exponent of its difference is smaller than one so that there is no differentiability \cite{review}). The volumes have a power-law behavior so in logarithmic time scale we can carry over the proof for the smooth case.  This proof provides us also with a generalization of the sum of the Lyapunov exponents in the inertial range.

The transition to multifractality for the density is less understood. The spectrum of the density is usually fit with a power law \cite{krit2007} though it is proposed that this law is only an approximation \cite{fractal}. In fact, we demonstrated that the spectrum of the concentration obeys a power law only approximately. This makes it plausible that the power law of the density spectrum is also only an approximation. If we use this approximation then the transition occurs when the spectrum decay exponent, considered as a function of ${\rm Ma}$, passes minus one from below (which seems to be underappreciated \cite{krit2007,fractal}). At this Mach number the concentration must also be multifractal. Indeed, the fluid density back reaction on the transporting velocity tends to arrest the formation of large densities and density gradients and, as a result, the density transition to multifractality occurs at a Mach number ${\rm Ma}_{cr}$ larger or equal (if the fields coincide) ${\rm Ma}_s$.
The behavior of the density at ${\rm Ma}<{\rm Ma}_{cr}$ is qualitatively similar to the small Mach number behavior where the density is proportional to the pressure of incompressible turbulence \cite{Lighthill,Montgomery,Matthaeus,bayly,Ristorcelli}. The gas particles form clusters and filaments at ${\rm Ma}>{\rm Ma}_{cr}$.

We cited in the Introduction the known fact that in all chaotic systems besides the natural measure (steady state concentration) there exist other steady state solutions of the continuity equation that are reached from initial conditions of measure zero \cite{grass}. Thus the possibility that the density and
the concentration differ must be considered as a real one. In order to facilitate the difference's identification in experiment we describe the situation that would hold if the density and the concentration are revealed to differ, ${\rm Ma}_s<{\rm Ma}_{cr}$. The difference is most striking at ${\rm Ma}_s<{\rm Ma}<{\rm Ma}_{cr}$ where the concentration is multifractal and the density is smooth. Both fields are transported along the same Lagrangian trajectories and obey the continuity equation with the same velocity. Motions of volumes of tracer and fluid particles are identical however their stability properties differ.
The density and the concentration are different stationary distributions of the same continuity equation. The density evolves from a measure zero (in functional space) initial conditions that are consistent with the velocity. These initial conditions are special: the resulting density fluctuations are finite at $\eta \to 0$ despite that the limit is infinite for typical initial conditions. For tracers the evolution of the same initial condition would be unstable: it is the interaction with the velocity that stabilizes the evolution of the density. The steady state concentration can be obtained by evolution of typical initial conditions. It gives the natural measure of the dynamics defined by Lagrangian evolution which is singular at $\eta\to 0$. At ${\rm Ma}>{\rm Ma}_{cr}$ both fields are multifractal however the fractal dimensions of the concentration are smaller than those of the density. The structures manifested by the two fields are different, cf. \cite{dust}. In fact the smallest scale of the multifractal can be somewhat different for gas and dust because the scaling mechanisms differ, cf. \cite{frisch}.

Our current understanding of the concentration's transition to multifractality is only moderately good. We do not have a good way for estimating ${\rm Ma}_s$ because that number is determined by the details of the strongly non-linear flow. It can be estimated from simulations \cite{krit2007,fractal} as ${\rm Ma}_s\simeq 4-6$. Probably the higher the compressibility of the stirring force the smaller ${\rm Ma}_s$. Further numerical studies are required. In contrast, we have a good qualitative understanding of how the transition occurs. We have developed a cascade model of the formation of the fluctuations of the concentration which is more quantitative than is usual in the theory of turbulence. That model describes the transition and fractal dimensions in terms of the properties of the flow.

Rigorous formulation for the fractal dimensions of the concentration is possible within the framework of the Kraichnan model. That model already proved highly useful in qualitative studies of turbulent transport and has become standard by now \cite{review}. It is by using this model that one can study the impact of the difference of scalings of solenoidal and potential components on the pair-correlation function in detail (in contrast, our derivation of multifractality for self-similar compressible flows does not need modelling). Our usage of this model is dictated uniquely by the derived Markov property of the distance and eddy diffusivity assumption. 
We found that pair correlations at distance $r$ are proportional to $(r+f)^h$ with $f$ and $h$ that depend on ${\rm Ma}$ reproducing the previously described behavior in all ranges of ${\rm Ma}$. In the multifractal phase the Kraichnan model estimates the scaling exponent of the pair-correlation function of the concentration, equivalent to the correlation dimension, as the ratio of the second order structure functions of the potential and the solenoidal components of the velocity. The ratio is scale-independent by ${\rm Ma}>{\rm Ma}_s$ and remarkably the common scaling exponent drops from the exponent. This prediction captures qualitatively the observed dependence of the scaling exponent on the compressibility of the forcing \cite{comprfr}. Quantitatively our study only yields an order of magnitude estimate of the exponent which is on general grounds is bounded between zero and three. Thus if our calculated exponent is small then the observed exponent must also be small and if our exponent is of order one then the observed exponent is of order one also. In contrast, we could make a quantitative prediction for the Navier-Stokes turbulence that the PDF of the distance between two tracers in the supersonic inertial range obeys a power-law at small distances with an exponent that is equal to the exponent of the steady state correlation function. This is the counterpart of the formula for the correlation dimension in terms of the statistics of the separation for smooth systems \cite{do,krzysztof}. Finally we demonstrated that the scaling of higher order correlation functions is anomalous due to the zero modes similarly to the incompressible case \cite{review}.


The transition of the density to multifractality is even less understood than that of the concentration
due to the strong non-linear interactions of the density and the velocity. We do not know what changes in turbulent velocity occur at ${\rm Ma}_{cr}$ and if they occur at all. So far no changes were observed. We can estimate from simulations \cite{krit2007,fractal} that ${\rm Ma}_{cr}\simeq 7-8$. We propose to look for changes in the velocity at ${\rm Ma}={\rm Ma}_{cr}$ in terms of the behavior of the Lagrangian trajectories. Indeed, we demonstrated by the study of the Kraichnan model that it is probable that separation of tracers in the supersonic inertial range undergoes a qualitative change at a Mach number ${\rm Ma}_r$ larger than ${\rm Ma}_s$. The separation at ${\rm Ma}<{\rm Ma}_r$ occurs similarly to incompressible turbulence obeying a modified Richardson law. A pair of tracers approach each other at most a finite number of times after which they are permanently
separated. In contrast, at ${\rm Ma}>{\rm Ma}_r$, trajectories of the two particles approach each other infinite number of times with probability one, due to the trapping effect of the compressibility \cite{phase,krho,we}. This transition does not cause a change in the behavior of the pair correlation functions of the concentration. We demonstrated that another transition discovered in \cite{phase}, where the tracer particles stick to each other with probability one, probably cannot occur in three-dimensional NS turbulence. It seems that the above recurrence transition could correspond to the density transition from a large-scale to a small-scale field. Thus we propose to study if there is a relation or equality between ${\rm Ma}_r$ and ${\rm Ma}_{cr}$.

More knowledge of the fractal dimensions $D(k)$ of the active fluid density may currently
be extracted from numerical simulations than corresponding information concerning the passive concentration. The statistics of the density in isothermal turbulence is observed to be lognormal, see e. g. \cite{krit2007,fractal}. For this type of statistics all fractal dimensions can be described as $D(k)=3-k\delta/2$ where $\delta$ determines the spectrum decay exponent $1-\delta$. This result has been derived previously in a different context \cite{fouxon}. We see that the formula for $D(k)$ is inconsistent at large $k$ giving negative dimension. The inconsistency of lognormal approximation to multifractality for $D(k)$ at large orders was observed by Mandelbrot in \cite{mandel} and by Frisch and Parisi in \cite{frischparisi}. The reason for the breakdown of the lognormal approximation at these orders is that the corresponding moments of the mass are determined by tails of the PDF of the logarithm of the density. The tails are determined by rare events for which the lognormal approximation breaks down. For avoiding misunderstanding we talk here of the PDF of finitely resolved density that determines the fractal dimensions. The single-point PDF of the density can be strictly lognormal without contradictions. 
We remark also that lognormality implies that a box counting dimension $D(0)$ equals the space dimension three which makes this dimension less interesting.

We have $\delta>0$ for ${\rm Ma}>{\rm Ma}_{cr}$ where all the dimensions that are describable by a lognormal approximation are strictly smaller than three, $3-D(k)=k\delta/2$. This is the reason why in defining the phase transition in the beginning of this Section we have not specified which of $D(k)$ determines the transition to multifractality. Similar fact holds for the concentration: if intermittency is neglected then all $D(k)$ become smaller than three at ${\rm Ma}_s$ where the flow becomes self-similar (intermittency would introduce the corresponding refinements). However, the dimensions $D(k)$
with large $k$, that are not describable by the lognormal approximation, could become smaller than three at ${\rm Ma}$ different from ${\rm Ma}_{cr}$. Since $D(k)$ is a non-increasing function of $k$ then they could become smaller than three at ${\rm Ma}<{\rm Ma}_{cr}$. Consideration of fractal dimensions of non-isothermal turbulence where lognormality does not hold \cite{vasquez,Passot1998,fed2015} is beyond the scope of this work.

If it is found that the statistics of the concentration and the density differ then this would make the measurement of the usually used Hentschel-Procaccia (HP) fractal dimensions of the density challenging. We would not have the representation of the multifractal support of the density as the limit of large number of discrete particles spread over the multifractal that is used in the definition \cite{hp}. Indeed, if we spread particles in space then after transients they would be located on the multifractal support of the natural measure and not the density. This and other problems however do not arise if the R\'{e}nyi dimensions are employed. These are more suitable for working with continuum fields and their use in numerical studies seems to be advantageous. These dimensions are derived from the usual spatial moments of the coarse-grained density and coincide with the HP dimensions where both are well defined. We provide the proof of the fundamental property that the information dimension in the supersonic inertial range describes the scaling of the mass up to points with zero total mass. This proof is needed because the flow is not smooth as in the usually considered situations.

We compare the statistics of the density and the concentration. The fields coincide at small Mach numbers where they are reducible in pseudo-sound regime to the statistics of the pressure of incompressible turbulence. 
Other regimes are possible, see e. g. \cite{bayly} for the theory and \cite{gotoh} for observations. The statistics in pseudo-sound regime is not lognormal since the pressure is not. For the positive tail, however, lognormality was observed \cite{dj}. When ${\rm Ma}$ increases the concentration and density can become different. The concentration
transits to multifractality when the scaling of the solenoidal and potential components of velocity becomes approximately equal. For the density
of isothermal turbulence the increase of ${\rm Ma}$ causes transition to lognormal statistics.  When the statistics is already lognormal, further increase of ${\rm Ma}$ induces transition to multifractality when the decay exponent of the density spectrum passes one from above. This transition occurs simultaneously with the concentration's transition or when the scaling of the solenoidal and potential components of velocity is already approximately equal and the concentration is multifractal.


We demonstrated that the collision kernel of the particles is significantly increased due to the transition to multifractality. The transition is thus expected to have dramatic consequences on formation of stars and other processes with coagulation of particles in high $Re$ and high ${\rm Ma}$ compressible turbulence.

Our derivation of the spectrum of the fractal dimensions of the density in terms of one parameter $\delta$ calls for the construction of a phenomenology of the turbulence that would provide us with the scaling exponents of the velocity. Indeed, it could seem that if we use the assumption that $\rho^{1/3}\bm v$ has scaling identical with incompressible velocity \cite{krit2007} then we could predict the scaling of the velocity in terms of $\delta$. We demonstrated that this is not straightforward if doable at all. Construction of a prediction for the velocity scaling that includes multifractality of density is one of the challenges for future work.

Particles transported by the flow have inertia that was neglected in our study, cf. \cite{dust}. The limit of zero inertia is singular in incompressible turbulence, see e. g. \cite{nature}. We briefly consider it in the compressible case. The limit is not singular in the inertial range as can be confirmed by using the Kraichnan model. In the viscous range this may be different because the sum of the Lyapunov exponents is no longer zero. When inertia is small the particle coordinate $\bm x(t)$ obeys \cite{maxey},
\begin{eqnarray}&&\!\!\!\!\!\!\!\!\!\!
\dot{\bm x}=\bm u(t, \bm x(t)), \ \  \bm u=\bm v-\tau_s\bm a,
\end{eqnarray}
where $\tau_s$ is the Stokes time that appears in the law of motion $\tau_s \ddot{\bm x}=\bm v(t, \bm x(t))-\dot{\bm x}$ and $\bm a=\partial_t\bm v+(\bm v\cdot\nabla)\bm v$ is the Lagrangian accelerations of the fluid. The assumption of linear friction holds provided the Reynolds number of the perturbation of the flow caused by the presence of the particle is small. It can be corrected \cite{dust}. In the leading order at small $\tau_s$ we have for the sum of the Lyapunov exponents $\lambda_i^p$ of the flow of the particles using the formula $\sum_{i=1}^3 \lambda^p_i=-\int_0^{\infty} \left\langle \nabla\cdot\bm u(0) \nabla\cdot\bm u(t)\right\rangle dt$ of \cite{ff},
\begin{eqnarray}&&\!\!\!\!\!\!\!\!\!\!
\sum_{i=1}^3 \lambda^p_i\approx
\tau_s\int_0^{\infty} \left(\left\langle \nabla\cdot\bm a(0) w(t)\right\rangle + \left\langle w(0) \nabla\cdot\bm a(t)\right\rangle\right) dt
\nonumber\\&&\!\!\!\!\!\!\!\!\!\!
-\int_0^{\infty} \left\langle w(0) \left(\delta \bm x(t)\cdot \nabla \right)w(t)\right\rangle dt,
\end{eqnarray}
where we used the result that the sum of the Lyapunov exponents of the fluid particles is zero. The last term describes the change in the correlation function due to the finite deviation $\delta \bm x(t)$ of the trajectory of the particle from the trajectory of the fluid particle, cf. \cite{grs}. Thus $\sum_{i=1}^3 \lambda^p_i\neq 0$ and particles' distribution is multifractal up to the smallest scales (finite size of the particles). This can produce non-negligible correction in the collision kernel of dust particles that are smaller than the viscous length.

Our consideration of the statistics of the concentration was independent of the origin of the velocity. Thus all our considerations go through also for transport of tracers by the velocity that comes from the solution of the magnetohydrodynamic equations (MHD). The concentration will become multifractal at ${\rm Ma}_s$ at which the scalings of solenoidal and potential components of the MHD velocity become similar. It is plausible that the fluid density is not lognormal for isothermal MHD since the magnetic stress tensor breaks the invariance with respect to multiplication of the density by a multiplicative constant. This question requires numerical studies.

Finally, the account of intermittency, necessary at very large Reynolds numbers and hardly doable with the presently existing knowledge or data, probably modifies the condition for the multifractality transition of the concentration. The condition of identical scaling of the solenoidal and the potential components of the flow becomes the condition of identical scaling of structure functions of the components with a given order.

\section*{Acknowledgments}

I. F. thanks S. Tcheremchantsev for helpful discussions. This work was supported by grant no. 366/15 of the Israel Science Foundation.

{}

\newpage
\begin{appendices}
\appendix
\section{Constraints due to density finiteness}\label{a}

Here we complete the study of Section \ref{setting} of the kinematic consequences of the condition that the flow does not generate infinite density. We start from demonstrating that zero sum of Lyapunov exponents implies zero rate of production of the Gibbs entropy \cite{ff,ruelle}, and vice versa. In order to show that we employ the tracers' mass conservation in the form $n(0, \bm x)d\bm x\!=\!n[t, \bm q(t, \bm x)]d\bm q(t, \bm x)$ (this consideration is more transparent in terms of tracers' concentration $n(t, \bm x)$),
\begin{eqnarray}&&\!\!\!\!\!\!\!\!\!\!\!\!\!\!\!
S(t)=-\int  n(0, \bm x)\ln n [t, \bm q(t, \bm x)]d\bm x=S(0)+ \nonumber\\&&\!\!\!\!\!\!\!\!\!\!\!\!\!\!\!\int_0^t \!\!dt'd\bm x n(0, \bm x) w [t', \bm q(t', \bm x)],\label{integral}
\end{eqnarray}
where we used Eq.~(\ref{solv}) for the concentration. For initial condition $n(0, \bm x)=1$ we find, using the independence of $\bm x$ of the sum of the Lyapunov exponents defined in Eq.(\ref{sums}) for ergodic systems,
\begin{eqnarray}&&\!\!\!\!\!\!\!\!\!\!\!\!\!\!\!
\lim_{t\to\infty}\frac{S(t)-S(0)}{t}=\sum_{i=1}^3\lambda_i.
\end{eqnarray}
However $S(t)<S(0)$ since the entropy is maximal for spatially constant $n$. We find the inequality $\sum_{i=1}^3\lambda_i\leq 0$ which can be seen as a form of the second law of thermodynamics \cite{ff,ruelle}. Thus compressible turbulent flow below the viscous scale is degenerate realizing the equality $\sum_{i=1}^3\lambda_i=0$ in the general inequality $\sum_{i=1}^3\lambda_i\leq 0$. Other useful observation is found by writing Eq.~(\ref{integral}) in the following form (we use interchangeably $\rho$ and $n$ since the conclusions hold for both of them),
\begin{eqnarray}&&\!\!\!\!\!\!\!\!\!\!\!\!\!\!\!
S(t)=S(0)+\int_0^t \langle \rho(t') w(t')\rangle dt'.
\end{eqnarray}
We find that Eq.~(\ref{per}) must hold in the steady state where $S(t)$ and $\langle \rho(t) w(t)\rangle$ are constant. We use that averaging of the continuity equation in steady state results in $\langle \rho w\rangle=-\langle \bm v\cdot\nabla\rho\rangle$.

{\it Anticorrelations of $w(t, \bm q(t, \bm x))$.}---The vanishing of the time average of $w(t, \bm q(t, \bm x))$ (see eq.(8)) means that the velocity divergence is anticorrelated along the Lagrangian trajectories $\bm q(t, \bm x)$. These anticorrelations can be given more detailed form. We use the identity \cite{ff,fouxon1},
\begin{eqnarray}&&\!\!\!\!\!\!\!\!\!\!\!\!\!\!\!\!\!\!\!
\langle w [t, \bm q(t, \bm x)]\rangle\!\equiv\!\!\int\!\! w [t, \bm q(t, \bm x)]d\bm x\!=\!-\!\!\int_0^{t}\!\!\langle w(0)w(t')\rangle dt'. \label{sd}
\end{eqnarray}
where $\langle w(0)w(t)\rangle=\int w(0, \bm x)w [t, \bm q(t, \bm x)]d\bm x$. The above relationship holds for time-independent 
or stationary, but not necessarily restricted to spatially homogeneous, flows, at $t>0$ or $t<0$. The term on the right hand side of Eq. ~{(\ref{sd})} in the limit $t\to\pm\infty$ may be calculated now by integrating Eq.~(\ref{sums}) over $\bm x$. This yields,
\begin{eqnarray}&&\!\!\!\!\!\!\!\!\!\!\!\!\!\!\!\!
-\sum_{i=1}^3\!\lambda_i\!=\!\lim_{t\to\infty}  \int_0^t\frac{dt'}{t}\int_0^{t'} dt''\!\langle w(0)w(t'')\rangle
\nonumber\\&&\!\!\!\!\!\!\!\!\!\!\!\!\!\!\!\!
=\int_0^{\infty}\langle w(0)w(t)\rangle dt=0,\label{va}
\end{eqnarray}
and similarly $\int_{-\infty}^0\!\langle w(0)w(t)\rangle dt\!=\!0$. 
We assumed that $\langle w(0)w(t)\rangle$ decays faster than $1/t^2$ so the integral of $t\langle w(0)w(t)\rangle$ is finite.

The anticorrelation property of the divergence may be demonstrated yet in another way. We observe that at large times $w(t, \bm q(t, \bm x))$ becomes a stationary process with zero average (where the average is the sum of the Lyapunov exponents). We introduce the asymptotic correlation functions,
\begin{eqnarray}&&\!\!\!\!\!\!\!\!\!\!\!\!\!\!
f_{\pm}(t)= \lim_{t'\to\pm\infty} \left\langle w(t', \bm q(t', \bm x))w(t'\!+\!t, \bm q(t'\!+\!t, \bm x))\right\rangle.\label{functionsd}
\end{eqnarray}
Starting again from Eq.~(\ref{solv}) we obtain that (here $\bm q(t)=\bm q(t, \bm x)$),
\begin{eqnarray}&&\!\!\!\!\!\!\!\!\!\!\!\!
\left\langle\!\ln^2\!\left(\frac{\rho(0)}{\rho(t, \bm q(t))}\right)\!\right\rangle
\!=\!\!\int_{0}^t\!\!dt_1dt_2\!\left\langle w(t_1, \bm q(t_1))w(t_2, \bm q(t_2))\right\rangle.\nonumber
\end{eqnarray}
In the limit of large $t$ the RHS becomes approximately $t\int_{-\infty}^{\infty} f_+(t') dt'$. Since the left hand side of the last equation is finite then we must have (we use that $t$ above can be negative),
\begin{eqnarray}&&\!\!\!\!\!\!\!\!\!\!\!\!\!\!
\int_{-\infty}^{\infty}\!\!\! f_{\pm}(t)dt=\!0.\label{functions}
\end{eqnarray}
Thus correlation functions of the stationary limiting processes $w(t, \bm q(t, \bm x))$ at $t\to\pm\infty$ have zero integrals. This brings a simple representation for Lagrangian differences of the density.

{\it The first moment of $\langle w(0)w(t)\rangle$ is the asymptotic logarithmic increment of the density.}---The difference in the fluid particle velocity at two different times (Lagrangian increments) is a much studied characteristics of turbulence, see e. g. \cite{lagrd}. We consider the Lagrangian increments of the density. We observe that the random variables,
\begin{eqnarray}&&\!\!\!\!\!\!\!\!\!\!\!\!\!
I=\int_{-\infty}^0 w(t, \bm q(t, \bm x))dt,\ \ I_+=\int_0^{\infty} w(t, \bm q(t, \bm x))dt.\label{df}
\end{eqnarray}
that involve infinite integration range are well-defined as $\langle I^2\rangle$ and $\langle I_+^2\rangle$ are finite by Eq.~(\ref{functions}). Using Eqs.~(\ref{sd})-(\ref{va}),
\begin{eqnarray}&&\!\!\!\!\!\!\!\!\!\!\!\!\!\!
\langle I\rangle=\lim_{T\to\infty}\int_{-T}^0 \langle w(0) w(t)\rangle (t+T)dt=\lim_{T\to\infty}\int_{-T}^0  t dt \nonumber\\&&\!\!\!\!\!\!\!\!\!\!\!\!\!\!\times
\langle w(0) w(t)\rangle
-T\int_{-\infty}^{-T} \left\langle w(0)w(t)\right\rangle dt.
\end{eqnarray}
Assuming that $\left\langle w(0)w(t)\right\rangle$ decays faster than $1/t^2$ the last integral vanishes at $T\to\infty$. Then using Eq.~(\ref{solv}) we find for the difference of $\langle \ln \rho\rangle$,
\begin{eqnarray}&&\!\!\!\!\!\!\!\!\!\!\!\!
\langle \ln \rho\rangle\!-\!\!\!\!\lim_{t\to-\infty}\langle \ln \rho(t, \bm q(t, \bm x))\rangle\!\!=\!\!\!\int_{-\infty}^0\!\!\!\! |t|
\langle w(0) w(t)\rangle dt\!=\!-\langle I\rangle,\nonumber\\&&\!\!\!\!\!\!\!\!\!\!\!\! \langle \ln \rho\rangle\!\!-\!\!\!\lim_{t\to\infty}\langle \ln \rho(t, \bm q(t, \bm x))\rangle\!\!=\!\!\!\int_0^{\infty}\!\!\! \!\! t
\langle w(0) w(t)\rangle dt\!=\!\langle I_+\rangle.\label{asympt}
\end{eqnarray}
In the integral for $\langle I_+\rangle$ the factor $t$ gives smaller weight to positive small time values of $\langle w(0) w(t)\rangle$ and larger weight to negative larger time values that exist due to the constraint $\int_0^{\infty} \langle w(0) w(t)\rangle dt=0$ (see Eq.~(\ref{va}) above) so $\langle I_+\rangle<0$. Similarly $\langle I\rangle>0$.
Therefore, the average difference of $\langle \ln \rho\rangle$ is the first moment of $\langle w(0) w(t)\rangle$ while the zeroth moment is zero.

\section{Pair-correlation function and RDF} \label{rdf}

Here we provide in more detail the definition of the RDF and the relation between the time averaging and averaging over the statistical ensemble of velocity fields. The concentration field of $N$ particles in the unit volume with coordinates $\bm x_i(t)$ is $n(t, \bm x)\!=\!\sum_{i=1}^N \delta(\bm x_i(t)\!-\!\bm x)/N$. We introduced the normalization factor so that $\langle n\rangle=1$ independently of $N$. The radial distribution function (RDF) $g(\bm r)$ is defined as,
\begin{eqnarray}&&\!\!\!\!\!\!\!\!\!\!\!\!\!\!\!
g(\bm r)\!=\!\langle n(0)n(\bm r)\rangle
\!=\!\!\!\lim_{N\to\infty}\!\frac{\sum_{i, k=1}^N \delta(\bm x_k(t)\!-\!\bm x_i(t)\!-\!\bm r)}{N^2}, \label{s}
\end{eqnarray}
where $N\to\infty$ is the continuum limit and the formula is readily verified using the definition $\langle n(0)n(\bm r)\rangle
\!=\!\int \!  n(t, \bm x)n(t, \bm x\!+\!\bm r) d\bm x$. Thus the RDF counts the fraction of pairs separated by given distance $\bm r$ and is normalized so that it tends to one at large $r$.

Since $f(\bm r)$ is a two-particle quantity then it can be derived from the problem of two particles only in the flow. We perform time averaging of Eq.~(\ref{s}). Ergodicity of the flow in the two-particle phase space (that is of the flow $(\bm v(\bm x_k), \bm v(\bm x_i))$  implies equality of time averages of all pairs giving,
\begin{eqnarray}&&\!\!\!\!\!\!\!\!\!\!\!\!\!\!
g(\bm r)=\left\langle\delta(\bm x_k(t)-\bm x_i(t)-\bm r)\right\rangle_t,
\end{eqnarray}
where the subscript $t$ stands for time-averaging. We can consider time average over arbitrarily large time $T$ as average over $N$ results of time averaging over time intervals $t_0(i, i+1)$ where $N=T/t_0$ and $i$ runs from zero to $N-1$. If $t_0$ is large then we can consider flows at disjoint time intervals as independent. Thus we can instead perform averaging over flows at different time intervals or equivalently we can find $g(\bm r)$ as average over ensemble of realizations of the flow. In this formulation we consider two particles put in the flow with some initial positions $\bm x_1$ and $\bm x_1+\bm r'$. We study the evolution of the distance $\bm r(t)=\bm q(t, \bm x_1+\bm r')-\bm q(t, \bm x_1)$ between them. Different realizations of the flow produce different evolution of $\bm r(t)$ defining the PDF of the distance $P(\bm r, \bm r', t)$ as given in Eq.~(\ref{propog}) of the main text. In that equation
the angular brackets stand also for averaging over the statistics of the flow. The limiting PDF is $\bm r'-$independent and gives $g(\bm r)$ in Eq.~(\ref{pdf}) of the main text. This
can be readily verified by inspection of the representation of $g(\bm r)$ as average over time averages over intervals with arbitrarily large length $t_0$.

\section{Kraichnan model} \label{krm}

We stressed in the main text that our usage of the Kraichnan model is understood as a model that reproduces the long-time asymptotic form of the propagator at least qualitatively. We consider here how the model is gauged so that the propagators produced by the NS flow and the model are similar. 

We consider the demand that the model reproduces the long-time behavior of the dispersion given by \cite{review},
\begin{eqnarray}&&\!\!\!\!\!\!\!\!\!\!\!
\left\langle \left(\bm r(t)\!-\!\bm r(0)\right)^2\right\rangle\!=\!2\int_0^t \!\!dt_1\int_0^{t_1}\!\! dt_2\left\langle  \delta \bm v(t_1)\! \cdot\! \delta \bm v(t_2)\right\rangle,
\end{eqnarray}
where $\delta \bm v(t)\equiv \bm v(t, \bm q(t, \bm x+\bm r))-\bm v(t, \bm q(t, \bm x))$ and we used,
\begin{eqnarray}&&\!\!\!\!\!\!\!\!\!\!\!
\bm r(t)\!=\!\bm q(t, \bm x\!+\!\bm r)\!-\!\bm q(t, \bm x)\!=\!\bm r(0)\!+\!\int_0^t \delta \bm v(t')dt'.
\end{eqnarray}
Taking the time derivative and using that eddies of scale $r$ have the correlation time $t_r\sim r/\delta v_r$ we have, 
\begin{eqnarray}&&\!\!\!\!\!\!\!\!\!\!\!
\frac{d}{dt}\left\langle \left(\bm r(t)\!-\!\bm r(0)\right)^2\right\rangle\!\sim \!\!\int_{-\infty}^t\!\!\! \left\langle  \delta \bm v(t)\! \cdot \!\delta \bm v(t')\right\rangle dt'
\!\sim\! \delta v_r^2 t_r\!\sim\! r \delta v_r,\nonumber 
\end{eqnarray}
where we do not distinguish the solenoidal and potential components of the flow considering at the moment the multifractal phase where the components scale similarly. We demand that the white noise in time velocity of the Kraichnan model $\bm u$ produces the same long-time growth of the dispersion as implied by the law above. We have for the dispersion in the white noise model,
\begin{eqnarray}&&\!\!\!\!\!\!\!\!\!\!\!
\frac{d}{dt}\left\langle \left(\bm r(t)\!-\!\bm r(0)\right)^2\right\rangle\!= \!\int_{-\infty}^t\!\!\! \left\langle  \delta \bm u(t)\! \cdot \!\delta \bm u(t')\right\rangle dt'\!\sim r^{\xi},
\end{eqnarray}
where we used that $K_{ik}(\bm r)\propto r^{\xi}$, see Section \ref{Kraic}. Thus we fix the value of $\xi$ by the demand that in the multifractal phase $r \delta v_r\sim r^{\xi}$. If the solenoidal component of the Navier-Stokes compressible turbulence has a spherically normalized spectrum proportional to $k^{-a}$ then the velocity scales in space as $r^{(a-1)/2}$ giving $\xi=(a+1)/2$.

The above consideration disregards the intermittency of the flow which is not a bad assumption as discussed in subsection \ref{inetr}. Thus if the growth of $r(t)$ at large times is self-similar with a good approximation (which it must be in the multifractal phase, cf. the incompressible turbulence case \cite{review}) then the Kraichnan model, where the growth is self-similar, will reproduce the law of growth of the distance up to a multiplicative constant. This overall constant of proportionality is however of less interest to us since it does not enter the scaling exponent of the pair-correlation function $\beta$. This exponent 
is roughly the ratio of the magnitudes of the potential and solenoidal components, see the main text. 

We see that the spatial scaling of $\bm u$ is different from the scaling of $\bm v$ which is so also in the incompressible case \cite{review} and below. Since zero correlation time results in effective Gaussianity \cite{risken} then the statistics is taken Gaussian with zero mean. The statistics is completely determined by the pair-correlation function,
\begin{eqnarray}&&\!\!\!\!\!\!\!\!\!\!\!
\left\langle u_i(t, \bm x)u_k(t', \bm x')\right\rangle=\delta(t-t')D_{ik}(\bm r),
\end{eqnarray}
where $\bm r=\bm x'-\bm x$. It is assumed that the statistics is stationary, spatially uniform and isotropic. Thus the Fourier transform of $D_{ik}(\bm r)$ has the following general form (${\hat k}=\bm k/k$),
\begin{eqnarray}&&\!\!\!\!\!\!\!\!\!\!\!
D_{ik}(\bm k)=f(k)\left(\delta_{ik}-{\hat k}_i{\hat k}_k\right)+h(k){\hat k}_i{\hat k}_k,
\end{eqnarray}
with arbitrary functions $f(k)$ and $h(k)$. In this model the symmetries imply that the solenoidal $\bm s$ and potential components $\bm p$ of the flow, $\bm u=\bm s+\bm p$, are independent,
\begin{eqnarray}&&\!\!\!\!\!\!\!\!\!
\langle s_i(t, \bm k)s_k(t', \bm k')\rangle\!=\!8\pi^3\delta(t'\!-\!t)\delta(\bm k\!+\!\bm k')f(k)\!\left(\!\delta_{ik}\!-\!{\hat k}_i{\hat k}_k\!\right)\nonumber,\\&&\!\!\!\!\!\!\!\!\!
\langle p_i(t, \bm k)p_k(t', \bm k')\rangle\!=\!8\pi^3\delta(t'\!-\!t)\delta(\bm k\!+\!\bm k')h(k){\hat k}_i{\hat k}_k. \label{77}
\end{eqnarray}
Thus $f(k)$ and $h(k)$ represent the spectra (not normalized spherically) of the solenoidal and potential components, respectively. We stress that these are not the spectra of the components, these are only their representations that have scaling different from the scalings of the spectra of the components of the NS flow. We fix the scalings of $f(k)$ and $h(k)$ by extension of the procedure that we used above for fixing $\xi$ in 
the multifractal phase. We demand that time integrals of the different time pair correlation functions of the solenoidal and potential components of the NS flow coincide with their counterparts for the Kraichnan model, see \cite{review}. This condition guarantees that the model reproduces the impact of these components on pair dispersion separately which is necessary for discussion of the concentration which is influenced by the components differently. Thus considering as previously that the solenoidal component of the Navier-Stokes compressible turbulence has a spherically normalized spectrum proportional to $k^{-a}$ we find that the solenoidal component 
of $\bm u$ must scale as $r^{(a+1)/2}$, resulting in $f(k)\sim k^{-3-(a+1)/2}$. Similarly if the spherically normalized spectrum of the potential component is proportional to $k^{-b}$ then $h(k)\sim k^{-3-(b+1)/2}$. These scaling relations refer to the supersonic inertial range. We observe that $f(k)$ and $h(k)$ depend on the temporal correlations of turbulence and not only on the spectra of the components of the turbulent flow that characterize the instantaneous statistics. Therefore the proportionality constants in $f(k)\sim k^{-3-(a+1)/2}$ and $h(k)\sim k^{-3-(b+1)/2}$ are non-trivial functionals of the spatio-temporal statistics of turbulence. The ratio of these constants, that defines $\beta$ as demonstrated in the main text, is roughly the ratio of the components' spectra at zero frequency. Thus $\beta$ can be considered as a ratio of magnitudes of potential and solenoidal components however providing this ratio in terms of instantaneous statistics of turbulence is impossible. This is in contrast with the scaling exponents for which the temporal behavior is fixed by the robust relation $t_r\sim r/\delta v_r$ implied by the NSE. 

We find in real space:
\begin{eqnarray}&&\!\!\!\!\!\!\!\!\!
\langle s_i(t, \bm x)s_k(t', \bm x')\rangle\!=\!\delta(t'\!-\!t)\left[\delta_{ik}f(r)+\nabla_i\nabla_kf_1(r)\right]\nonumber\\&&\!\!\!\!\!\!\!\!\!
=\delta(t'\!-\!t)\left[\frac{r_ir_k}{r}(f_1'/r)'-\delta_{ik}\frac{f_1'+r(f_1)''}{r}\right],
\end{eqnarray}
where $f(r)$ is the inverse Fourier transform of $f(k)$,
\begin{eqnarray}&&\!\!\!\!\!\!\!\!\!\!\!\!\!\!\!
f(r)\!=\!\!\int \!\! f(k)\exp(-i\bm k\!\cdot\!\bm r)\frac{d\bm k}{8\pi^3}\!=\!\!\int_0^{\infty}\!\!\frac{f(k)\sin(kr)k dk}{2\pi^2 r}.
\end{eqnarray}
We introduced the function $f_1(r)$,
\begin{eqnarray}&&\!\!\!\!\!\!\!\!\!
f_1(r)\equiv \int_0^{\infty} \frac{f(k)dk}{2\pi^2} \frac{\sin(kr)}{kr},\ \
f(r)=-\frac{(rf_1)''}{r}. \label{dl}
\end{eqnarray}
We introduce the function $u_0(r)$,
\begin{eqnarray}&&\!\!\!\!\!\!\!\!\!\!\!\!\!\!
u_0r^2\!=\!\frac{2f_1'}{r}\!+\!v_0,\ \ \frac{f_1'\!+\!r(f_1)''}{r}\!=\!-v_0\!+\!\frac{(r^4u_0)'}{2r},
\end{eqnarray}
where the constant $v_0$ is taken so that $u_0(0)$ has regular Taylor expansion at the smallest $r$ where the viscosity smoothens the flow,
\begin{eqnarray}&&\!\!\!\!\!\!\!\!\!
v_0=-2f_1''(r=0)=\frac{2f(r=0)}{3},
\end{eqnarray}
cf. \cite{arxiv}. The contribution of the potential component to the pair correlation function in real space is given by:
\begin{eqnarray}&&\!\!\!\!\!\!\!\!
\langle p_i(t, \bm x)p_k(t', \bm x')\rangle\!=\!-\delta(t'\!-\!t)\left(\frac{\delta_{ik} h_1'(r)}{r}+\frac{r_ir_k (h_1'/r)'}{r}\right),\nonumber\\&&\!\!\!\!\!\!\!\! h_1=\int_0^{\infty} \frac{h(k)dk}{2\pi^2} \frac{\sin(kr)}{kr}, \ \ h=-\frac{(rh_1)''}{r }=-h_1''-\frac{2h_1'}{r}.\nonumber
\end{eqnarray}
Adding up both contributions we find:
\begin{eqnarray}&&\!\!\!\!\!\!\!\!\!
\langle u_i(t, \bm x)u_k(t', \bm x')\rangle\!=\!\nonumber\delta(t'\!-\!t)\left[\frac{\left[(r^2u_0)'-2(h_1'/r)'\right]r_ir_k}{2r}\right.\nonumber\\&&\!\!\!\!\!\!\!\!\!\left.+v_0\delta_{ik}
-\frac{\left[(r^4u_0)'+2h_1'(r)\right]\delta_{ik}}{2r}\right].
\end{eqnarray}
Finally, we introduce,
\begin{eqnarray}&&\!\!\!\!\!\!\!\!\!
u=u_0+\frac{h_1''(r)-h_1''(0)}{r^2},\ \ c=-\frac{h'}{r},\ \ V_0=v_0-h_1''(0),\nonumber
\end{eqnarray}
where $u(r)$ and $c(r)$ that have regular Taylor expansion in the viscous range and $-3h_1''(0)=h(0)$. Using these functions and $V_0$ we reproduce the correlation function in the form that was used in \cite{arxiv} and is given by Eqs.~(\ref{mdl})-(\ref{tensor}) from the main text.

We consider in more detail the form of the functions above in the real space. We can easily see from $f(k)\propto k^{-3-(a+1)/2}$ that,
\begin{eqnarray}&&\!\!\!\!\!\!\!\!
f(0)-f(r)=\theta_1 r^{(a+1)/2},\ \ r\ll L.\label{dd}
\end{eqnarray}
It is seen by performing inverse Fourier transform of the first of Eqs.~(\ref{77}),
\begin{eqnarray}&&\!\!\!\!\!\!\!\!\!
\left\langle \left(s_i(t, \bm r)-s_i(t, 0)\right)\left(s_i(0, \bm r)-s_i(0, 0)\right)\right\rangle
\\&&\!\!\!\!\!\!\!\!\!
=\!\delta(t)\int\!\! f(k)\left(1\!-\!\exp\left(i\bm k\!\cdot\!\bm r\right)\right)\frac{d\bm k}{2\pi^3}\!=\!4\delta(t)\left(f(0)\!-\!f(r)\right),\nonumber
\end{eqnarray}
that $\theta_1$ is a positive constant characterizing the magnitude of the solenoidal component, cf. \cite{krs}. Similarly we have from $h(k)\propto k^{-3-(b+1)/2}$ that,
\begin{eqnarray}&&\!\!\!\!\!\!\!\!
h(0)-h(r)=\theta_2 r^{(b+1)/2},\ \ r\ll L,
\end{eqnarray}
where $\theta_2$ is a positive constant characterizing the magnitude of the potential component. We assumed that both $a$ and $b$ change between one and three. Indeed, the decay exponent of the spectrum of the solenoidal component changes between about the Kolmogorov value $5/3$ at small Mach numbers to probably the Burgers equation's value $2$ at large Mach numbers. Similarly, the decay exponent of the spectrum of potential component changes between about the incompressible turbulence's pressure spectrum exponent's value of $3$ at ${\rm Ma}\ll 1$ to the same Burgers equation's value $2$ at large Mach numbers.

We have from the definitions,
\begin{eqnarray}&&\!\!\!\!\!\!\!\!\!
\frac{c}{ru}=-\frac{h'}{2(f_1'(r)/r-f_1''(0))+h_1''(r)-h_1''(0)}.\label{cru}
\end{eqnarray}
We observe from Eqs.~(\ref{dl}) and (\ref{dd}) that,
\begin{eqnarray}&&\!\!\!\!\!\!\!\!\!
(rf_1)''=r[f(0)-f(r)-f(0)]=\theta_1 r^{(a+3)/2}-rf(0).\nonumber
\end{eqnarray}
We find integrating this equation twice and demanding regularity of $f_1$ at small $r$ implied by the definition in Eq.~(\ref{dl}) that,
\begin{eqnarray}&&\!\!\!\!\!\!\!\!\!
f_1=\frac{4\theta_1 r^{(a+5)/2}}{(a+5)(a+7)}-\frac{r^2 f(0)}{6}+c_f,
\end{eqnarray}
where $c_f$ is a constant. This gives,
\begin{eqnarray}&&\!\!\!\!\!\!\!\!\!
\frac{f_1'}{r}-f_1''(0)=\frac{2\theta_1 r^{(a+1)/2}}{a+7}.
\end{eqnarray}
Similarly we have,
\begin{eqnarray}&&\!\!\!\!\!\!\!\!\!
(rh_1)''=r[h(0)-h(r)-h(0)]=\theta_2 r^{(b+3)/2}-rh(0),\nonumber\\&&\!\!\!\!\!\!\!\!\!
h_1=\frac{4\theta_2 r^{(b+5)/2}}{(b+5)(b+7)}-\frac{r^2 h(0)}{6}+c_h,
\end{eqnarray}
where $c_h$ is a constant. This gives,
\begin{eqnarray}&&\!\!\!\!\!\!\!\!\!
h_1''(r)-h_1''(0)=\frac{(b+3)\theta_2 r^{(b+1)/2}}{b+7}.
\end{eqnarray}
We find from Eq.~(\ref{cru}) using the formulas above that,
\begin{eqnarray}&&\!\!\!\!\!\!\!\!\!
\frac{c}{ru}=\frac{(b+1)(b+7)(a+7)\Gamma' r^{\Delta-1} {\tilde L}^{-\Delta}}{8(b+7)+2(b+3)(a+7)\Gamma' (r/{\tilde L})^{\Delta}},\nonumber
\end{eqnarray}
where $\Delta=(b-a)/2$. We introduced dimensionless constant $\Gamma'=\theta_2 {\tilde L}^{\Delta}/\theta_1$ with scale ${\tilde L}$ of order $L$. This scale is defined as the effective upper cutoff of the inertial range. The equation above, derived from the asymptotic power laws in the inertial range, holds below ${\tilde L}$ so that Eq.~(\ref{frm}) from the main text is true at $r\ll L$. This constant gives the ratio of the structure functions of the potential and solenoidal component at the scale $L$. Since at these scales the structure functions are approximately equal to the dispersion of the respective velocity component then $\Gamma'$ is roughly the ratio of magnitudes of the potential and solenoidal components.

\section{Supercritical transport} \label{krms}

Here we confirm the prediction that the scaling exponent of the pair-correlation function of the concentration determines the scaling of $P(r, r', \tau)$ at small $r$ also at finite $r'$ in accord with Eq.~(\ref{predictions}). We denote by $P(r, \bm r', |t|)$ the PDF  $P(\bm r, \bm r', |t|)$ averaged over all directions of $\bm r$. Isotropy implies that $P(r, \bm r', |t|)$ is independent of the direction of $\bm r'$ so $P(r, \bm r', |t|)=P(r, r', |t|)$. The angle-averaged PDF obeys a closed equation whose solution can be written as \cite{phase,krho},
\begin{eqnarray}&&\!\!\!\!\!\!\!\!\!\!\!
P(r, r', \tau)\!=\!\frac{(r r_0)^{(\beta-\xi-1)/2}\Gamma(1\!-\!b_0)}{ |\tau|^{b_0} (2\!-\!\xi)^{2b_0}}
I_{-b_0}\left(\frac{2(rr')^{(2-\xi)/2}}{(2-\xi)^2|\tau|}\right)\nonumber\\&&\!\!\!\!\!\!\!\!\!\!\!
\exp\left(-\frac{r'^{2-\xi}}{(2\!-\!\xi)^2|\tau|}\right)f_s(r, \tau) 
,\label{green}
\end{eqnarray}
where $I_{-b_0}$ is the modified Bessel function of the first kind of index $-b_0$. It is readily seen that $P(r, r'\to 0, \tau)=f_s(r, \tau)$ as claimed previously. Quite similarly we have,
\begin{eqnarray}&&\!\!\!\!\!\!\!\!\!\!\!
P(r, r', \tau)\!\sim \exp\left(-\frac{r'^{2-\xi}}{(2\!-\!\xi)^2|\tau|}\right)f_s(r, \tau),
\end{eqnarray}
at small $r$. This confirms Eq.~(\ref{l}):
\begin{eqnarray}&&\!\!\!\!\!\!\!\!\!\!\!\!\!\!
P(r, L, t_L)L^3\sim f_s(r, c'_0t_L)L^3\sim \frac{(c'_0t_L)^{b_0-1}}{r^{\beta}}L^3\sim \left(\frac{L}{r}\right)^{\beta},\nonumber
\end{eqnarray}
where we used $(2-\xi)(b_0-1)=\beta-3$ and restored dimensional time by multiplying with $c'_0$.

For future reference we bring the formula for the moments $\langle r^k(t)\rangle=4\pi \int_0^{\infty} P(r, r_0, t)r^{2+k}dr$. Integration using Eq.~(\ref{green}) gives \cite{phase},
\begin{eqnarray}&&\!\!\!\!\!\!\!\!\!\!\!
\langle r^k(t)\rangle=\frac{\Gamma(k/(2-\xi)+1-b_0)}{\Gamma(1-b_0)} \left((2-\xi)^2|t|c_0'\right)^{k/(2-\xi)}\nonumber\\&&\!\!\!\!\!\!\!\!\!\!\!
F\left(-\frac{k}{2-\xi}, 1-b_0, -\frac{r_0^{2-\xi}}{(2-\xi)^2|t|c_0'}\right),\label{moments}
\end{eqnarray}
where $F(a, b, z)$ is confluent hypergeometric function. This holds for all the convergent moments that obey $k>\beta-3$ where $\beta-3<0$. This formula is provided in \cite{phase} only for $k\geq 0$. Moments of negative order become relevant in the curious hypothetical situation of weakly compressible flow with identical scaling of the solenoidal and potential components. In that case if inertial range is large we could get large fluctuations of the concentration however small compressibility is: the factor $(L/r)^{\beta}$ can get large for small fixed $\beta$ if $r$ is suitably small.

\end{appendices}

\end{document}